\documentclass[iop,twocolappendix]{emulateapj}
\slugcomment{Accepted to be published in ApJ}

\usepackage{amsmath}
\usepackage{aas_macros}
\usepackage{natbib}
\usepackage{xspace}
\usepackage[varg]{txfonts}
\usepackage{graphicx}
\usepackage{nicefrac}
\usepackage{tablefootnote}
\usepackage{multirow}
\usepackage{soul}
\usepackage{float}
\usepackage[T1]{fontenc}
\usepackage{lmodern}
 
\newcommand{\Msun}{\ensuremath{\,{\rm M}_\odot}\xspace}

\defcitealias{Zapartas+2017}{Z17}

\newcommand{\pievariable}{0.1622}
\newcommand{\onedvariable}{0.1622}

\begin{document}

\title{Predicting the Presence of Companions for Stripped-Envelope Supernovae: \\The Case of the Broad-Lined Type Ic SN 2002ap}

\author{E. Zapartas\altaffilmark{1}}
\author{S.\,E. de Mink\altaffilmark{1}}
\author{S.\,D. Van Dyk\altaffilmark{2}}
\author{O.\,D. Fox\altaffilmark{3}}
\author{N. Smith\altaffilmark{4}}
\author{K.\,A. Bostroem\altaffilmark{5}}
\author{A. de Koter\altaffilmark{1,6}}
\author{A.\,V. Filippenko\altaffilmark{7}}
\author{R.\,G. Izzard\altaffilmark{8}}
\author{P.\,L. Kelly\altaffilmark{7}}
\author{C.\,J. Neijssel\altaffilmark{9}}
\author{M. Renzo\altaffilmark{1}}
\author{S. Ryder\altaffilmark{10}}

\altaffiltext{1}{Anton Pannekoek Institute for Astronomy, University of Amsterdam, Science Park 904, 1098 XH, Amsterdam, The Netherlands.}
\altaffiltext{2}{Infrared Processing and Analysis Center, California Institute of Technology, 1200 E. California Boulevard, Pasadena, CA 91125, USA.}
\altaffiltext{3}{Space Telescope Science Institute, 3700 San Martin Drive, Baltimore, MD 21218, USA.} 
\altaffiltext{4}{Steward Observatory, University of Arizona, 933 N. Cherry Ave, Tucson, AZ 85721, USA.} 
\altaffiltext{5}{Department of Physics, University of California, Davis, CA 95616, USA.} 
\altaffiltext{6}{Institute of Astronomy, KU Leuven, Celestijnenlaan 200 D, B-3001 Leuven, Belgium} 
\altaffiltext{7}{Department of Astronomy, University of California, Berkeley, CA 94720-3411, USA.}
\altaffiltext{8}{Institute of Astronomy, Madingley Road, Cambridge, CB3 0HA, United Kingdom.}
\altaffiltext{9}{School of Physics and Astronomy, University of Birmingham, Edgbaston, Birmingham B15 2TT, UK.}
\altaffiltext{10}{Australian Astronomical Observatory, 105 Delhi Rd, North Ryde, NSW 2113, Australia.}

\begin{abstract}

Many young, massive stars are found in close binaries. Using population synthesis simulations we predict the likelihood of a companion star being present when these massive stars end their lives as core-collapse supernovae (SNe).  We focus on stripped-envelope SNe, whose progenitors have lost their outer hydrogen and possibly helium layers before explosion.  We use these results to interpret new {\it Hubble Space Telescope} observations of the site of the broad-lined Type~Ic SN 2002ap, 14 years post-explosion.
For a subsolar metallicity consistent with SN 2002ap, we expect a main-sequence companion present in about two thirds of all stripped-envelope SNe and a compact companion (likely a stripped helium star or a white dwarf/neutron star/black hole) in about 5\% of cases. About a quarter
of progenitors are single at explosion (originating from initially single stars, mergers or disrupted systems). All the latter scenarios 
require a massive progenitor, inconsistent with earlier studies of SN 2002ap. 
Our new, deeper upper limits exclude the presence of a main-sequence companion star $>8$--$10\Msun$, ruling out about 40\% of all stripped-envelope SN channels.  The most likely scenario for SN 2002ap includes nonconservative binary interaction of a primary star initially $\lesssim 23 \Msun$. Although unlikely ($<$1\% of the scenarios), we also discuss the possibility of an exotic reverse merger channel for broad-lined Type~Ic events.
Finally, we explore how our results depend on the metallicity and the 
model assumptions and discuss how additional searches for companions can constrain the physics that governs the evolution of SN progenitors.

\end{abstract}



\section{Introduction}\label{intro}

Massive stars end their lives when their cores collapse under their own weight and form either a neutron star or black hole \citep[e.g.,][]{Baade+1934,Bethe+1979,Woosley+2002}. A core-collapse supernova (SN) will result if the outer layers are successfully ejected.  Stripped-envelope SNe refer to a subset of core-collapse SNe, whose progenitor stars have lost a significant fraction of their outer hydrogen and helium envelopes.  They may be classified as Type IIb, Ib, Ic, Ibn, and Ic-BL.  The first three subtypes empirically exhibit a sequence of progenitors that are increasingly stripped \citep{Filippenko1997}. Type Ibn SNe have narrow lines likely related to the presence of helium-rich circumstellar material \citep[e.g.,][]{Pastorello+2007,Foley+2007,Hosseinzadeh+2017}. Type Ic-BL SNe exhibit very broad lines (BL) indicative of ejecta with high kinetic energy and are sometimes observed in connection with long gamma-ray bursts \citep[GRBs;][]{Galama+1998, Woosley+2006a}.

Traditional single-star evolution models require stripped-envelope SN progenitors to lose their outer envelopes via stellar winds.  The amount of mass lost in these models correlates with the initial mass, metallicity, and rotation rate of the progenitor \citep{Heger+2003, Eldridge+2004, Georgy+2009,Langer2012}. Less massive single stars evolve to become red supergiants (RSGs), which still have massive and extended hydrogen envelopes at core collapse, giving rise to Type II SNe. More massive single stars experience stronger winds, losing their entire hydrogen envelopes and becoming Wolf-Rayet (WR) stars, which eventually may explode as stripped-envelope SNe \citep{Maeder+1982}.

Over the past decade, the single-star paradigm has begun to shift \citep[e.g.,][and references therein]{De-Marco+2017}.  The fraction of massive stars that form in close binary systems is substantially larger than previously considered. Interaction of the primary star with a close binary companion (secondary star) can therefore play an important role in the evolution of the progenitors of stripped-envelope SNe \citep{Kobulnicky+2007, Chini+2012, Kiminki+2012, Sana+2012, Sana+2013, Kobulnicky+2014, Dunstall+2015, Almeida+2017,Moe+2016}. This idea dates back to the earliest computer simulations that showed how a primary star can lose its outer layers through Roche-lobe overflow (RLOF) and transfers mass to its companion \citep[e.g.,][]{Morton1960, Kippenhahn+1967, Podsiadlowski+1992, Woosley+1994, Nomoto+1995}, with many recent studies focusing on these processes \citep[e.g.,][]{Eldridge+2008, Claeys+2011, Gotberg+2017,Yoon+2017}. 

Observationally, single-star models predict massive, luminous WR star progenitors for SNe Ib/c, but no such progenitors have been detected in pre-explosion images.  A series of progenitor nondetections \citep[e.g.,][]{Van-Dyk+2003,Maund+2005,Maund+2005a,Smartt+2009,Eldridge+2013,Van-Dyk2016} seemingly argues against the single-star model (although the nondetections can be explained if WR progenitors become optically faint at the very end of their lives or if they are obscured and highly reddened by mass loss in the very late phases of their evolution; see \citealt{Yoon+2012a}, \citealt{Eldridge+2013}, and \citealt{Tramper+2015}).  These nondetections are, however, consistent with the binary scenario, 
in which the expected progenitors are lower-mass helium giants that can more easily elude detection.  The first and only detection of a progenitor of a SN~Ib/c (iPTF13bvn) further suggests that the progenitor is a helium giant \citep{Cao+2013,Groh+2013b, Bersten+2014, Fremling+2014, Eldridge+2015, Kuncarayakti+2015,Folatelli+2016, Eldridge+2016}. 

In addition, the rapidly increasing sample of stripped-envelope SNe is building up statistics \citep[e.g.,][]{Drout+2011, Modjaz+2016, Liu+2016, Fremling+2016}. The ejecta masses tend to be $\sim2$--5\Msun \citep[][]{Ensman+1988, Drout+2011, Taddia+2015, Lyman+2016}, suggesting low progenitor masses. This is also consistent with the binary scenario, in which the progenitors are the stripped helium cores of stars having initial masses of about 8\Msun\ and 
higher.  Furthermore, about a third of all core-collapse SNe are observed to
have stripped progenitors \citep[e.g.,][]{Smartt+2009, Smith+2011, Li+2011}. Single-star models fail to explain the high relative rates of stripped-envelope SNe and a binary channel is required to explain this discrepancy \citep{Podsiadlowski+1992,De-Donder+1998, Eldridge+2008, Yoon+2010, Smith+2011}.  If wind mass-loss rate estimates are reduced due to the effect of clumping on wind diagnostics, this will 
exacerbate the problems with attributing stripped-envelope SNe to wind mass loss alone \citep{Smith2014}.

\subsection{Searching for Surviving Binary Companions}

Post-explosion searches at the SN site for a former companion to the the star that exploded can offer important clues about the progenitor system.  The number of events for which this is feasible is limited because the SN has to occur within a reasonable {\sl Hubble Space Telescope\/} ({\sl HST}) sensitivity window (up to $\sim15$\,Mpc).  The most compelling case of a detected companion is for the Type IIb SN 1993J \citep{Maund+2004,Fox+2014}. Further possible detections have been discussed for the Type IIb SN 2011dh \citep{Folatelli+2014, Maund+2015} and SN 2001ig \citep{Ryder+2006}, as well as for the Type Ibn SN 2006jc \citep{Maund+2016}.

Nondetections of companions can also be useful to test the theoretical predictions.  Deep upper limits on the Type Ic SN 1994I, 20\,yr after explosion, constrain the mass of a possible main-sequence (MS) companion to $\lesssim 10 \Msun$ \citep{Van-Dyk+2016}.  \citet{Crockett+2007} derive a limit of $\lesssim 20 \Msun$ for the mass of a possible MS companion in the case of the Type Ic-BL SN 2002ap,
for which we present deeper limits in this paper.  Limits on companions have also been discussed for the Type IIP SN~2005cs \citep{Maund+2005,Li+2006}, Type IIP SN~2008bk \citep{Mattila+2008}, Type IIb SN~2008ax \citep{Crockett+2008,Folatelli+2015}, and even for SN remnants in our Galaxy \citep[Crab, Cas~A;][]{Kochanek2017} and the Large Magellanic Cloud \citep[SN 1987A;][]{Graves+2005,Kochanek2017}.  Moreover, a core-collapse SN may disrupt a binary system, ejecting its companion \citep{Blaauw1961,Hoogerwerf+2001}. Some runaway stars (stars with velocities of around tens of $\rm{km\,s^{-1}}$ compared to their surrounding populations) can be tentatively linked with SN remnants and pulsars \citep{Dufton+2011,Tetzlaff+2013,Tetzlaff+2014,Dincel+2015,Boubert+2017}, suggesting that they were the companions of the SN progenitor at the moment of explosion. 

\subsection{The Case of SN 2002ap} 

SN 2002ap was discovered in Messier 74 (M74; NGC 628) and classified as a Type Ic-BL (``hypernova'') owing to its large kinetic energy and broad spectral features \citep{Kinugasa+2002,Meikle+2002, Gal-Yam+2002,Mazzali+2002,Foley+2003}.  Type Ic-BL SNe are the only known SN subclass associated with some long-duration GRBs (\citealt{Woosley+2006a, Cano+2017}), although no GRB was detected in association with SN 2002ap.  While the progenitor and central engine remain ambiguous, the proposed scenarios include massive star ($\gtrsim25\Msun$) 
progenitors and a central engine powered by either a black hole or a magnetar \citep{Woosley+2006}.

\citet{Mazzali+2002} and \citet{Mazzali+2007} infer a kinetic energy of $\sim 4 \times 10^{51}$ ergs, ejecta mass $M_{\rm ej} \approx 2.5 \Msun$, and a nickel production of $M_{\rm Ni} \approx 0.1 \Msun$. The ejecta mass and the kinetic energy are quite high for normal core-collapse SNe, but lower than those of the prototypical Type Ic-BL SN~1998bw (which was associated with GRB 980425). Assuming a remnant of $\sim 2.5 \Msun$, \citet{Mazzali+2002} estimate a carbon-oxygen core progenitor of $\sim 5 \Msun$, which is inconsistent with an initially very massive star of $>30\Msun$.

A high-quality set of pre-explosion ground-based images place limiting absolute magnitudes for the progenitor of $M_B \geq -4.2 \pm 0.5$\,mag and $M_R \geq -5.1 \pm 0.5$\,mag \citep{Crockett+2007}, which is the deepest so far for a SN~Ic progenitor. These limits also indicate that the single-star scenario is unlikely, instead pointing toward evolutionary channels with  a lower-mass primary star interacting with a companion. \citet{Crockett+2007} constrain any possible binary companion to either a MS star of $<20\Msun$ or a compact remnant (neutron star, black hole, or even white dwarf).

\subsection{Scope of this Paper}

In this work, we present a theoretical investigation of the probability of a companion to be present at the explosion site of stripped supernovae. The direct incentive for this study comes from new UV observations of the explosion site of SN 2002ap that we obtained with {\sl HST\/} Wide-Field Camera 3 (WFC3). The data is taken about 14\,yr after explosion, at which time the SN had faded sufficiently to search for the presence of a surviving companion.  Comparing our new deep upper limits with simulations allows us to not only constrain the properties of a potential companion but also explore the initial conditions and evolution history of the progenitor system.

For this purpose we make use of the grid of simulations presented in \citet[][hereafter \citetalias{Zapartas+2017}]{Zapartas+2017}.  These are population synthesis simulations that span the multidimensional parameter space of initial properties that determine the evolution of binary systems. These simulations also explore variations in the uncertain assumptions concerning the initial conditions and the treatment of the physical processes. This allows us to assess the robustness of our findings and investigate how future observations can help to constrain the physics that governs the evolution of SN progenitors.

We expand on the work by \citet{Kochanek2009}, who presented estimates for various properties of the possible companion stars of core-collapse SNe progenitors, but who did not explicitly treat the effects of binary interaction phases as we do here. Further related recent work has been presented by \citet{Liu+2015} and \citet{Moriya+2015}, who mainly focus on implications of the presence of a companion for the light curve.  Besides these studies we refer to the multitude of theoretical studies on the impact of binary interaction on the evolution of supernova progenitors and their companions, which we cited above.

The outline of our paper is as follows. Section \ref{sec:obs} presents our observations. Section \ref{sec:model} describes our theoretical simulations and assumptions. In Section \ref{sec:res1} we discuss our general predictions for companions of stripped SNe and in Section \ref{sec:res2} we discuss the specific implications for the case of SN 2002ap. 
Section \ref{sec:discussion1} explores the robustness of the results against model uncertainties, Section \ref{sec:comparison} compares with earlier studies, and \ref{sec:dis} presents possible implications. We summarize our conclusions in Section \ref{sec:con}. 

\begin{figure*}[t]
\begin{center}
\includegraphics[width=0.35\textwidth]{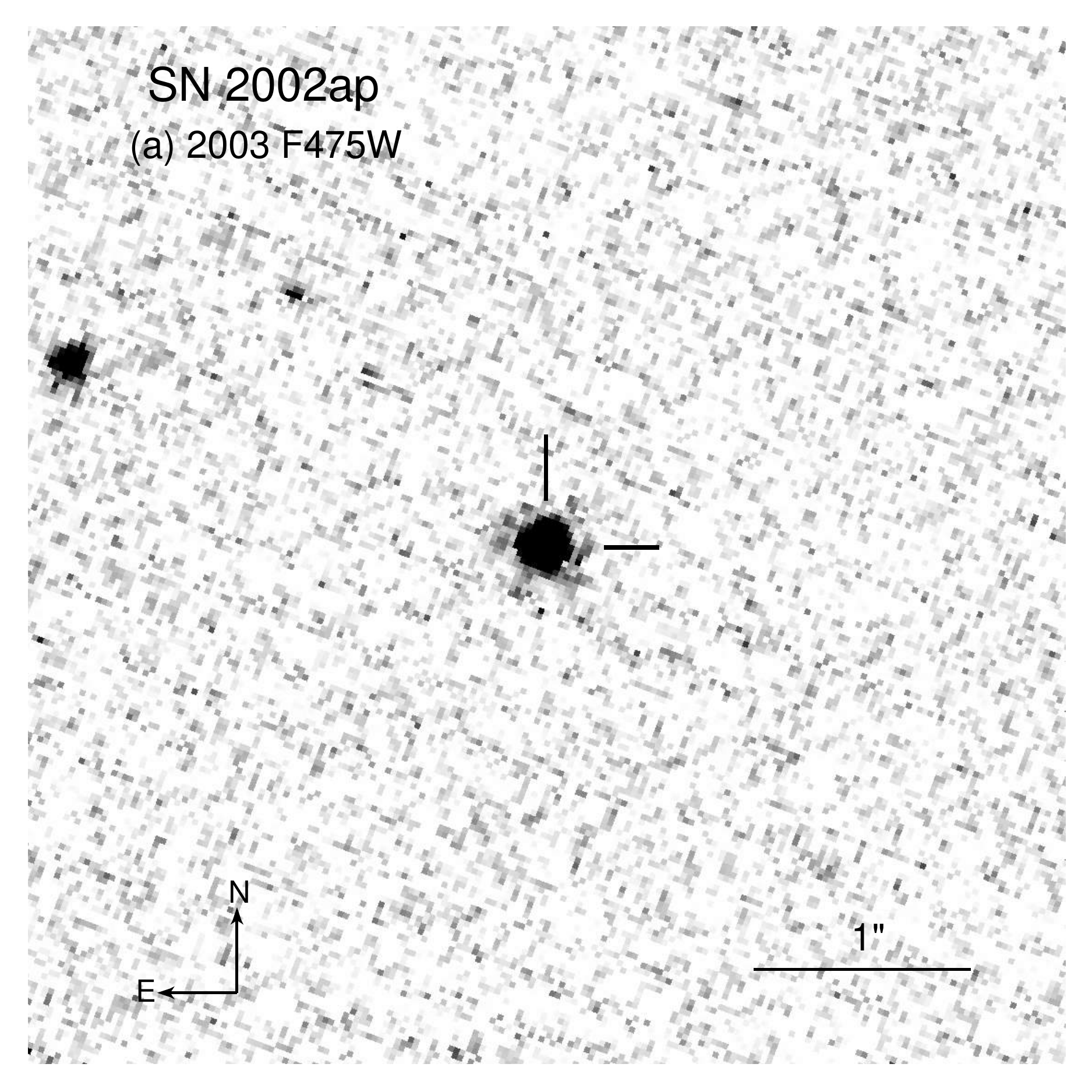}\includegraphics[width=0.35\textwidth]{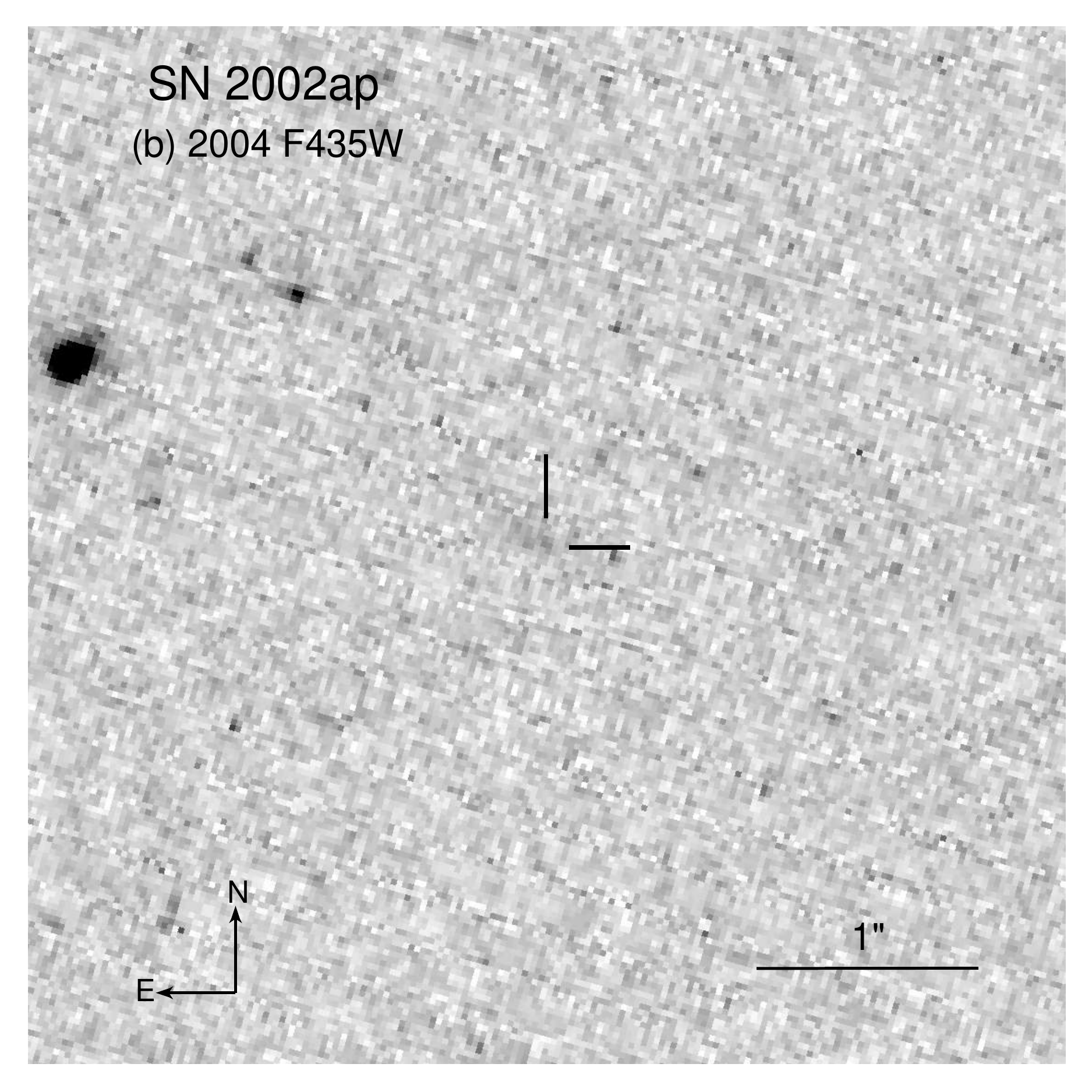}\\
\includegraphics[width=0.35\textwidth]{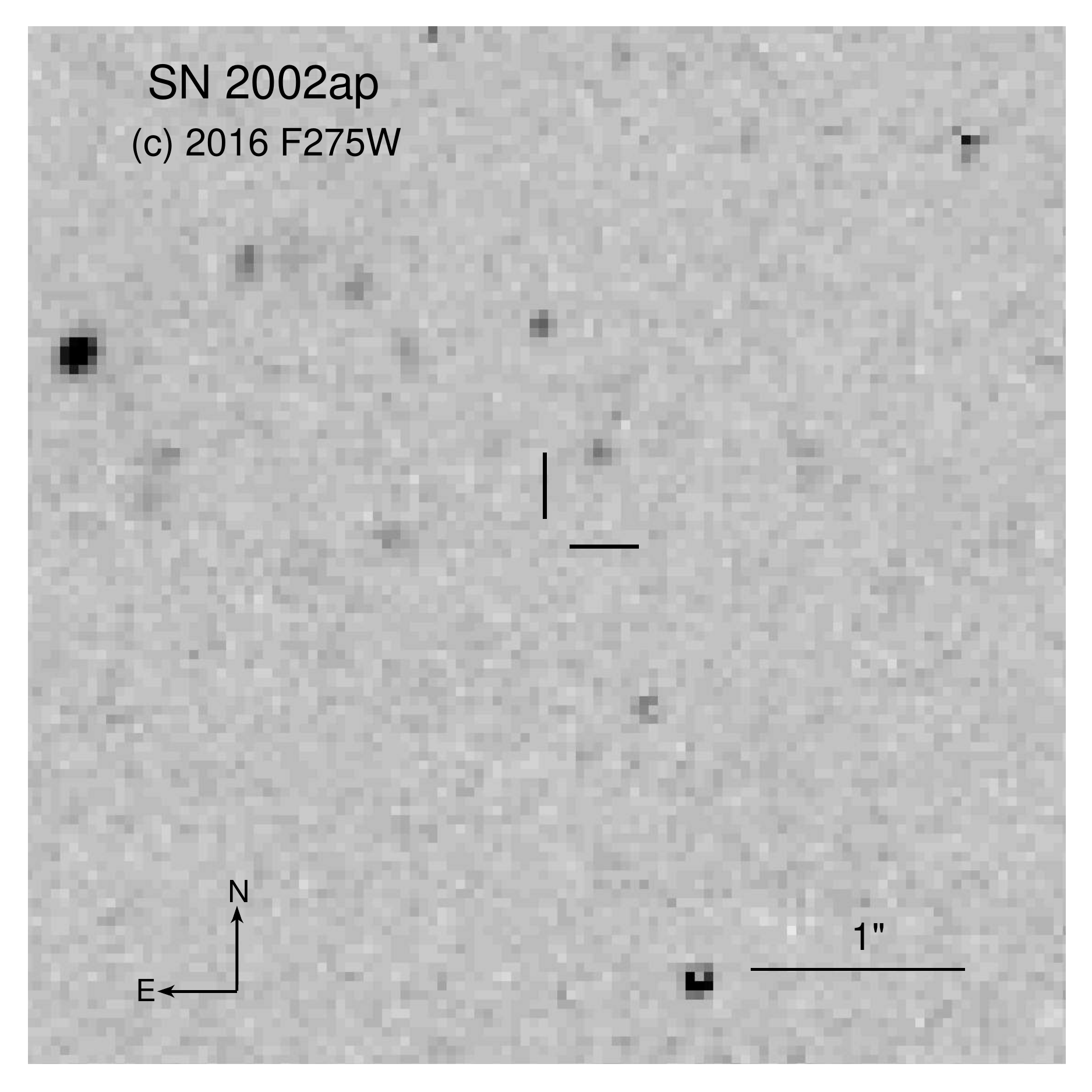}\includegraphics[width=0.35\textwidth]{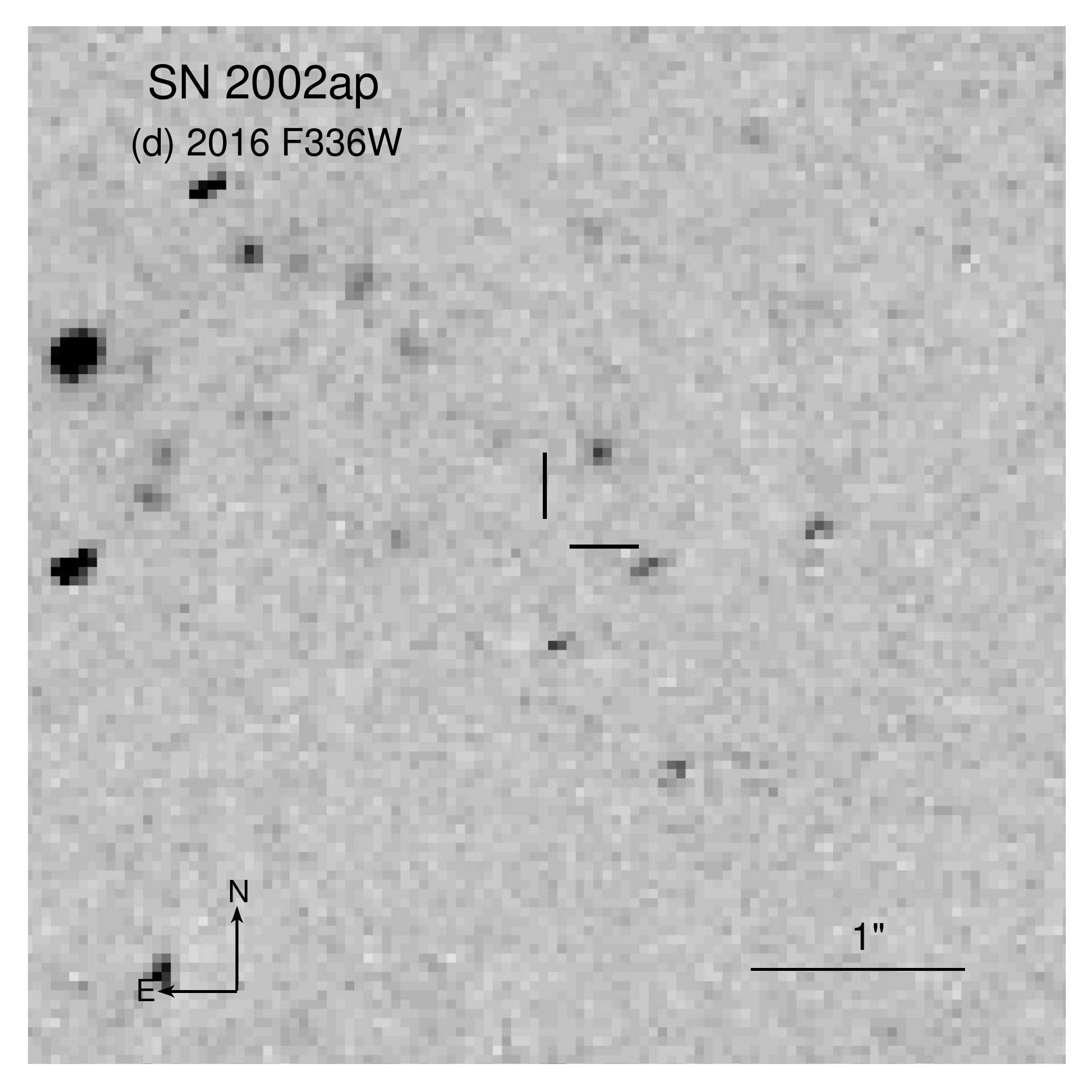}\\
\caption{
{\it HST\/} images of the SN 2002ap site with year and filter labeled. The two panels on the bottom (c, d) show our new deep limits (GO-14075), with no visible star near the SN site. Some unmasked cosmic-ray hits are still visible in the images.  \label{fig1}
}
\end{center}
\end{figure*}

\section{Observations}
\label{sec:obs}

SN 2002ap was observed with the {\it HST\/} WFC3/UVIS channel as part of program GO-14075 (PI O.~Fox) in bands F275W and F336W on 2016 February 16 UT. The total exposure time is 2778\,s in each band. The individual UVIS {\tt flc\/} frames in all bands are obtained from the {\it HST} Archive, following standard pipeline processing.  The frames in each band then have cosmic-ray hits masked and are combined into mosaics by running them through AstroDrizzle in PyRAF.  We locate the position of the SN in the new image mosaics using {\it HST\/} Advanced Camera for Surveys (ACS) High Resolution Channel (HRC) images of the SN from 2003 January (PI R.~Kirshner, GO-9114) when the SN was still relatively young, and also ACS/HRC images from 2004 July and August (PI A.~Filippenko, GO-10272) when the SN was significantly fainter. Specifically, we employ the {\tt IRAF\/} tasks {\tt GEOMAP\/} and {\tt GEOXYTRAN\/} to align the 2004 F435W image (total exposure time 840\,s) to the 2003 F475W image (120\,s), using 11 fiducial stars in common, with a $1\sigma$ uncertainty of 0.30 pixel. We then align the 2016 F336W image to the 2004 image using 20 stars in common, with a $1\sigma$ uncertainty of 0.25 pixel. These images are shown in Figure \ref{fig1}. 

Source detection and photometry are performed on the individual frames in each band using DOLPHOT v2.0 \citep{Dolphin2000}.  We run this code with values of the DOLPHOT input parameters {\tt FitSky\/} and {\tt RAper\/} set to 1 and 4 , respectively, appropriate for relatively empty, uncrowded fields, and also set to 3 and 8, respectively, more appropriate for complex backgrounds in galaxies. We also set {\tt InterpPSFlib=true} and {\tt SkipSky=1}, and use the TinyTim point-spread function (PSF) library.  Aperture corrections are applied. No source is detected at the position of SN 2002ap in the 2016 images, as can be seen in Figure~\ref{fig1}. A $\approx 9\ \mu$Jy (at 3.6 $\mu$m) source is $\approx 1{\farcs}4$ from the SN position \citep{Berger+2002}.

We determine the detection limit for any potential surviving companion in the 2016 data by inserting into the individual {\tt flc\/} frames an artificial star at the exact SN position with progressively fainter brightness, using Dolphot. We run these artificial star tests for both combined sets of {\tt FitSky\/} and {\tt RAper\/} values. We find overall agreement between these two sets, with upper limits of 25.8 (5.1$\sigma$) and 27.4 (3.1$\sigma$)\,mag for {\tt FitSky=1}, and 25.8 (4.0$\sigma$) and 27.3 (3.0$\sigma$) for {\tt FitSky=3}, at F275W and F336W, respectively. We adopt 25.8\,mag at F275W and 27.3\,mag at F336W.

\subsection{Metallicity Estimate}
\label{sec:metallicity}

SN 2002ap occurred $280''$ ($\gtrsim 14$ kpc)  SW  of the nucleus of M74 
\citep{Nakano+2002}, a massive (log$(M/{\rm M}_\odot) = 11.52 \pm 0.05$\,dex) early-type spiral galaxy \citep{Kelly+2012}.  \citet{Kelly+2012} find that the local host-galaxy environments of SNe Ic-BL are blue and metal poor in comparison to hosts of other core-collapse SNe.  The host-galaxy environment of SN 2002ap has an exceptionally blue color, $u -z = -0.16$\,mag, compared to those of other SNe Ic-BL, consistent with a young population of stars and little reddening from dust.  In fact, among the SN Ic-BL host environments studied by \citet{Kelly+2012}, only the explosion environment of SN 2007ce had a similarly blue color ($u -z = -0.14$\,mag).  \citet{Williams+2014} conclude that available archival {\it HST} images are not sufficiently deep to constrain the local star-formation history, and an ongoing program is currently obtaining additional imaging of the site (GO-14768; PI: Williams).

\citet{Modjaz+2008} assume a linear abundance gradient and use literature measurements of M74 to infer a local abundance of $\log_{10} {\rm [O/H]} + 12 = 8.62_{-0.05}^{+0.05}$\, using the bright line diagnostics from \citet{Kewley+2002}, 
  $8.56_{-0.05}^{+0.05}$\, from \citet{McGaugh1991}, and 
  $8.38_{-0.05}^{+0.05}$\, from \citet{Pettini+2004}, whereas \citet{Berg+2013} and \citet{Pilyugin+2014} find $\log_{10} {\rm [O/H]} + 12 \approx 8.20$ and $\approx 8.25$, respectively, at the radius of SN~2002ap.  The \citet{Pettini+2004} local abundance is close to the average value of other SNe Ic-BL discovered by both targeted and untargeted SNe \citep[Figure 8 of][]{Sanders+2012}, and it is more metal-rich than the environments of most long-duration GRBs \citep{Modjaz+2008}.  Given the solar oxygen abundance of \citet{Asplund+2009} ($8.69 \pm 0.05$, Z$_\odot = 0.014$), the local metallicity of the SN site is $Z \approx 0.3$--1\, Z$_\odot$.  Thus, even allowing for the inconsistency between various abundance calibrations, subsolar metallicity is appropriate for the environment of SN 2002ap. We assume a metallicity of $Z = 0.0055$ in our standard simulation, matching the estimate by \citet{Pilyugin+2014}, although we also run models varying this value, as discussed in Section \ref{sec:discussion1}.

\subsection{Constraints on the Upper Mass for a Main-Sequence Companion}
To interpret the upper limits on the apparent brightness of a binary progenitor companion in the bands, in which we observed, we transform them to upper limits on absolute brightness (i.e., luminosity). We assume a distance to the host galaxy, M74, of 10.19\,Mpc from the tip of the red giant branch determination by \citet[with a random error in the measurement of $\pm 0.14$ and a systematic of  $\pm 0.56$, corresponding to a distance modulus of $30.04 \pm 0.04$ (random) $\pm 0.12$ (systematic)\.mag]{Jang+2014}. The Galactic foreground extinction along the line of sight to SN 2002ap has been estimated by \citet{Schlafly+2011} to be $A_V=0.197$\,mag ($E(B-V)=0.064$\,mag, assuming $R_V=3.1$); following the reddening law of \citet{Cardelli+1989}, this corresponds to $A_{\rm F275W}=0.395$ and $A_{\rm F336W}=0.321$\,mag. 

Extinction local to the SN within the host galaxy or produced by the SN itself is more difficult to assess. In their analyses of SN 2002ap nebular spectra, \citet{Mazzali+2007} and \citet{Maurer+2010a} found no evidence for excess dust, at least to 395 days past explosion. In fact, both studies assume a rather low value for the total reddening of $E(B-V)=0.09$\,mag, not dissimilar from the Galactic foreground estimate above. One could potentially determine how much dust is at the SN site from late-time data obtained using the {\sl Spitzer Space Telescope}. Unfortunately, the explosion site is not contained in post-cryogenic mission data from 2014--2016 (i.e., epochs comparable to that of our {\sl HST\/} imaging). However, a source of about $ 9\ \mu$Jy (at 3.6 $\mu$m) is around $ 1{\farcs}4$ from the SN position \citep{Berger+2002}.
Thus, all existing indications point toward little or no local reddening, so we assume only the contribution from the Galactic foreground. Future observations with the far more sensitive James Webb Space Telescope will provide a basis to assess this assertion.

\begin{figure*}[t]
\begin{center}
\includegraphics[width=0.95\textwidth]{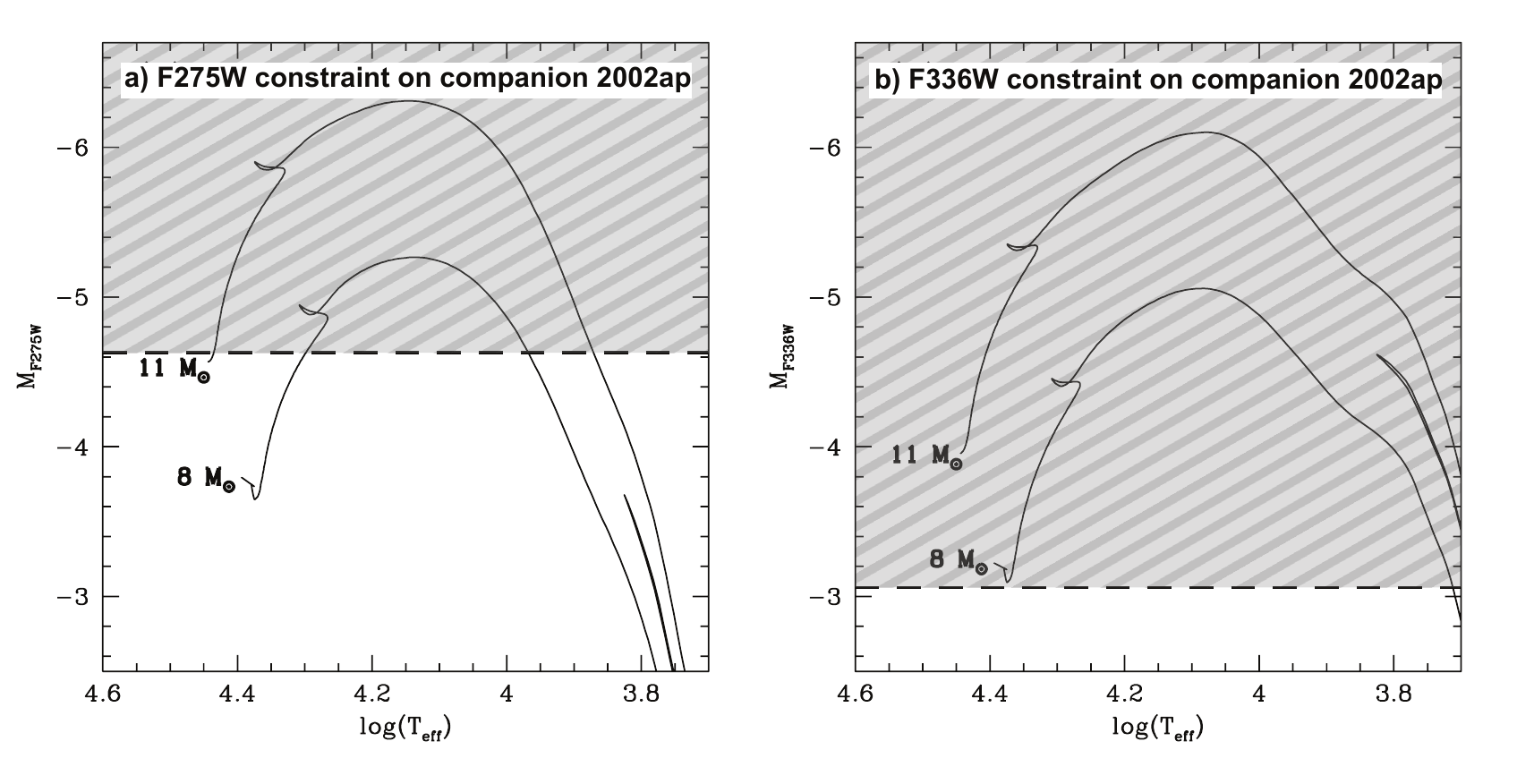}
\caption{Deep {\it HST} photometry places upper limits on the F275W (a) and F336W (b) flux at the location of the explosion site of SN 2002ap. Evolutionary tracks by MIST \citep{Choi+2016} for metallicity $Z=0.008$ are also shown for comparison. Typically, we expect a possible companion to still be unevolved and lie relatively close to the ZAMS.  The upper limit in the F336W band is the most constraining; it excludes the presence of a surviving companion (or very nearby star) with a mass in excess of about 8\Msun (shaded area).  \label{fig:lim}
}
\end{center}
\end{figure*}

These resulting upper limits to a companion's absolute brightness are shown in Figure~\ref{fig:lim}. We can evaluate these limits, comparing them with stellar evolutionary tracks including rotation by Mesa Isochrones and Stellar Tracks \citep[MIST; ][]{Choi+2016, Paxton+2011, Paxton+2013, Paxton+2015}, calculated at a subsolar metallicity of [Fe/H] = $-0.25$ ($Z=0.008$), also shown in the Figure.

For a typical system, we expect that the companion is less evolved and still resides on the MS, as argued by \citet{Claeys+2011} and as we also show in Section~\ref{sec:res1}. The reason is that the exploding star is typically the initially more massive primary, which evolves on a shorter evolutionary timescale than the secondary.  Stars spend about 90\% of their lifetime on the MS. Only in binary systems with an initial mass ratio that is very close to unity may we expect a more evolved companion at the moment of explosion of the primary.  Furthermore, if the binary system has experienced mass transfer, we expect that the companion has gained mass. Such stars adapt their structure to their new mass and are thought to increase the size of their core, which mixes fresh hydrogen to the central regions where nuclear burning takes place. This rejuvenates the star and makes it appear younger than it is. We therefore expect a typical companion to reside relatively close to the zero-age main sequence (ZAMS); see also \citet{Benvenuto+2013} and \citet{Bersten+2014}.  

Comparing the ZAMS of various tracks to the observational limits in each band, we find that the more restrictive limits imposed at F336W constrain the mass of a MS star to $\lesssim 8 \Msun$; the less restrictive limit at F275W allows for a more massive star, of mass $\lesssim 11 \Msun$.

In principle, our data also provide limits to a possible companion star that is not in the MS phase at the moment of explosion (e.g., a stripped helium star or a giant star).  However, for the reasons mentioned above, scenarios including post-MS companions are much less probable because the companion is typically expected to be close to the ZAMS.
In addition, as stripped helium stars are typically expected to be mostly bright at shorter wavelengths than our observed bands \citep[e.g.,][]{Gotberg+2017}, the inferred limits would be less restrictive than for a MS companion star. Our data also cannot put very tight constraints on a possible red supergiant companion, which would emit mostly at longer wavelengths than our observations owing to their lower temperatures (shown as a drop in the luminosity in the observed bands of post-MS evolutionary tracks on the right part of Figure~\ref{fig:lim}).  However, we could consider limits from pre-explosion images of previous studies at longer wavelengths that put tighter constraints on an evolved star at the SN site. For example,\citep{Crockett+2007} mentions that a supergiant more massive than about $8 \Msun$, either the SN progenitor or a possible post-MS companion, would have been visible in their pre-explosion images. Still, because of the very low possibility of channels with post-MS companions, even tight observational limits on a red supergiant companion exclude only a very limited range of all the possible evolutionary scenarios, so we do not focus on these cases.
Moreover, other types of possible companions (white dwarfs, neutron stars, and black holes) would not have been visible in our images and thus cannot be constrained with our data. For all these reasons, we express our limits as constraints only on the mass of a possible MS companion.

The uncertainty in our limits is dominated by those in our measurement, not by uncertainties in the distance to the host galaxy or in the Galactic foreground extinction.
A signal-to-noise ratio of about $ 3$ corresponds to a measurement uncertainty of $\approx 1/3 \approx 0.35$\,mag.
Thus, a conservative error of $0.4$\,mag in Figure~\ref{fig:lim}  would correspond to an error in our inferred limits of at most 1--2$\Msun$, thus driving our upper limit at F336W to $\sim 10$\Msun for a possible MS companion. We note that this limit on a surviving companion is deeper than what was previously applied by \citet{Crockett+2007}.

\section{Simulations}
\label{sec:model}

\citet[][hereafter \citetalias{Zapartas+2017}]{Zapartas+2017} created a theoretical framework to interpret the observations of core-collapse SNe in a probabilistic way. We extend these simulations to generate statistical predictions concerning the presence and properties of a companion to a stripped-envelope SN at the moment of explosion.  We follow a similar approach as presented by \citet{Van-Dyk+2016} for the Type Ic SN 1994I, but here we use the full grid of simulations created by \citetalias{Zapartas+2017}.
Below we briefly describe the code and discuss our assumptions for the initial conditions and physical processes.  We refer to  \citetalias{Zapartas+2017} and references therein for a more extensive description of the grid and the adopted assumptions.

\subsection{Code} 
Our code follows the evolution of millions of single and binary stars until one or both stars explode as a core-collapse SN. For the simulations we use the binary population synthesis code {\tt binary\_c} (version 2.0, SVN revision 4105). This is a code developed by \citet{Izzard+2004,Izzard+2006,Izzard+2009} with updates described by \citet{de-Mink+2013} and \citet{Schneider+2015}.  The code relies on the rapid evolutionary algorithms by \citet{Tout+1997} and \citet{Hurley+2000,Hurley+2002}, which in turn are based on a grid of stellar models computed by \citet{Pols+1998}.

This rapid code enables exploration of the extensive multidimensional parameter space of initial properties that determine the evolution of binary systems. The output yields statistical predictions for the distribution of the properties of stripped-envelope SNe and enables us to investigate the robustness of our predictions against the variations in the input assumptions and model uncertainties.  A caveat is that, in order to explore so many scenarios, the treatment of the evolution and interaction phases is necessarily approximated. In contrast to  full evolutionary calculations that solve the stellar structure equations, the code does not follow the interior chemical structure of the progenitor stars in detail. The consequences of these limitations are discussed in Section~\ref{sec:helium}. 

\subsection{Initial Conditions}\label{sec:model_init_cond}

We simulate the evolution of over 3 million binary systems for our default simulation, varying the initial primary masses, mass ratios, and orbital periods on a regular grid of $150 \times 150 \times 150$ models. We also simulate the evolution of $10^4$ single stars. We assume a population containing a mix of single and binary systems, with a binary fraction of 70\%, following \citet{Sana+2012}.

Single stars or primary stars in binary systems (the initially more massive ones) are assumed to be born with an initial mass $M_1$ according to the \citet{kroupa2001} initial mass function (IMF; $\mathrm{d}N/\mathrm{d}M_1 \propto (M_1)^{\alpha}$, with $\alpha=-2.3$ in the upper mass range relevant to our study).
We consider primary masses between 3 and 100\Msun. This safely includes the lower mass limit for binary stars to produce core-collapse SNe (\citetalias{Zapartas+2017}). Higher mass systems may exist \citep{Crowther+2016}, but they are too rare to matter for statistics of core-collapse SNe. Companion masses ($M_2$) are chosen such that the mass ratio, $q \equiv M_2/M_1$, is distributed uniformly between 0.1 and 1 (i.e., $\mathrm{d}N/\mathrm{d}q  \propto  q^{\kappa}$, where $\kappa = 0$), consistent with observations \citep[e.g.,][]{Sana+2012,Duchene+2013}. 

We place the primary and secondary star in a binary system with an initial orbital period, $P$, assuming a power-law distribution of $\mathrm{d}N/\mathrm{d}{\rm log}_{10}P \propto ({\rm log}_{10}P)^{\pi}$, with $\pi = -0.55$ for $M_1 > 15\Msun$, consistent with \citet{Sana+2012}. We account for binary systems in the range of $0.15 \le \log_{10} P \, ({\rm days}) \le 3.5$. In systems with lower-mass primaries we adopt a standard \citet{Opik1924} law which is a flat distribution in the logarithm of the period. To limit the grid dimensions, we assume that the orbits are circular at birth, although we account for the effects of eccentric orbits later in the evolution, for example in case a system remains bound after the SN explosion of one of the stars. This assumption does not have an important effect on our results \citep[cf.][]{Hurley+2002,de-Mink+2015}.  We do not include the possible influence that more distant companions in a triple system may have on the orbital dynamics and chances of interaction \citep[e.g.,][]{Moe+2016,Toonen+2016}.

Our standard simulation adopts a metallicity of $Z=0.0055$, chosen to match the most probable metallicity of the explosion site of SN~2002ap (see Section \ref{sec:metallicity}).
Results for other metallicities, initial distributions, and binary fractions are provided in the Appendix and discussed in Section~\ref{sec:discussion1}.

\begin{figure*}\centering
\includegraphics[width=0.65\textwidth]{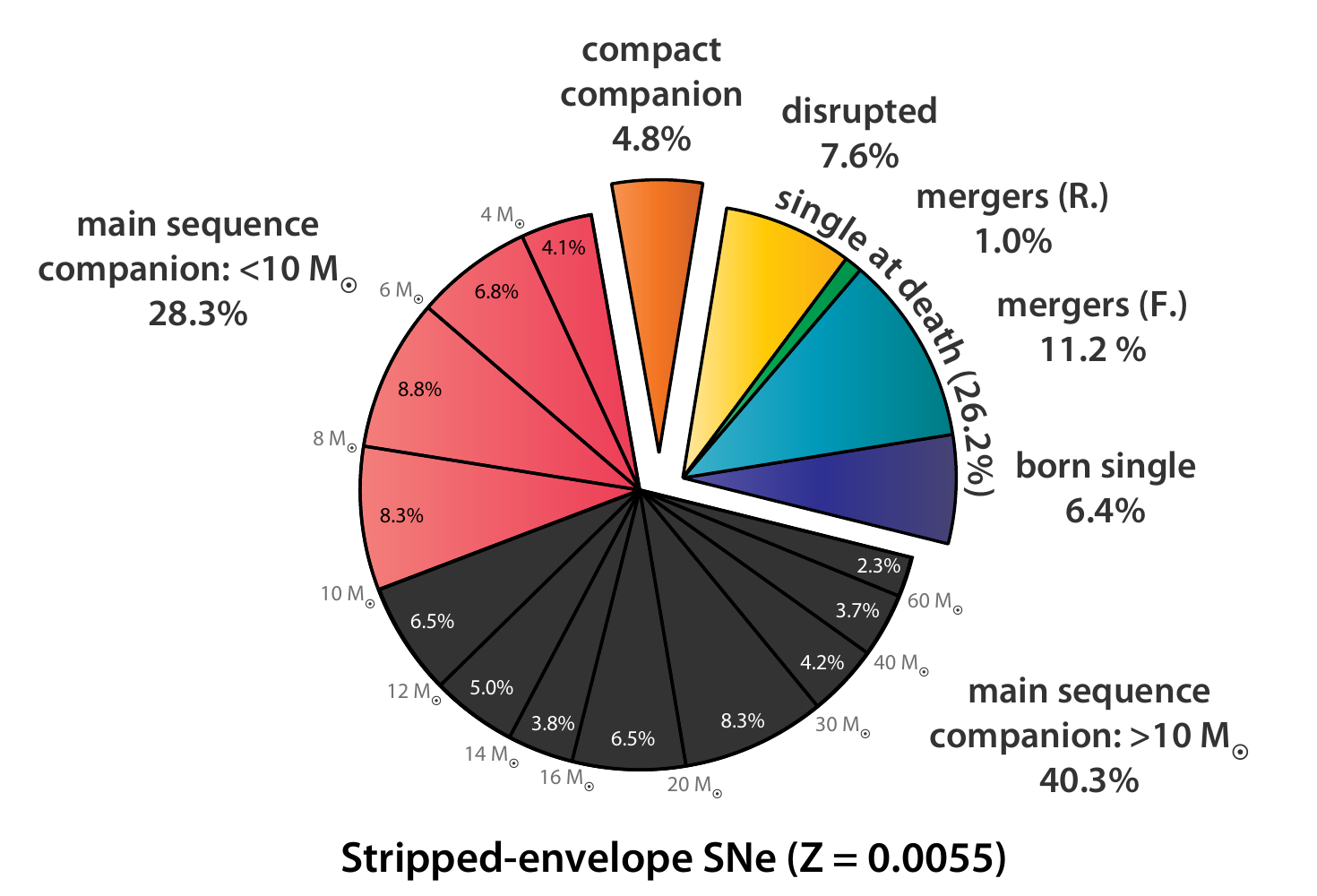}
\caption{Expected progenitors of stripped-envelope SNe and their possible companions at the moment of explosion.  These results are derived from a population synthesis simulation including a realistic mix of single and binary stars, with assumptions specified in Section~\ref{sec:model}.  Red and black segments indicate progenitors with main sequence (MS) companions at the time of explosion, with dividing lines 
showing the contribution of MS companions within a specified mass interval. Black segments indicate those possibilities ruled out by the upper limits on SN 2002ap provided in this paper.
Stripped-envelope SNe that are not expected to have a companion at explosion are divided in groups of ``disrupted'' systems (yellow; SNe resulting from a star that was once the secondary star in a system disrupted by the explosion of the primary), ``mergers'' (green and turquoise; former binary systems that experienced merging induced by RLOF of either the primary [forward, or F.] or the secondary [reverse, or R.]), and ``born single'' (dark blue; stars for which we assumed there was no companion at birth). Channels with compact companions (a white dwarf, neutron star, black hole, or compact helium star) at explosion are show in orange.
Other possibilities, such as the presence of a giant companion, contribute less than 0.5\% and are not shown.}\label{fig:pie}. 
\end{figure*}

\subsection{Physical Assumptions}

We follow the evolution of each simulation from the onset of central hydrogen burning until the final fate as a compact remnant (white dwarf, neutron star, black hole) using the evolutionary algorithms for single stars provided by \citet{Hurley+2000}. We account for mass loss through stellar winds following \citet{Vink+2000,Vink+2001} and \citet{Nieuwenhuijzen+1990}, implemented as described by \citet{de-Mink+2013}. For stars that have lost their hydrogen envelopes, we adopt the WR mass-loss prescription by \citet{Hamann+1995}, reduced by a factor of 10 to account for the effect of wind clumping \citep[e.g.,][]{Yoon+2005}.
We do not account for a reduction of the wind by clumping  in other evolutionary phases in our standard model, but we consider this when discussing variations in the physics assumptions (Section \ref{sec:discussion1}).

We model the effects of tides on the stellar spins and orbit, following \citet{Zahn1977} and \citet{Hut1980,Hut1981} as described by \citet{Hurley+2002}. When a star fills its Roche lobe, we compute the mass-transfer rate from the donor star by removing as much mass as needed for the star to remain inside its Roche lobe, never exceeding the thermal timescale of the donor. 
We limit the rate at which a companion accretes mass from a donor star to ten times the thermal rate of the accreting star \citep{Schneider+2015}. In the case of nonconservative mass transfer, we assume that mass is lost from the system via a fast wind originating from the accretor or its accretion disk and takes away the specific orbital angular momentum of the accreting star \citep{van-den-Heuvel1994}. The evolution of stars that have lost their envelope are approximated using models of pure helium stars \citep{Pols+1998, Hurley+2002, Claeys+2014}. 

When a star accretes mass during its MS evolution, we assume that the interior structure and size of the convective core adapt to its new mass \citep{Braun+1995,Dray+2007}. This rejuvenates the star as fresh fuel is mixed into the central regions. To account for this we use the algorithm by \citet{Tout+1997}, as updated by \citet{de-Mink+2013} and \citet{Schneider+2015} to treat massive stars more appropriately. 

RLOF can lead to the formation of contact systems or the onset of common envelope (CE) evolution when mass transfer is unstable \citep{Hurley+2002} or because the accretor swells up and fills its Roche lobe \citep{Neo+1977, de-Mink+2007}. Mass transfer in binaries with mass ratios more extreme than some assumed critical mass ratios, $q_{\rm crit}$, lead to contact or CE, as detailed by \citet{Hurley+2002} and \citet{de-Mink+2013}. Stars without a well-defined core-envelope structure, such as MS stars, are assumed to merge as a result of contact. Giant-like stars, such as red giants and stars crossing the Hertzsprung gap, are assumed to enter a CE phase.  This process is treated by adopting the \citet{Webbink1984} energy balance prescription, using an efficiency parameter ($\alpha_{\rm CE}$) which is chosen to be unity in our standard simulation. The binding energy of the envelope is taken from \citet{Dewi+2000,Dewi+2001} and \citet{Tauris+2001}. We assume that the companion does not accrete during the inspiral process. 
When the two stars merge as a result of contact or because the system fails to eject the CE, we follow the further evolution of the merger as described by \citet{Hurley+2002} and \citet{de-Mink+2013}. In these cases, the star will not have a companion at the moment of explosion (unless it was initially in a triple system, which, as mentioned in Section \ref{sec:model_init_cond}, we do not consider in our simulations).

We account for the possible disruption of the binary system by mass loss resulting from the SN \citep{Blaauw1961} and the natal kick of the compact object. We assume the compact remnant gets a birth kick in a random direction, drawing a scalar velocity from a Maxwellian distribution characterized by a one-dimensional (1D) root-mean square of $\sigma = 265$\,km\,s$^{-1}$ \citep{Hobbs+2005}. 
If the system remains bound, we continue to follow the evolution accounting for the eccentricity of the orbit. 
We terminate the evolution of a system, binary or single star, when all its components become compact final remnants (white dwarfs, neutron stars, or black holes).

In Section~\ref{sec:discussion1} and in the Appendix we also provide results for different physical assumptions.

\subsection{Description of the Numerical Experiment}\label{subsec:experiment}

Using the setup described above, we simulate the evolution of a population of single and binary stars and identify all systems that produce one or two core-collapse SNe. In this work, we only consider cases where the progenitor star has lost its hydrogen envelope prior to the explosion (i.e., stripped envelope).  Our results therefore apply to all H-poor and H-absent subtypes: IIb, Ib, Ibn, Ic, and Ic-BL. SNe~IIb are included even though a small trace of hydrogen may still be present at the surface at the moment of explosion. We do not explicitly distinguish between the various subtypes since our simulations do not allow us to follow the detailed chemical structure or the explosion engine reliably (see Section~\ref{sec:dis} for a discussion).

We divide our experiment in two parts. First, we investigate the possibility of a companion star being present at the moment of explosion for the general case of a stripped-envelope SN (Section~\ref{sec:res1}).  We then explore the implications of deep photometric constraints that exclude the presence of a bright companion. Specifically, we investigate how the new upper limits for a companion to SN 2002ap constrain possible progenitor channels (Section~\ref{sec:res2}).   To allow for uncertainties in the derived limit,  as discussed in Section \ref{sec:obs}, and to keep our findings more general such that they can be applied to future cases, we use the more conservative value of 10\Msun as an upper limit for the mass of a possible MS companion. We discuss the implications for the allowed progenitor channels. 

\section{Companions of Stripped SNe or Absence Thereof}
\label{sec:res1} 

For the general case of a stripped-envelope SN, our simulation predicts the likelihood of potential companions at the moment of explosion.  Our results are summarized in Figure~\ref{fig:pie}.  Most probable is the presence of a MS companion, which is expected in 68\% of cases according to our standard simulation. We discuss these cases and the channels that produce them in Section~\ref{sec:mscomp}.  We estimate a 26\% chance that the stripped SN is single at the moment of explosion. The absence of a companion may result because the progenitor was initially born as a single star, or because a binary system evolved into a single star through a merger, or because the system was disrupted by a previous SN, as discussed in Section~\ref{sec:nocomp}. In about 5\% of cases we expect compact or other types of companions, discussed in Section~\ref{sec:othercomp}. 

\subsection{Main-Sequence Companions}\label{sec:mscomp}

\begin{figure}
\includegraphics[width=0.5\textwidth]{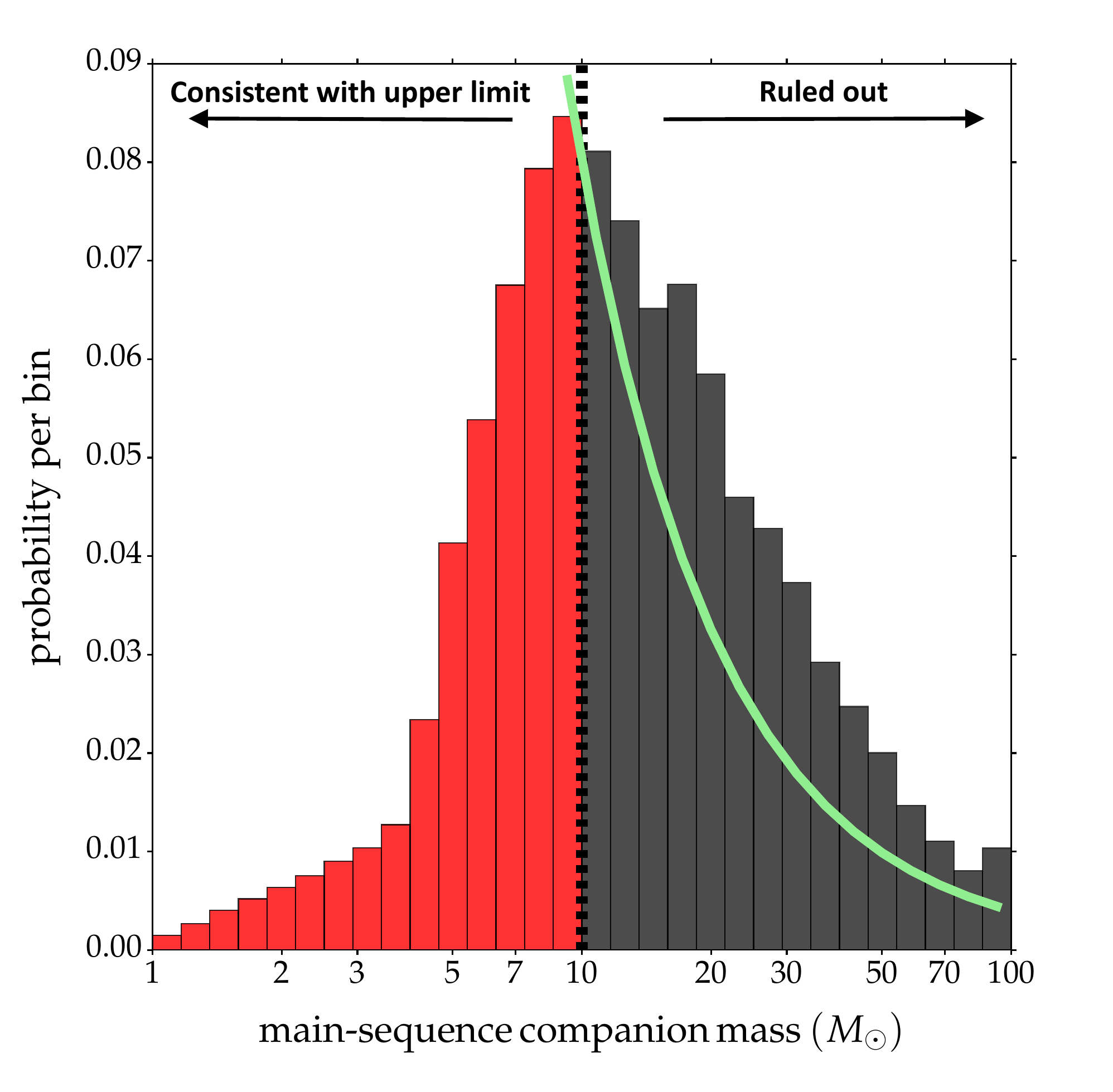}
\caption{Predicted distribution of masses of MS companions of stripped-envelope SNe. See also Figure \ref{fig:pie}, where the same color coding is used  (black indicates those possibilities ruled out by the upper limits on SN 2002ap provided in this paper). The vertical black dashed line shows the threshold of $10\Msun$, which is the upper limit derived from our new {\it HST} data for SN 2002ap. The green line follows the $\alpha=-2.3$ slope of a \citet{kroupa2001} IMF, positioned so as to match the probability per bin at $10\Msun$.} \label{fig:mass_dist}
\end{figure} 

Our simulation reveals that the majority of stripped SNe result from binary systems in which the primary star (i.e., the initially more massive star) explodes after having lost its hydrogen-rich envelope during a phase of RLOF. In most of these cases, the secondary star (i.e., the initially less massive star) is still present at the moment of explosion, resides relatively close to the ZAMS \citep{Claeys+2011,Benvenuto+2013,Bersten+2014}, and is therefore relatively compact. The explosion therefore has little effect on the secondary, except in rare configurations entailing little separation \citep{Tauris+1998, Moriya+2015, Liu+2015, Rimoldi+2016}. 

Figures~\ref{fig:pie} and~\ref{fig:mass_dist} show the distribution of masses of the MS companions at the moment of the primary star explosion for the standard simulation.  The mass distribution peaks around $9\Msun$ and drops off quickly at lower masses, but shows an extended tail to high masses. The shape of the distribution still largely reflects the main features of the initial mass distribution of the secondaries. 
We draw from a flat mass ratio distribution, and thus the initial secondary mass distribution has the same slope as that of primaries that follow a \citet{kroupa2001} IMF (depicted in Figure~\ref{fig:mass_dist}). However, binary interaction has modified the shape of the distribution.  Mass accretion shifts the mass distribution of the companions upward.  The dropoff at low companion masses results from the fact that the minimum mass for the primary to explode through this channel is  about 8\Msun, combined with the fact that the most extreme mass-ratio systems lead to a merger instead of stripping of the primary star.   

Two main channels of binary evolution contribute to the distribution of stripped-envelope SNe in Figure~\ref{fig:mass_dist}. The first channel consists of systems where the mass transfer occurs semiconservatively and the secondary accretes a substantial fraction of the mass lost from the primary. The second channel involves nonconservative mass transfer in which the secondary star does not accrete significantly. 

In cases with semiconservative mass transfer, binary systems have initial orbital periods less than roughly a few hundred days. In these systems the primary star fills its Roche lobe either as a result of expansion during its MS evolution \citep[``Case A'';][]{Morton1960} or during the rapid expansion phase after it ignites hydrogen-shell burning \citep[``early Case B'';][]{Kippenhahn+1967}. In these systems mass transfer is stable, if the mass ratio is not so extreme to trigger contact or CE. The secondary gains a substantial fraction of the mass stripped from the primary. 

In this scenario, the secondary star evolves on its own evolutionary timescale. The secondary usually has not yet completed central-hydrogen burning when mass transfer starts, so it has not yet developed a very steep internal density gradient between its core and envelope. As it accretes mass, the secondary tends to adjust its internal structure to the new mass, which may be substantially higher than its initial mass. Detailed evolutionary models show that the convective core grows and transfers fresh hydrogen to the central regions. This effectively rejuvenates the accreting star \citep{Braun+1995, Dray+2007, Schneider+2014}. After mass transfer ceases, the star quickly regains thermal equilibrium and closely resembles a younger single ZAMS star of the same (new) mass \citep{Hellings1983,Hellings1984}, although it may be spinning rapidly \citep[e.g.,][]{Packet1981,de-Mink+2013}.  
The increased mass accelerates the evolutionary timescales of the secondary, but usually not enough to overcome the rejuvenating effect of mass gain. The secondary is therefore less likely to catch up with the evolution of the primary star. The possibility of this happening and thus the secondary exploding first, possibly in a stripped SN, with the primary as a naked helium star companion, is discussed in Section \ref{sec:othercomp}.

The second binary channel that contributes to stripped SNe with MS companions consists of systems with initially wider separations, where the primary star can expand to giant dimensions before filling its Roche lobe \citep[``late Case B'' or ``Case C'';][]{Lauterborn1970}. Mass transfer in such systems is often unstable and leads to a CE phase. If the envelope is ejected successfully, the result is a naked helium core, which would be the progenitor of the stripped-envelope SN. The secondary still resides on the ZAMS with little change in mass.

\subsection{Single at Death: Mergers, Disrupted Systems, and Initially Single Stars \label{sec:nocomp}}

In about 26\% of the cases, we expect a stripped-envelope SN to have no companion at the moment of explosion (see Figure~\ref{fig:pie}).  The single-star scenario has a diverse group of progenitors.  About $6\%$ of all stripped-envelope SNe originate from stars that were \emph{born single} --- that is, stars assumed to have formed without a companion. Only the most massive single stars have winds that are strong enough to remove the hydrogen envelope.  This is especially true at the low metallicity that we assume for the case of SN 2002ap ($Z=0.0055$) because of the metallicity dependence of line-driven winds \citep[e.g.,][]{Vink+2001}. For this metallicity, single stars have to be initially above $\sim 35 \Msun$ to become WR stars and explode as stripped-envelope SNe, according to our model assumptions. Since MS winds are clumpy, mass-loss rate reductions will tend to even further raise the initial mass above which single stars can shed their hydrogen envelope, making an even smaller contribution to stripped-envelope SNe \citep{Smith2014}.  Results for different metallicity and for variations of the adopted wind mass-loss rate are provided in the Appendix. 

Our standard simulation indicates that $7.6\%$ of all stripped-envelope SNe originate from secondary stars that were once members of a binary system. In these cases, the explosion of the primary \emph{disrupted} the system. We expect that this subgroup explodes in relatively isolated locations because the disruption of the system can impart a significant spatial velocity onto the secondary (possibly of the order of tens of $\rm{km\,s^{-1}}$), which allows it to travel far from its birth location. The fastest of these are usually referred to as runaway stars \citep[e.g.,][]{Zwicky1957,Blaauw1961,Hoogerwerf+2001}. 

Only a subset of the ejected secondaries give rise to a stripped-envelope SN since they do not have their own companion to help remove the envelope. These ejected secondaries must therefore be significantly massive stars that have winds strong enough to remove the hydrogen envelope. There is also a contribution to this group from secondaries that were partially stripped during reverse RLOF (back toward the primary star) before the disruption of the system.

A significant fraction (11.2\%) of stripped-envelope SNe arise from merger systems. In these scenarios, the progenitor started as a binary system, but the secondary has been swallowed by the primary prior to the explosion. We refer to these as \emph{forward mergers}. The merger process itself removed part of the hydrogen envelope. The remainder is lost as a result of stellar winds.

We also find a small contribution ($\sim1$\%) from more exotic merger channels. This number can increase if we vary the physical assumptions. These exotic mergers typically involve two or more phases of mass transfer. The primary loses its envelope first during an episode of forward mass transfer. Later, when the secondary starts to expand, a phase of reverse mass transfer is initiated.  By this stage of reverse mass transfer, the original primary is a naked helium star or a white dwarf.  This phase can lead to a merger where the secondary engulfs the remnant of the primary, which we refer to a \emph{reverse merger}. If the remaining hydrogen envelope is not sufficiently massive, it will be removed by winds. The outcome of these mergers and whether they indeed lead to core-collapse SNe is highly uncertain (see \citetalias{Zapartas+2017} for a more general discussion on reverse mergers).

\subsection{Compact or Giant Companions \label{sec:othercomp}}
The simulations show that about 5\% of stripped-envelope SNe have a compact companion at the moment of explosion.  In these cases, the secondary star explodes as the SN and the compact companion is the remnant of the primary.  In half of these cases the companion is a stripped helium star but it may also be, although less likely, a young white dwarf, a neutron star or a black hole.

A very small contribution ($\sim0.5$\%) of stripped-envelope SNe have companions that are giants (not shown in Figure~\ref{fig:pie}).  The giant phases are short lived. This scenario requires fine tuning of the initial system parameters for the evolution of the two stars to be sufficiently synchronized to have the secondary star in a giant phase at the moment the primary explodes. The rejuvenating effect of mass exchange further reduces the chance of having a giant companion at the moment of explosion.  For further details we refer to \citet{Claeys+2011}, who draw similar conclusions and provide a discussion of the uncertainties.  

\begin{figure*}
\includegraphics[width=\textwidth]{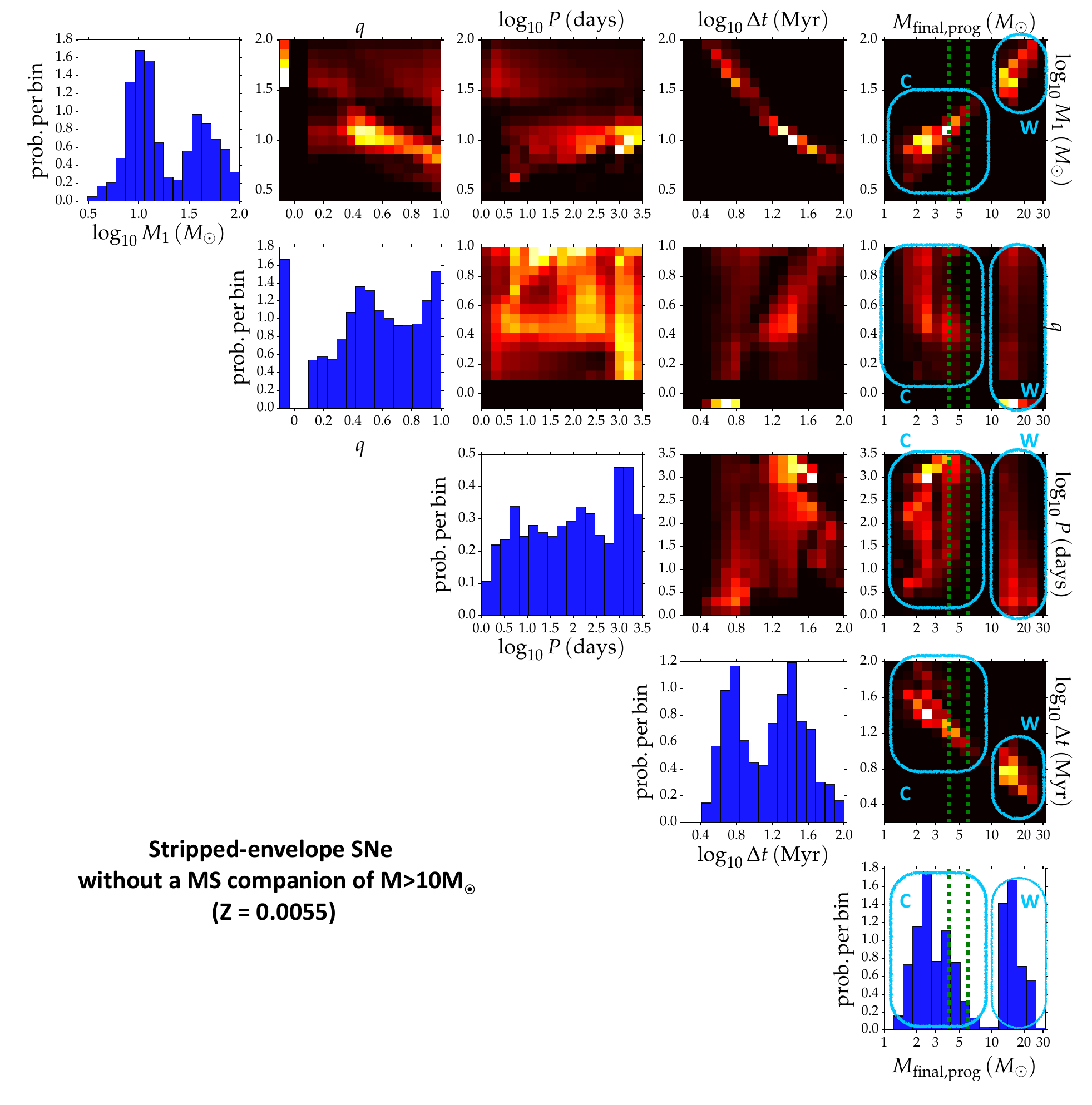}
\caption{Population synthesis predictions of stripped SNe excluding those that have a MS companion of mass $>10$ \Msun. Diagonal panels show the normalized 1D distributions of the initial primary mass ($M_1$) in solar masses; the initial mass ratio ($q = M_2/M_1$); the initial orbital period ($P$) in days; the total lifetime of the progenitor system until SN ($\Delta t$) in Myr; and the final total mass of the progenitor star at the moment of the SN ($M_{\mathrm{final,prog}}$) in solar masses. The correlation between these parameters is illustrated in the two-dimensional (2D) distributions, in which the lighter the color shown, the higher the probability density per bin. Single stars are assigned an initial $q=-0.1$ (and a period out of the depicted boundaries) for the purpose of presenting them in this figure. Approximate regions, outlined with light blue in the rightmost panels, encircle the two groups in $M_{\mathrm{final,prog}}$, and show their correlation with the other parameters: (i) those that are stripped of their hydrogen envelope mainly by mass transfer to a companion (group C), with relatively low mass at the moment of explosion; and (ii) those for which the dominant process of stripping is wind mass loss (group W), with relatively high mass before exploding as a SN. The two vertical green dashed lines indicate the boundaries of $M_{\mathrm{final,prog}} = 4$--6 $\Msun$, around the value of $\sim 5\Msun$ for SN 2002ap found by \citet{Mazzali+2002}.}\label{fig:triangle}
\end{figure*}

\section{Absence of a Bright Companion: Interpreting SN~2002\lowercase{ap}}\label{sec:res2} 

The observations of SN 2002ap presented in Section~\ref{sec:obs} exclude the presence of a MS companion with a mass greater than 8\Msun.  We round this limit up to a more conservative 10\Msun, as discussed in Section \ref{subsec:experiment}.

This limit alone excludes $\sim40$\% of the possible stripped-envelope SN progenitors in our simulations (see Figure~\ref{fig:pie}).  For comparison, the constraints derived earlier by \citet{Crockett+2007} provided an upper limit of about 20\Msun for a possible companion, thereby excluding only 18.5\% of the possibilities. The new limit still allows for channels with a lower-mass MS companion, a compact companion, or no companion at all (all discussed in Section \ref{sec:res1}).  In Section~\ref{sec:lim_comp} we show that the allowed progenitor population is bimodal and we discuss its characteristics in more detail. In Section~\ref{sec:lim_prog} we consider the impact of additional observational constraints for SN 2002ap.  

\subsection{The Bimodality for Stripped SN Progenitors Without a Bright Companion at Explosion}\label{sec:lim_comp}

A diverse set of progenitors produce stripped-envelope SNe without a bright companion at the moment of explosion (Section \ref{sec:res1}). We provide an overview of their key properties in a matrix diagram, Figure~\ref{fig:triangle}.  Panels on the diagonal of the matrix show histograms with 1D distributions. The remaining panels in the upper right provide density maps of the 2D distributions, showing the correlations between pairs of parameters. 

The first three panels on the diagonal display the initial parameters that characterize the progenitor system. The first panel in the upper left shows the distribution of the initial mass of the primary star, $M_1$, or simply the initial mass in the case of a single star. In the second panel on the diagonal we show the distribution of the initial mass ratios, $q = M_2/M_1$. To visualize the contribution of single stars, we depict them as systems with a negative mass ratio, $q=-0.1$. The third entry gives the distribution of the initial orbital periods, $P$.  For clarity, single stars are not shown on this row. The remaining two panels on the diagonal show the total lifetime of the progenitor system until the SN explosion, $\Delta t$ (row 4), and the final pre-explosion mass of the progenitor of the stripped-envelope SN, $M_{\mathrm{final,prog}}$ (row 5).  

One of the most striking features is that two groups of progenitors can be distinguished.   This is most clearly seen in the distribution of the final pre-explosion mass of progenitor, but it can also be observed in the distribution of initial primary mass. The first group is characterized by high pre-explosion masses ($ M_{\mathrm{final,prog}}\gtrsim 10\Msun$). Strong stellar-wind mass loss is largely or fully responsible for the removal of the hydrogen envelope. We refer to this group as Group W to emphasize the role of winds.   The second group consist of progenitors where mass stripping by RLOF onto a companion is the main process responsible for the removal of the hydrogen-rich envelope.  We refer to this group as Group C to emphasize the role of the companion.  These groups are highlighted with approximate light-blue outlines in several of the panels in Figure~\ref{fig:triangle}. 
Although a bimodality is already present in the full progenitor population of stripped-envelope SNe, here we focus only on channels allowed by our observational constraints on a companion.
 In this case  Group C is approximately $1.6$ times more probable than group W.
 
\begin{description}
\item[ {\bf Group W (wind stripped)} ] 
The classic single-star progenitors form a subset of Group W. These all have initial masses larger than about 36\Msun\ in our  simulations. (Note that the standard simulation presented here is for a subsolar metallicity, $Z=0.0055$.) Group W also contains massive binary progenitors that contain at least one massive star above $\gtrsim 23$\Msun (see Figure~\ref{fig:triangle}). The progenitors resulting from binaries consist of massive mergers or massive secondaries of binary systems that were disrupted by the explosion of the primary. No companions are expected at the moment of explosion. 

\item[{\bf Group C (companion stripped)}] All of the progenitors in Group C are part of a binary system at birth, with $M_1 \lesssim 23$ \Msun. Most of them still have a companion at the moment of explosion, which is either a MS or a compact remnant. These scenarios represent the red ($\sim 28.3\%$) and orange ($\sim 5\%$) segments of Figure~\ref{fig:pie}, respectively. There is also a small contribution from more exotic channels. These include reverse mergers, in which the secondary star engulfs the stripped remnant of the primary ($\sim 1\%$ in Figure~\ref{fig:pie}). Another small contribution comes from some stars expelled from disrupted systems. These were secondaries that were stripped of their hydrogen-rich envelope during a reverse CE, \emph{before} being ejected by the SN of the primary. They are not massive enough to remove the envelope through winds after the disruption.
\end{description}

This bimodality in evolutionary channels translates into a diversity of various properties, several of which can, in principle, be inferred from observations.   In Table~\ref{table:ibc_bimodality} we provide the median values of several key properties for the two groups. These include the median initial and final pre-explosion mass of the progenitor and the expected ejecta mass. The latter depends on the mass of the compact remnant that is left behind, and here we assume a 2\Msun compact object. We further provide the typical pre-explosion mass-loss rates of the progenitors as given by our models.  Most notable is the difference in time delay between birth and explosion, a property that can in principle be inferred from age dating the surrounding stellar population.  The progenitors of group W all have short lifetimes with a median $\Delta t \approx 5$\,Myr.  Group C consists of lower-mass progenitors with longer lifetimes, with a median $\Delta t \approx  20$\,Myr. 

\begin{table*}[t!]
\caption{Properties of the ``wind stripped'' (W) and the ``companion stripped'' (C) groups.
\tablenotemark{$\alpha$}
} \label{table:ibc_bimodality}
\centering
\begin{tabular}{llccc}
\hline \hline 
 Quantity & Symbol [Unit]&   Group W  &   Group C \\
 && ``wind stripped'' & ``companion stripped''  \\
\hline  	
 Initial mass of the progenitor& $M_{1}$  [\Msun]   & 48   &  12   \\ 
 Mass of the progenitor at the momnet of SN& $M_{\mathrm{final,prog}}$ [\Msun]   & 14  & 3.5 \\
 Ejecta mass (assuming a 2\Msun remnant)&$M_{\mathrm{ej}}$ [\Msun]  & $\sim 12$  & $\sim 1.5$ \\
Wind-mass-loss rate of the progenitor &$\dot{M}$ [\Msun yr$^{-1}$] & $\gtrsim 10^{-6}$ & $\lesssim 10^{-6}$ \\
Time delay between birth and explosion&$\Delta t$ [Myr]   & 5.2  & 19.4   \\
&& No companion: 		&	Probably main-sequence  \\
Companion at explosion  && single,  forward merger,	&	 or compact companion \\
 &&  disrupted	&                    (unless  reverse merger or disrupted)   	\\
\hline 
\end{tabular}
\tablenotetext{1}{Values shown are approximate median values for stripped-envelope SNe without a MS companion of $>10 \Msun$, for our standard simulation at metallicity $Z=0.0055$. }
\end{table*}

\subsection{Additional Constraints from the Derived Ejecta Mass}\label{sec:lim_prog}

Up to this point we only considered constraints on the progenitor evolutionary scenarios coming from the upper mass limits on the presence of a companion. There are further observational constraints on SN 2002ap that provide additional clues about the nature of the progenitor. Here we consider findings by \citet{Mazzali+2002}, who estimate the ejecta mass from the photometric light curve combined with the spectral data and find $M_{\rm ej} \approx 2.5\Msun$. They argue that the event is consistent with a final progenitor mass of about 5\Msun. This estimate is subject to uncertainties, especially the question of whether the compact object left behind is a neutron star or a black hole. Nevertheless, we consider the implications of this additional constraint on the allowed progenitor scenarios discussed above.

In Figure~\ref{fig:triangle} we highlight the approximate constraints on the final progenitor mass in the panels in the rightmost column. Boundaries between 4 and 6\Msun, to encompass the 5\Msun estimate of \citet{Mazzali+2002}, are plotted with vertical green dashed lines. 
About $6\%$ of all stripped-envelope SNe have final progenitor masses in this range. 

This additional constraint provided by the progenitor final mass limits the progenitor of SN 2002ap to the higher-mass end of group C (cf.\ Section~\ref{sec:lim_comp}), thus having lost its envelope through binary interactions.  In most ($\sim$90\%) of these cases, the initial primary mass is $M_1 \approx 13$--23 $\Msun$ and the secondary companions have significantly lower initial mass, $q \lesssim 0.6$ (see Figure~\ref{fig:triangle}). Their initial orbital periods are longer than several tens of days, with a preference for very wide orbits of $P \gtrsim 1000 $ days.  These systems undergo highly nonconservative Case B or Case C mass transfer, possibly involving a CE phase. The companion does not accrete a significant amount of mass.   Conservative episodes of mass transfer are excluded, as they would result in massive companions bright enough to be observed in our survey.

There is also a less probable chance ($\sim$10\%) of the progenitor of SN 2002ap being an explosion of the secondary star, originating from binary systems of almost equal mass. For these we expect a bound compact companion to be present at the moment of explosion, either a white dwarf ($\sim 6\%$) or, less likely, a neutron star or black hole ($\sim 1\%$) or a stripped helium star ($\sim 1\%$). In 1\% of the cases the primary SN has already disrupted the system and the secondary does not have a bound companion.

A further minority channel for the progenitor of SN~2002ap concerns reverse mergers, in which the secondary engulfs the stripped remnant of the primary. Their contribution is not significant according to our simulations ($\sim 1\%$). However, they are of potential interest as they provide a way to produce the rapidly rotating systems that are probably required for SNe Ic-BL. We discuss these scenarios further in Section~\ref{sec:reverse}.

\section{Model Uncertainties and Variations} \label{sec:discussion1}

Our simulations are subject to various uncertainties, including the initial conditions and the treatment of the physical processes. We investigate the robustness of our findings by varying several of our main assumptions within an extensive grid of simulations computed and presented by \citetalias{Zapartas+2017} (see their Tables~1 and 2 for a full overview). We compute the probability of the various types of companions expected for stripped-envelope SNe. We limit the following discussion to those variations that have a significant impact on our results.  An overview is given in Table~\ref{table_uncert} in the Appendix.  Figures \ref{fig:metallicity} to \ref{fig:initial_conditions} show pie charts and the mass distributions of a possible MS companion for each unique set of assumptions.

Note that in the analysis of these uncertainties, the reference simulation assumes solar metallicity \citep[$Z=0.014$;][]{Asplund+2009} because of choices made by \citetalias{Zapartas+2017} (Model 00 of \citetalias{Zapartas+2017}; fifth panel in Figure~\ref{fig:metallicity}). This metallicity is higher than we assumed in our standard simulation in this paper, where we adopted a metallicity more appropriate for SN 2002ap ($Z=0.0055$).  Except for the metallicity, all other assumptions are the same.

\subsection{Effect of Metallicity}

Metallicity affects mass loss through stellar line-driven winds, with stronger winds at higher metallicity \citep[e.g.,][]{Vink+2001}. This is the main impact of metallicity variations on our results. Also, at lower metallicity the minimum mass for a core-collapse SN slightly decreases \citep[e.g.,][]{Pols+1998}, an effect taken into account in our models. Metallicity can also influence how stars interact in binary systems, but these effects are more subtle. For example, \citet[][]{Gotberg+2017} argue that the effect of metallicity on the opacity in the stellar interior influences how a star responds to Roche-lobe stripping, ultimately impacting the amount of hydrogen that remains. This may be important for the relative ratios of different subtypes of stripped-envelope SNe, but we do not consider those here.

In our low-metallicity simulations, stripping by a binary companion is the dominant mechanism for producing stripped-envelope SNe.  For higher metallicities, however, there is an increasing contribution of progenitor channels that rely on stripping of the envelope through stellar winds. These include single stars, mergers, and disrupted systems.  The increase can be seen in Figure \ref{fig:metallicity} of the Appendix.  

Metallicity also affects the ratio of stripped-envelope SNe to the total number of all core-collapse SNe.  We find that about $34\%$ of all SNe are stripped-envelope SNe in our standard simulation of $Z=0.0055$, but this increases to $\sim40\%$ at solar metallicity with $Z=0.014$. The observed value of about 36\% at roughly solar metallicity \citep{Smith+2011} supports the idea that wind mass-loss rates should be reduced to account for clumping. 

At the same time, for higher metallicities, the relative impact of lower-mass binaries, that produce stripped SNe with low-mass MS companions $(\lesssim 10 \Msun)$, decreases compared to systems with higher-mass MS companions.
This is based on the fact that some SN progenitors which are initially just above the minimum mass for SN and get stripped from a low mass companion, eventually are not massive enough to collapse and instead form a white dwarf, both because of the stronger mass loss and the slight increase of the minimum mass threshold for SN in higher metallicities. Thus, systems with low-mass MS companions $(\lesssim 10 \Msun)$ present at the moment of explosion become less prominent at higher metallicities.

In the case of SN~2002ap, the metallicity at the SN site is subsolar, but the exact value is uncertain (Section~\ref{sec:metallicity}).  The variations of $Z=0.004$ and $Z=0.008$, which can be seen as approximate boundaries for the uncertain metallicity boundaries of the SN~2002ap site, only result in small changes (see Figure~\ref{fig:metallicity}; Models 41 and 42 of \citetalias{Zapartas+2017}). 

\subsection{Variations in the Physical Assumptions}\label{sec:uncert_phys}

Varying the  mass-transfer efficiency, governed by parameter $\beta$ (Figure~\ref{fig:physical_assumptions}; Models 1--3 of \citetalias{Zapartas+2017}), directly impacts how much the companion accretes and how massive it will be at the moment of the explosion. This parameter is very poorly constrained. Certain observed post-interaction systems show support for highly conservative mass transfer ($\beta \approx 1$), while others clearly require a highly nonconservative one  ($\beta \approx 0$); see \citet{de-Mink+2007}, and references therein. An increase in $\beta$ results in a shift toward higher MS companion masses with fully conservative mass transfer, implying that a typical companion star is about 20\Msun. This is excluded by our observational data. The mass-transfer efficiency also affects the relative contribution of the binary channels. In particular, the contribution of stripped-envelope SN progenitors created via reverse mergers increases with conservative mass transfer. This is because  the mass gainer reaches a higher mass and is more likely to engulf the primary remnant (owing to the extreme mass ratio) and because the eventually more massive merger product is more efficient at ejecting its remaining hydrogen envelope. In the extreme nonconservative case ($\beta = 0$), the channels leading to stripped SNe with evolved companions (that are not shown in the pie charts) become significant relative to the other channels.

Another uncertainty arises from the question of how much angular momentum is lost from the binary system during nonconservative mass transfer. This depends on the mechanism by which mass is lost from the system. Observations \citep[e.g.,][]{Smith+2002} show evidence for both fast outflows from the accretor, which is our standard assumption, as well as mass ejection into a circumbinary disk, which corresponds to Model 5 of \citetalias{Zapartas+2017}. This is described by parameter $\gamma$.  Angular momentum loss results in a smaller orbit, which increases the contribution of mergers and decreases the contribution of channels where we expect a companion to be present at the moment of explosion, especially of low mass (Figure~\ref{fig:physical_assumptions}; Models 4 and 5 of \citetalias{Zapartas+2017}).

We also explore variations in the assumptions concerning the birth kick of the compact objects formed during a SN explosion, governed by parameter $\sigma$. The birth kick for neutron stars is constrained by the spatial velocities measured for pulsars \citep[e.g.,][]{Hobbs+2005}. However, the birth kick for black holes is a matter of  vivid current debate \citep{Repetto+2012,Janka2016,Mandel2016}.  Lower SN kicks mostly increase the importance of channels involving compact companions at the expense of the channels consisting of disrupted systems. They have little effect on the other evolutionary channels and no effect in the mass distribution of MS companions. Even when we assume all systems are disrupted at the moment of the SN explosion due to very high kicks, we find a maximum contribution of only about 12\% of all stripped-envelope SNe occur through this channel (Figure~\ref{fig:physical_assumptions}; Models 10 and 11 of \citetalias{Zapartas+2017}).

We assumed a typical value of unity for CE efficiency, $\alpha_{\mathrm CE}$, in our standard simulation \citep[e.g.,][]{Webbink1984, Iben+1984,Hurley+2002}, but there are also studies, mostly focused on lower mass systems, suggesting either higher values \citep[e.g.,][]{Nelemans+2000,Nelemans+2001,De-Marco+2011} or lower values \citep[e.g.,][]{Zorotovic+2010,Toonen+2013,Portegies-Zwart2013,Camacho+2014}.
In our variation with $\alpha_{\mathrm CE}=5$, where there is an increase of the number of systems that successfully eject their envelope and prevent coalescence, we find a larger contribution of stripped SNe with the presence of a compact companion (i.e., helium star, white dwarf, or more rarely a neutron star or a black hole).  Also, in the case of MS companions, the possibility of low-mass ones slightly increases as they are less likely to merge.  A low efficiency of $\alpha_{\mathrm CE}=0.2$ increases the contribution of mergers (Figure\ref{fig:physical_assumptions}; Models 13 and 16 of \citetalias{Zapartas+2017}).

We also examine the assumptions concerning which binaries undergo stable or unstable mass transfer.  This is governed in our simulations with critical mass ratio parameters, $q_{\rm crit}$, that effectively set the limiting mass ratio leading to unstable mass transfer. We independently vary the parameter for Case A mass transfer (which does not have a significant effect) and for Case B, when the star is crossing the Hertzsprung gap (HG). Increasing this parameter (i.e., $q_{\rm crit, HG}$ reaching closer to 1) leads to a larger fraction of systems entering contact or a CE phase, with many of them eventually merging, and decreases the contribution of channels with MS companions, especially of low mass (Figure~\ref{fig:physical_assumptions}; Models 22 and 23 of \citetalias{Zapartas+2017}).

Varying the efficiency of wind mass loss, $\eta$, affects the importance of channels that are stripped of their hydrogen envelope by wind, in a similar way to metallicity variations. The binary stripping mechanism dominates even more at low wind mass-loss efficiencies, which are suggested by some studies that take into account the effect of wind clumping \citep[for a review, see][]{Smith2014}. In Figure~\ref{fig:physical_assumptions} (including Models 25 and 26 of \citetalias{Zapartas+2017}), we show our quantitative results for wind mass-loss efficiencies of $\eta=1/10$, $1/3$, and $3$.

Finally, there are studies suggesting that the internal structure of the progenitor affects the potential final result of core collapse, in some cases leading to fallback of matter without a bright detectable transient \citep[e.g.,][]{OConnor+2011,Ugliano+2012}. Although the outcome of a core collapse seems to be very sensitive to the initial mass \citep[e.g.,][]{Sukhbold+2016}, to account for this uncertainty in a simplified way, we vary the maximum single-star mass for a SN, $M_{\max, \rm cc}$ (and equivalently for binary products with the same core mass), considering the possibility that the most massive stars do not produce an observable explosion. Channels that involve massive progenitors (single stars, massive mergers, and disrupted systems) become less important for lower $M_{\max, \rm cc}$. Also, the number of MS companions with very high masses ($\gtrsim 50$\Msun) decreases, but the overall mass distribution is not affected much because most of these companions come from lower-mass systems anyway (Figure~\ref{fig:physical_assumptions}; Models 28 and 29 of \citetalias{Zapartas+2017}).

\subsection{Variations in the Initial Conditions}

We vary the slope of the initial mass function, $\alpha$, between $-1.6$ and $-3.0$ \citep{kroupa2001,Kroupa+2003}. 
A flatter IMF favors more massive stars with stronger winds.  This increases the contribution of stripped-envelope SNe from Group W, which includes single stars, mergers, and disrupted systems (Figure~\ref{fig:initial_conditions}; Models 32--34 of \citetalias{Zapartas+2017}). The distribution of companion masses is also affected, with massive companions more likely with a shallower IMF.

The initial distribution of mass ratios, which is assumed to be a power law ($\mathrm{d}N/\mathrm{d}q \propto q^{\kappa}$), 
directly affects the distribution of companion masses at the moment of explosion (Figure~\ref{fig:initial_conditions}; Models 35 and 36 of \citetalias{Zapartas+2017}). A distribution that favors systems with unequal masses at birth (e.g., $\kappa=-1$) implies a larger fraction of low-mass companions to the progenitors of stripped-envelope SNe.  It also leads to a larger contribution of mergers and a smaller contribution of disrupted systems.

An initial period distribution that favors very close binaries even more (e.g., $\pi=-1$ in our assumed power-law distribution of $ \mathrm{d}N/\mathrm{d}\log_{10}P \propto (\log_{10}P)^{\pi}$), strongly increases the contribution of mergers and of systems that undergo conservative mass transfer, lowering the channels  with MS companions (Figure~\ref{fig:initial_conditions}; Models 37 and 38 of \citetalias{Zapartas+2017}). 

Changing the initial binary fraction, $f_{\rm bin}$, equally across the whole mass range (Figure~\ref{fig:initial_conditions}; Models 45 and 46 of  \citetalias{Zapartas+2017}) only scales our results by enhancing or decreasing the relative contribution of single stars. The shape of the distribution of companion masses and the relative contributions of the various binary progenitors are not affected. Drastically reducing the initial binary fraction from our standard 0.7 to 0.3, such that the majority of stars are single at birth, still implies that the majority of stripped SNe are produced from binary channels. Following a mass-dependent binary fraction (Figure~\ref{fig:initial_conditions}; Model 47 of  \citetalias{Zapartas+2017}) which favors binarity in massive stars (as shown in Equation 5 of \citetalias{Zapartas+2017}, which is based on \citealt{Moe+2013}) slightly increases the contribution of massive progenitors stripped through winds and at the same time favors slightly more massive MS companions.

\subsection{Potential Degeneracies and Higher-Order Effects}
We stress that in our uncertainty analysis we vary only one parameter at a time with respect to a reference simulation, which is the main simulation of \citetalias{Zapartas+2017}, assuming solar metallicity (panel 5 in Figure \ref{fig:metallicity}). We did not investigate 
higher-order effects when varying multiple uncertain parameters simultaneously, because of the computational cost. These effects are potentially important. For example, as we discussed above, both lowering the metallicity and the wind mass-loss efficiency decrease the contribution of wind-stripped SN progenitors. Lowering them both simultaneously (in other words, accounting for wind clumping at subsolar metallicity) should cause an even stronger effect.
Indeed, we run a test simulation assuming both $\eta = 1/3 $ \citep[e.g.,][]{Smith2014} and $Z=0.0055$ and find similar but stronger trends than varying each of these two parameters independently. In fact, the results are similar to the simulation in which we assume $\eta = 1/10 $ (but with same metallicity as the reference simulation, $Z=0.014$). Specifically for SN 2002ap, this would make massive single progenitor channels even less likely, further implying an origin from a ``companion-stripped,'' not very massive progenitor.

\section{Comparison with Earlier Studies}
\label{sec:comparison}

Previous papers have examined the possibility of the presence of a companion next to a core-collapse SN. \citet{Kochanek2009} estimated the fraction of core-collapse SNe that have a binary companion at the moment of explosion to be $\gtrsim 50\%$~based on initial binary fraction and distributions but without considering binary interactions. This result is consistent with our findings.

Recently, \citet{Moriya+2015}, also using the {\tt binary\_c} code, calculated that about $60\%$ of SNe~Ib and Ic are in binaries at the moment of explosion.  This result is similar to our findings.  \citet{Liu+2015} used the same {\tt binary\_c} models as in \citet{Moriya+2015}, but find higher companion masses compared to ours, despite adopting similar assumptions for the initial distributions.
We believe this to have occurred due to different assumptions on the mass and/or angular momentum losses during the RLOF phase (parameters $\beta$, $\gamma$). We have discussed the impact of these free parameters on our results in Section~\ref{sec:uncert_phys}.

\citet{Crockett+2007} discuss SN 2002ap and exclude the presence of a bright progenitor of SN 2002ap in pre-explosion images, thereby favoring a binary scenario. They also constrain any possible binary MS companion to be less massive than $20\Msun$.  We agree with their general conclusions, but our observational data provide a more stringent constraint of 8\Msun~for a MS companion star. For instance, the example discussed by  \citet{Crockett+2007} of an initially $20\Msun$~primary and a $14\Msun$~secondary evolving through CE evolution is excluded by our new  upper limits.

The possibility of the companion being a compact star is also mentioned by \citet{Crockett+2007} (mainly discussing the case of a neutron star or a black hole).  Our simulations show that channels involving the compact remnant of the primary as a companion are indeed possible (about $10\%$ of our parameter space for SN~2002ap), but it is more likely that the companion is a naked helium star or a white dwarf. Neutron star and black hole companions are disfavored because of natal kicks that are likely to unbind the system. They only remain bound to their companion under favorable directions and magnitudes of the kick.

\citet{Crockett+2007} argue for a Case B mass-transfer scenario for the progenitor of SN~2002ap. We find Case C mass transfer to be roughly equally probable.  Case C systems are more likely to successfully eject the envelope because of the larger orbital energy available and the lower binding energy of the envelope. This is especially true for lower-mass companions, to which we are restricted by the new observational data.

Case C mass transfer is considered by \citet{Crockett+2007} as more unlikely, arguing against it based on two important arguments. First, there is no evidence of interaction of SN 2002ap with a circumstellar medium (CSM) or of extra extinction toward the SN site, which may be expected if the explosion occurs shortly after the ejection of a CE \citep[e.g.,][]{Margutti+2017}. In our simulations, we find a typical time delay for Case C of $10^4$--$10^5$\,yr between ejection of the envelope and explosion. It is not clear that the CSM can remain close enough to the system for this long such that it would be detectable.  Second, \citet{Crockett+2007} argued that the time between the stripping of the hydrogen layer in Case C mass transfer and the explosion may not be long enough for standard mass-loss rates also to strip the star of its helium layers.  We discuss this valid point further in Section \ref{sec:helium}.

\section{Discussion and Implications}
\label{sec:dis}

In this section we discuss further observational tests to constrain our models and the specific complication of removing the helium layer in Type Ic and Type Ic-BL SNe. We then speculate on the bimodality of the progenitor populations and the exotic merger channels, which provide interesting channels to power a SN~Ic-BL engine with angular momentum.

\subsection{Systematic Searches for Companions}

Larger samples of stripped-envelope SN observations and their possible companions would contribute to a better understanding of the evolutionary histories of their progenitors. Eventually, comprehensive statistical comparisons of these observations with theoretical predictions may constrain uncertain physical processes that play a role in single and binary stars.  For example, our models assuming highly conservative accretion predict that the distribution of masses of MS companions peaks near 20\Msun (variation with $\beta =1$ in Figure~\ref{fig:physical_assumptions}). Such companions would be detectable in nearby events \citep[e.g.,][]{Crockett+2007}.  If observational searches systematically fail to detect companions, this would begin to rule out such models and have important implications on the final products of stripped-envelope SN progenitor systems, including gravitational wave sources \citep{LIGO+2016_mainPaper}.  

Unfortunately, there are only a limited number of recent stripped-envelope SNe that are sufficiently nearby to perform a similar analysis as we have presented for the SN~2002ap.  A different approach is to systematically search for companions in SN remnants within the Local Group, as originally proposed by \citet{van-den-Bergh1980}.  Their proximity allows for both deep searches and accurate characterization of the local stellar population \citep[e.g.,][]{Williams+2014}.  Already, 77 SN remnants have been identified in the Magellanic Clouds \citep[e.g.,][ and references therein]{Badenes+2010} and 245 SN  remnants in M31 and M33 \citep[e.g.,][ and references therein]{Jennings+2014, Elwood+2017}. A homogeneous analysis of these samples providing constraints on the presence of companions would be valuable. Such a study is not available at present, but several individual SN remnants have been studied in depth and we discuss them below.

\citet{Dufton+2011} discuss a possible case of a massive MS former companion to the progenitor of the pulsar PSR J0537$-$6910. This pulsar is located in 30 Doradus in the Large Magellanic Cloud. They propose the very rapidly rotating O9-type runaway star VFTS102 as the candidate companion. However, the location of VFTS102 appears to be outside the remnant. Proper-motion measurements are needed to test the hypothesis of a common origin. 

Possibly the most convincing case of a companion detection is that associated with SN remnant S147. \citet{Dincel+2015} report the discovery of a main-sequence B0.5~V-type runaway star (HD37424) with a mass of 13\Msun inside the remnant.  The authors argue that the trajectory of the star can be traced back to the position of the remaining central compact object, PSR J0538+2817. \citet{Boubert+2017} confirm this candidate companion using an independent Bayesian method, which takes into account both the kinematic and the photometric properties of stars around the SN remnant combined with the expected properties of runaway stars computed by {\tt binary\_c}. This method also helps constrain the initial configuration of the possible binary system and its evolutionary history until the explosion. In the case of HD37424, they find that it was most likely initially a $7 \pm 2 \Msun$ star that accreted mass and became a runaway of $10.38 \pm 1.04 \Msun$, a bit less massive than what \citet{Dincel+2015} report. 

With the same method, \citet{Boubert+2017} identify three new possible former companions of SN progenitors in other nearby remnants. The first is a massive B-type star showing emission lines (referred to as Be-type) of around $11\Msun$ in the HB21 SN remnant. The emission features may be explained by prior mass accretion onto the star by the SN progenitor \citep{Harmanec1987,Pols+1991}.  Another massive B5~V-type star ($\sim 6 \Msun$) may be connected  with the Monoceros Loop remnant, although the young age of the nearby association, of which the remnant is possibly a member, argues against this scenario. Finally, a lower-mass A-type star ($\sim 1.7 \Msun$) seems to be associated with the Cygnus Loop remnant. \citet{Boubert+2017} discuss different possible channels in which the star either did not accrete mass during its evolution or lost most of its initially higher mass owing to mass transfer onto the SN progenitor.

There are a few more cases of possible low-mass MS companions. \citet{Tetzlaff+2013} investigate the flight paths of seven neutron stars in connection with the nearby Antlia SN remnant \citep{McCullough+2002}. They argue that the pulsar PSR J0630-2834 and the runaway star HIP\,47155 are both possibly associated with the remnant. The runaway star is an A-type dwarf, $\lambda$\,Bootis star \citep{Houk1982,Paunzen2001}. If this is indeed the progenitor companion, it would be at the low end of the mass distribution that we predict for MS companions.  Another low-mass MS star, the G0~Ia runaway star  HIP~13962, has also been suggested as a former companion to the young pulsar PSR J0826+2637 \citep{Tetzlaff+2014a}. The most intriguing case of a low-mass MS companion concerns the SN remnant RCW 103. \citet{Pizzolato+2008} propose that the X-ray source 1E~161348-5055 is a neutron star in close orbit with a low-mass MS star.  \citet{Tetzlaff+2014} discuss a  possible common origin for PSR J0152-1637 and the runaway star  HIP 9470.  However, the runaway star itself is a single-lined spectroscopic binary.  A common origin with the pulsar implies that the progenitor must have been a triple system. We have not considered  triple systems in our simulations, but many stars are found in multiple systems \citep{Moe+2016}, so this remains a possibility.

Finally, it is worth noting that upper limits to the detection of a companion to the Cas~A progenitor have been claimed to disfavor MS stars above a few $\Msun$  \citep[][]{Kochanek2017}; Cas~A is thought to be the remnant of a Type IIb SN based on light-echo spectra \citep{Krause+2008,Rest+2008}. These findings can be reconciled with our simulations if Cas~A is either the result of a disrupted system, merger, or true single star, or if Cas~A formed a large amount of dust that may obscure a companion.

\subsection{The Absence of Helium in SNe~Ic and SNe~Ic-BL}\label{sec:helium}

Stripped-envelope SNe of Type Ic and Ic-BL do not show signs of helium in their spectra.  How the helium layers are removed (or how helium can be hidden from observation) remains an open question \citep[e.g.,][and references therein]{Shivvers+2016, Dessart+2016, Liu+2016, Modjaz+2016, Yoon+2017}.   

For the case of SN 2002ap, the removal of the helium layer presents several complications to our analysis, which suggests a low-mass binary progenitor scenario.  In general, helium layers can also be partially removed through binary interaction when a low-mass helium star ($M\lesssim 3\Msun$), having already being stripped of its hydrogen-rich envelope, fills its Roche lobe again after completing central helium burning \citep[e.g.,][]{Habets1986,Gotberg+2017}.  This scenario provides a satisfactory explanation for the Type Ic SN~1994I \citep{Nomoto+1994,Van-Dyk+2016}, but is not expected for helium stars with higher masses, such as for a helium core of $\sim 7\Msun$ for SN~2002ap \citep{Mazzali+2002}. Furthermore, for scenarios in which the progenitor does not have a nearby companion at the moment of explosion (single stars, mergers, and disrupted systems), the helium layers could not have been stripped by binary mass transfer.

An additional possibility for the removal of helium is via enhanced mass loss in the late phases of evolution (see \citealt{Smith2014}, for a review). There is no clear consensus concerning the mechanism that may be responsible for such extreme mass loss, but synchronization with the time of core collapse points to instabilities arising during late burning phases.  Energy from the latest nuclear burning sequences may deposit heat in a star's envelope, driving sudden mass loss or swelling the star to trigger binary interaction \citep{Arnett+2011,Quataert+2012,Smith+2014}.  This appears to be a plausible explanation for the origin of SNe IIn, and similar processes might also eject a significant fraction of the more tightly bound helium layers, as observed in SNe Ibn.

If mass loss occurs shortly before the explosion, one would expect evidence of a dense CSM in radio and X-ray observations.  No such evidence is found for SN~2002ap \citep{Berger+2002,Sutaria+2003,Bjornsson+2004,Soria+2004,Margutti+2017}.  \citet{Crockett+2007} argue that this implies that any major mass-loss event must have happened at least 500--1200\,yr prior to the explosion, otherwise signals would have been seen, assuming mass ejected with a velocity of $\sim 100$\,km\,s$^{-1}$ by binary interaction. The same argument holds for enhanced mass loss during late burning phases, although possible higher velocities of the lost mass shorten the excluded time difference between the mass-loss event and the explosion. The absence of any evidence of extra extinction toward the SN~2002ap site can also be an indication of no recent major mass-loss event. 

Because SNe~Ic-BL are rare, more exotic explanations for the absence of helium may apply.  Some of our formation channels involve mergers; we discuss these in Section \ref{sec:reverse}. 

\subsection{Progenitor Bimodality and SN~Ic-BL Ejecta Masses}\label{sec:bimodality_discussion}

The ejecta mass inferred for SN~2002ap by  \citet{Mazzali+2002} is about $2.5\Msun$. This is comparable to that of other SNe~Ic-BL, including SN~2004aw and SN~2003jd, both of which lack detected GRBs \citep{Taubenberger+2006,Valenti+2008}, but also SN~2006aj \citep{Mazzali+2006a}, which did have an associated GRB.  However, some SNe~Ic-BL have estimated ejecta masses of  $\sim (8$--15) $\Msun$, including SN~1997ef \citep{Mazzali+2000}, SN~1998bw\citep{Iwamoto+1998}, SN~2003dh \citep{Mazzali+2003}, and SN~2003lw \citep{Mazzali+2006}. The last three of these SNe have associated GRBs. It is not clear at present whether there is a continuum of events with different ejecta masses or we are instead dealing with two classes of SNe~Ic-BL (see \citealt{Smartt+2009,Modjaz+2016} for a discussion).  

If we assume that SNe~Ic-BL are a uniform subset all stripped-envelope SNe in our simulations,  
it is tempting to speculate about two classes of SNe Ic-BL given the bimodality in the progenitor population found in our simulations above.  As discussed in Section~\ref{sec:lim_comp}, we find that the progenitors can be divided in two distinct groups, one ``wind-stripped'' and one ``companion-stripped'' (Groups W and C, respectively).  We expect Group C to have low ejecta masses of a few $\Msun$ (see Table~\ref{table:ibc_bimodality}).  Type Ic-BL SNe only constitute a small fraction of the stripped-envelope SNe that we have modeled, and we still lack understanding of the physical conditions necessary for these energetic phenomena.  A further caveat is the difficulty of predicting the final masses of the compact remnants and thus the ejecta masses of the explosions \citep[e.g.,][]{OConnor+2011,Ugliano+2012,Fryer+2012,Nadezhin1980, Lovegrove+2013, Piro2013,Sukhbold+2014,Sukhbold+2016}.

Nevertheless, it appears worthwhile to further investigate a possible connection between the bimodality in the progenitor scenarios and the explosion properties of Type Ic-BL SNe.  Our simulations indicate several trends that can be observationally tested (see Table~\ref{table:ibc_bimodality}).  For example, Group C is characterized by longer delay times. This parameter can be measured by age-dating the surrounding population of stripped-envelope SNe \citep{Leloudas+2011,Murphy+2011,Williams+2014,Jennings+2014}.

Group C is also expected to have lower stellar wind mass-loss rates prior to the explosion, which can be measured with radio and X-ray observations post-explosion. Radio and X-ray observations that can probe the nearby surroundings of the SN may provide valuable answers about the progenitor history just before the explosion \citep[e.g.,][]{Margutti+2014,Margutti+2017}. In the case of SN~2002ap, \citet{Berger+2002} infer a mass-loss rate of the progenitor of $\sim 5 \times 10^{-7} \Msun~\mathrm{yr}^{-1}$, consistent again with a binary stripped progenitor of lower mass than a WR star. The technique of ``flash spectroscopy'' \citep[early spectra due to recombination of possible CSM ionized by the shock breakout flash; e.g.,][]{Gal-Yam+2014} may also provide valuable information about the progenitor mass-loss rate shortly prior to the SN
\citep[e.g.,][for Type IIb SN 1993J and 2013cu, respectively]{Benetti+1994,Gal-Yam+2014}.

\subsection{Exotic Mergers as Progenitors of Some SNe~Ic-BL} \label{sec:reverse}

Type Ic-BL SNe are rare events, comprising only $\sim1$\% of all core-collapse SNe \citep[e.g.,][]{Smith+2011}. The physical requirements that are essential for such explosions are not known, but several have argued that a large amount of angular momentum is necessary \citep[e.g.,][]{Woosley1993,MacFadyen+1999,MacFadyen+2001,Dessart+2008}.  In this respect, it is interesting to highlight that a small fraction of the stripped-envelope SN progenitors in our simulations arise from exotic evolutionary channels involving a reverse merger. For our standard simulation of $Z=0.0055$, they constitute around 1\% of stripped-envelope SNe (Figure \ref{fig:pie}), but this fraction increases for higher metallicities (see channel ``mergers (R.)'' in Table \ref{table_uncert}). In these systems, the evolving secondary star engulfs the stripped remnant of the primary, which can be a helium-burning compact star or a white dwarf. The hydrogen-rich envelope of the secondary is either lost before merging in a previous mass-transfer episode or immediately after it, in which case the remaining hydrogen envelope is very thin.

The outcome of such merger events is highly uncertain, but it likely results in an evolved, rapidly rotating stellar object.  Qualitatively, the merged star will already have completed its central hydrogen-burning phase and possibly even contain a helium-exhausted core. In such a scenario, the remaining lifetime of these merged stars will be short.  With little time for angular momentum loss through stellar winds, a rapidly rotating star will be present at the moment of core collapse. An accretion disk may form around the neutron star or black hole that can feed the compact object, as proposed in the collapsar scenario by \citet{Woosley1993} for long GRBs. Interestingly, \citet{Tout+2011} suggest a similar scenario of reverse merger between an oxygen/neon white dwarf and the compact core of an evolved secondary as a possible progenitor of long GRBs. These mergers form only a minority channel in our simulations, but GRBs are rare compared to core-collapse SNe, and the connection between GRBs and Type Ic-BL SNe makes them worth considering as a possible progenitor channel.

\section{Summary}
\label{sec:con} 

In this study, we present theoretical predictions for the presence of binary companions to stripped-envelope SNe (IIb, Ib, Ibn, Ic, and Ic-BL) based on binary population synthesis simulations. 
We compare our predictions with our new {\it HST} observations in which we search for a companion at the explosion site of the Type Ic-BL SN~2002ap. No companion was found, and the data provide new deep upper limits excluding the presence of a MS companion more massive than about 8\Msun.  We use SN~2002ap as a case study and interpret new and existing constraints in the theoretical framework that our simulations provide.  Our main findings are the following. 

\begin{itemize}
\item According to our standard simulation (for subsolar metallicity, $Z=0.0055$), 68\% of all stripped-envelope SNe are expected to have a MS companion at the moment of explosion. The SN progenitor loses its hydrogen envelope due to binary interactions with a companion, which accretes part of the transferred mass in some cases.
\item Around one in four stripped-envelope SNe have no companion nearby, either because they were born as single stars or they originate from a binary system that merged or was disrupted by a prior SN.  
\item In the remaining cases of stripped-envelope SNe, we expect a compact companion (most likely a stripped helium star or else a white dwarf/neutron star/black hole).  Companions that are giants are very rare.
\item We investigate the distribution of masses in the case of MS companions, which is broad and peaks at about $ 9\Msun$ in our standard simulation. If we assume a conservative upper limit of $ 10 \Msun$ for a MS companion of SN~2002ap, approximately $40\%$ of all stripped SN channels are ruled out. 
\item We find a bimodal distribution of the final progenitor mass of the stripped-envelope SNe without a MS companion more massive than $ 10 \Msun$, as in the case of SN~2002ap. We identify two groups, ``wind-stripped'' (W) and ``companion-stripped'' (C), to indicate the main mechanism responsible for removal of the hydrogen-rich envelope. Wind-stripped progenitors are either massive single stars, mergers, or massive stars ejected from a binary system.  Companion-stripped progenitors originate from binary systems with initial primary mass below about $ 23 \Msun$. We speculate about a link between this bimodality and the apparent spread in the ejecta masses of Type Ic-BL SNe. We discuss how observations of the surrounding population of stripped-envelope SNe (and of SNe~Ic-BL in particular) may test this hypothesis.
\item Our results are consistent with the progenitor of SN~2002ap being the initially more massive star of a binary system. The initial primary mass, $M_1$, is roughly 13--23$\Msun$ and the initial mass ratio $M_2/M_1 \lesssim 0.6$. The progenitor experienced nonconservative Case B or Case C mass transfer, possibly involving CE evolution, with its companion. These conclusions are generally similar to those of \citet{Crockett+2007}, but our new deeper limits on a possible companion call for initially less-massive secondary stars.
\item Our predictions of expected companions and of their properties can be compared with other observational searches for companions to stripped-envelope SNe, constraining the possible evolutionary scenarios of their progenitors.
We show results for variations in the assumed metallicity, in the initial conditions of the population, and in physical parameters governing stellar and binary evolution.  A statistically significant sample of constraints on companions to stripped SNe, for example coming from SN remnants in the Local Group,  may allow us to test the physics of stellar evolution and of binary interaction. 

\end{itemize}

\section*{Acknowledgments}
We thank Jon Mauerhan, Isaac Shivvers, Jeff Silverman, Heechan Yuk, and WeiKang Zheng for contributing to the observing proposal. We are grateful to Ylva G\"otberg, Maryam Modjaz, Silvia Toonen and Jacco Vink for very useful discussions.
This work is based on observations made with the NASA/ESA {\it Hubble Space Telescope}, obtained at the Space Telescope Science Institute (STScI), which is operated by the Association of Universities for Research in Astronomy, Inc., under NASA contract NAS 5-26555.  It is also based in part on observations made with the {\it Spitzer Space Telescope}, which is operated by the Jet Propulsion Laboratory (JPL), California Institute of Technology, under a contract with NASA.  Support was provided by NASA through grants GO-14075 and AR-14295 from STScI. We thank Andrew Dolphin for his patient advice on how best to implement artificial star tests in Dolphot.
E.Z. is supported by a grant of the Netherlands Research School for Astronomy (NOVA).  S.dM. acknowledges support by a Marie Sklodowska-Curie Action (H2020 MSCA-IF-2014, project BinCosmos, ID 661502).  A.V.F.'s group is also grateful for generous financial assistance from the Christopher R. Redlich Fund, the TABASGO Foundation, and NSF grant AST-1211916. 
N.S. is grateful for support from NSF grants AST-1312221 and AST-1515559.
R.G.I. thanks the STFC for funding his Rutherford Fellowship under grant ST/L003910/1, and Churchill College, Cambridge, for his fellowship and access to their library.
%

\bibliographystyle{aasjournal}
\bibliography{my_bib}

\appendix

\begin{table*}[p!]
  \caption{Variations of the metallicities, physical assumptions and initial conditions considered.\tablenotemark{a} \label{table_uncert}}
  \centering
   {\small
  \begin{tabular}{llccccccc}
  \hline \hline 
Model &Description  & MS comp.   & MS comp. & compact & born  & disrupted & mergers& mergers\\  
\citetalias{Zapartas+2017} &&$>10\Msun$&$<10\Msun$&comp.&single&& (F.)& (R.)\\
\hline
&  & \multicolumn{7}{c}{[------------------------------------------------\%-----------------------------------------------]}\\  
{\bf 00} &  {\bf Reference simulation of \citetalias{Zapartas+2017} } &      {\bf  35.7 } &   {\bf    20.3} &  {\bf   2.8 }&  {\bf   9.3} &  {\bf   11.6} &  {\bf   16.4} &  {\bf   3.0} \\
  &{\bf(solar metallicity, $Z=0.014$)}&&&&&&&\\
  
\multicolumn{5}{l}{\it \bf Metallicity} \\

39 & $Z=0.0002$                                               &  $     35.1$ &     33.1 &    9.5 &    5.6 &    6.0  &    7.5 &    0.8  \\
40 & $Z=0.001$                                                 &  $     35.3$ &     33.8 &    8.4 &    5.4 &    6.1  &    8.4 &    0.7  \\
41 & $Z=0.004$                                                 &  $     37.3$ &     30.7 &    6.8 &    5.3 &    6.5  &    10.4 &    1.0  \\
-- & $Z=0.0055$ {\bf(standard in this work)}       &  $     40.3$ &     28.3 &    4.8 &    6.8 &    7.6  &    11.2 &    1.0  \\
42 & $Z=0.008$                                                 &  $     36.7$ &     25.7 &    4.6 &    7.4 &    9.1  &    13.8 &    1.4  \\
43 & $Z=0.02$                                                   &  $     31.4$ &     18.4 &    3.8 &    9.7 &    12.8  &    17.6 &    5.3  \\
44 & $Z=0.03$                                                   &  $     26.2$ &     14.0 &    2.9 &    13.7 &   16.0  &    21.3 &    5.0  \\
  
\multicolumn{5}{l}{\it \bf Physical assumptions} \\
01 &  mass transfer efficiency, $\beta=0$                        &  $     23.8$ &     31.8 &    1.8 &   11.0 &     6.7 &    16.5 &    0.7  \\ 
02 &  mass transfer efficiency, $\beta=0.2$                      &  $     39.1$ &     23.8&     0.5 &   10.6 &     9.0 &    15.2&    0.7 \\
03 &  mass transfer efficiency, $\beta=1.0$                      &  $     40.2$ &      5.5&     2.6 &    7.7 &    14.8 &    14.2 &    14.2 \\
04 &  angular momentum loss,  $\gamma=0$                         &  $     37.1$ &     18.8 &    2.6 &    9.5 &    12.3 &    16.0 &    2.8  \\
05 &  angular momentum loss,  $\gamma=\gamma_{\rm disk}$         &  $     29.4$ &      9.8 &    0.6 &   14.3 &    13.3 &    29.1 &    2.4  \\
10 &  birth kick of compact remnant,  $\sigma = 0$           &  $     35.0$ &     19.6&    16.4 &    9.3 &     0.0 &    16.0 &    2.9 \\
11 &  birth kick of compact remnant,  $\sigma  =\infty$             &  $     35.5$ &     19.7 &    2.5 &    9.4 &    12.4 &    16.3 &    2.9  \\
13 &  common envelope efficiency,   $\alpha_{\rm CE} = 0.2$     &  $     35.1$ &     16.7 &    0.7 &    9.3 &    11.5 &    20.0 &    5.9  \\
16 &  common envelope efficiency,   $\alpha_{\rm CE} = 5.0$     &  $     32.4$ &     22.9 &    9.9 &    8.5 &    10.6 &    11.2 &    2.7  \\
22 & critical mass ratio, $q_{\mathrm{crit,HG}} = 0.25$          &  $     35.5$ &      23.9&    2.8 &    9.1 &    11.2 &    13.8 &    2.9  \\
23 & critical mass ratio, $q_{\mathrm{crit,HG}} = 0.8$           &  $     27.7$ &     13.7 &    3.2 &   11.4 &    10.8 &    29.1 &    3.1  \\
-- &  stellar wind mass-loss efficiency $ \eta= 0.1$           &  $     40.1$ &     30.6 &    3.6 &    5.7 &     6.7 &    11.4 &    1.1  \\
25 &  stellar wind mass-loss efficiency, $ \eta= 0.33$           &  $     42.0$ &     25.7 &    3.5 &    6.2 &     7.3 &    13.3 &    1.1  \\
26 &  stellar wind mass-loss efficiency, $ \eta= 3.0$            &  $     24.0$ &     10.7 &    1.8 &   17.7 &    18.8 &    20.2 &    6.4  \\
28 &  $M_{\max, \rm cc}\tablenotemark{b} =35$    &  $     42.3$ &     24.9 &    3.3 &    6.0 &  7.4  &    11.9 &    3.6  \\
29 & $M_{\max, \rm cc}\tablenotemark{b} =20$     &  $     50.7$ &     34.8 &    4.5 &    0.0 &   0.2 &    5.2 &    3.6  \\

\multicolumn{5}{l}{\it \bf Initial conditions} \\
32 & initial mass function, $\alpha =-1.6$                      &  $     34.5$ &     12.2 &    1.6 &    14.1 &    16.2  &    18.6 &    1.6  \\
33 & initial mass function, $\alpha =-2.7$                      &  $     35.4$ &     24.2 &    3.9 &    7.0 &    9.2  &    15.6 &    4.0  \\
34 & initial mass function, $\alpha =-3.0$                      &  $     34.5$ &     26.8 &    4.8 &    5.3 &    7.5  &    15.6 &    4.8  \\
35 &  initial mass ratio distribution, $\kappa=-1$              &  $     26.6$ &     24.1 &    1.6 &    12.4 &    8.0  &    24.9 &    1.8  \\
36 &  initial mass ratio distribution, $\kappa=+1$              &  $     40.8$ &     15.6 &    3.9 &    8.0 &    14.1  &    12.5 &    3.9  \\
37 &  initial period distribution,   $\pi=+1$                   &  $     39.1$ &     26.8 &    2.3 &    8.7 &    12.7  &    7.5 &    1.2  \\
38 &  initial period distribution,   $\pi=-1$                   &  $     32.4$ &     12.1 &    2.3 &    11.1 &    11.4  &    25.7 &    4.4  \\
45 &  binary fraction, $f_{\mathrm{bin}} = 0.3$                 &  $     25.3$ &     14.0 &    2.0 &    36.3 &    8.3  &    11.4 &    2.1  \\
46 &  binary fraction, $f_{\mathrm{bin}} = 0.99$                &  $     39.4$ &     21.9 &    3.1 &    0.3 &    12.9  &    18.2 &    3.3  \\
47 & mass dependent binary fraction \tablenotemark{c}, $f_{\mathrm{bin}} (M_1)$   &  $     36.2$ &     18.7 &    2.6 &    10.0 &    12.4  &    16.3 &    2.8  \\  
\hline
  \end{tabular}
  \tablenotetext{1}{Fractions of possible evolutionary channels and companions of stripped SNe to the total number of them. Terms have the same meaning as in Figure \ref{fig:pie}. First column shows the model number corresponding to Table 2 of \citetalias{Zapartas+2017}. We use the standard simulation of \citetalias{Zapartas+2017} as a reference, which assumes a solar metallicity of $Z=0.014$ (Model $00$, panel 5  in Figure \ref{fig:metallicity}), and we change our assumed parameters one by one. For an overview of the variations we refer to Tables 1 and 2 of \citetalias{Zapartas+2017}.   }
\tablenotetext{2}{Maximum single-star equivalent birth mass for a core-collapse SN.} 
\tablenotetext{3} {Favoring binarity in massive stars (Equation 5 of \citetalias{Zapartas+2017}, based on \citealt{Moe+2013})}
  }
\end{table*}

\section{Variations in Assumptions}

\begin{figure*}
\begin{center}
\includegraphics[width=\textwidth]{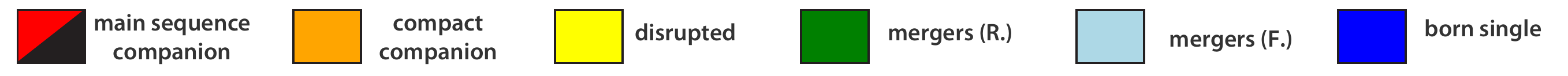}
\noindent\makebox[\linewidth]{\rule{0.8\paperwidth}{0.4pt}}\\
\includegraphics[width=\pievariable\textwidth]{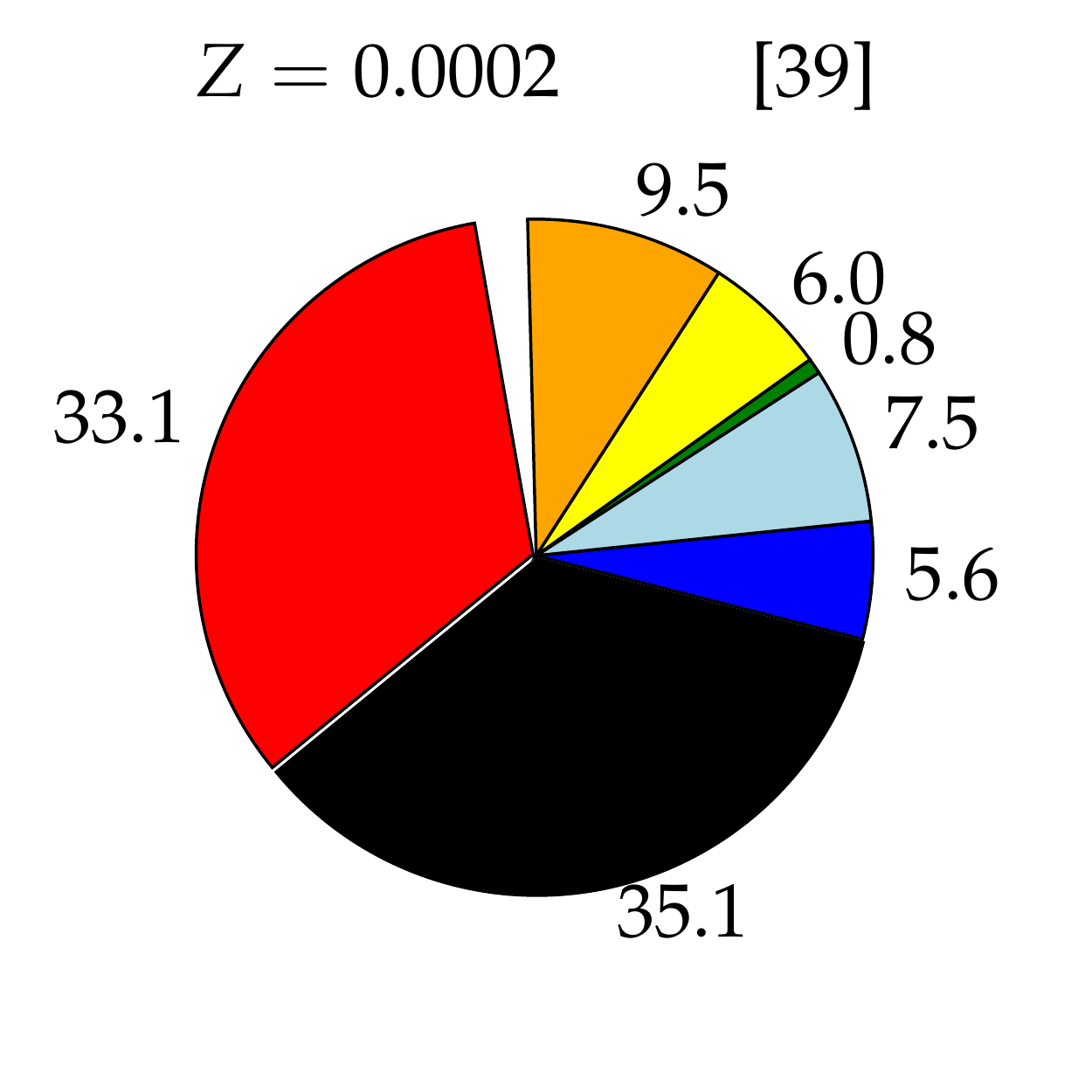}%
\includegraphics[width=\pievariable\textwidth]{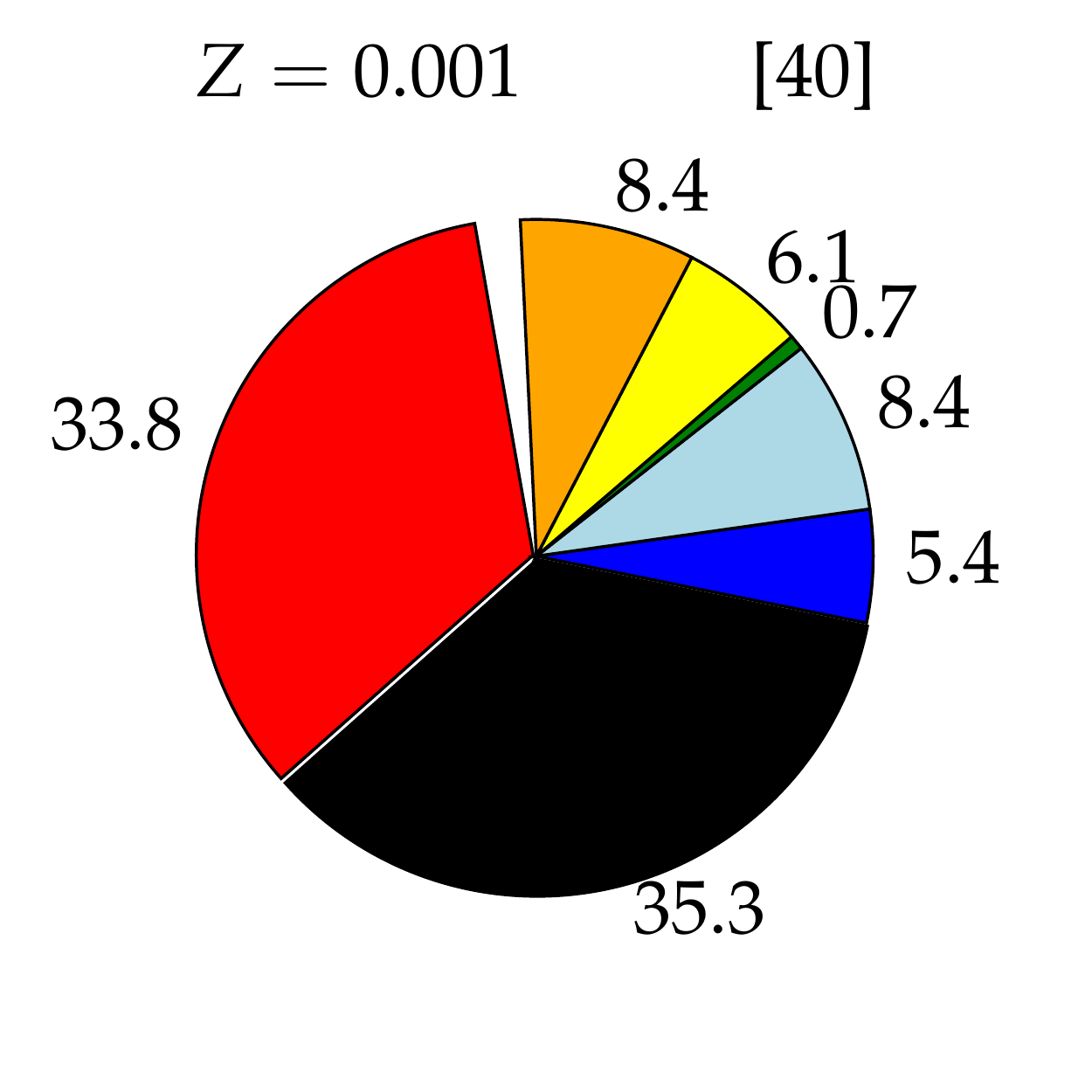}%
\includegraphics[width=\pievariable\textwidth]{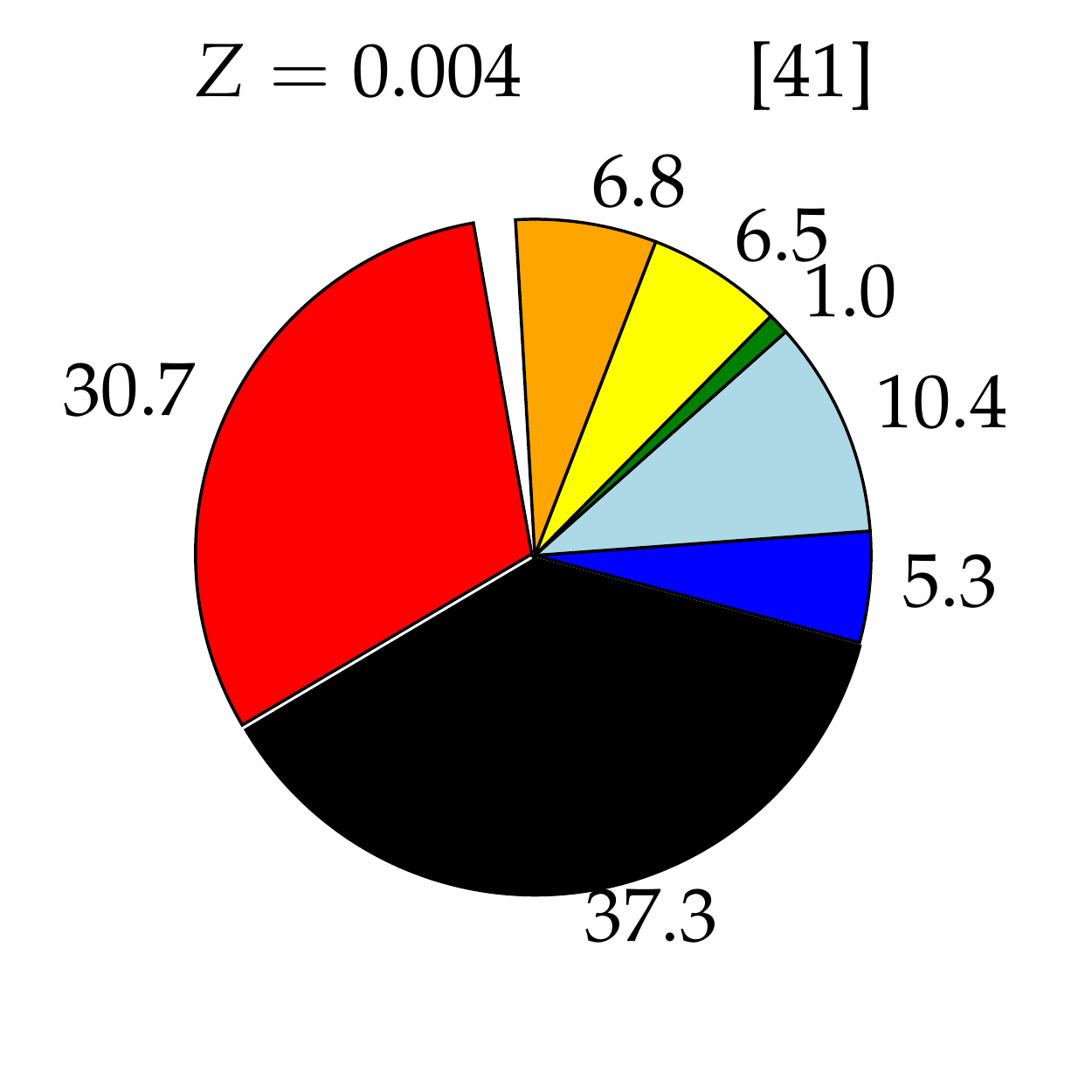}
\includegraphics[width=\pievariable\textwidth]{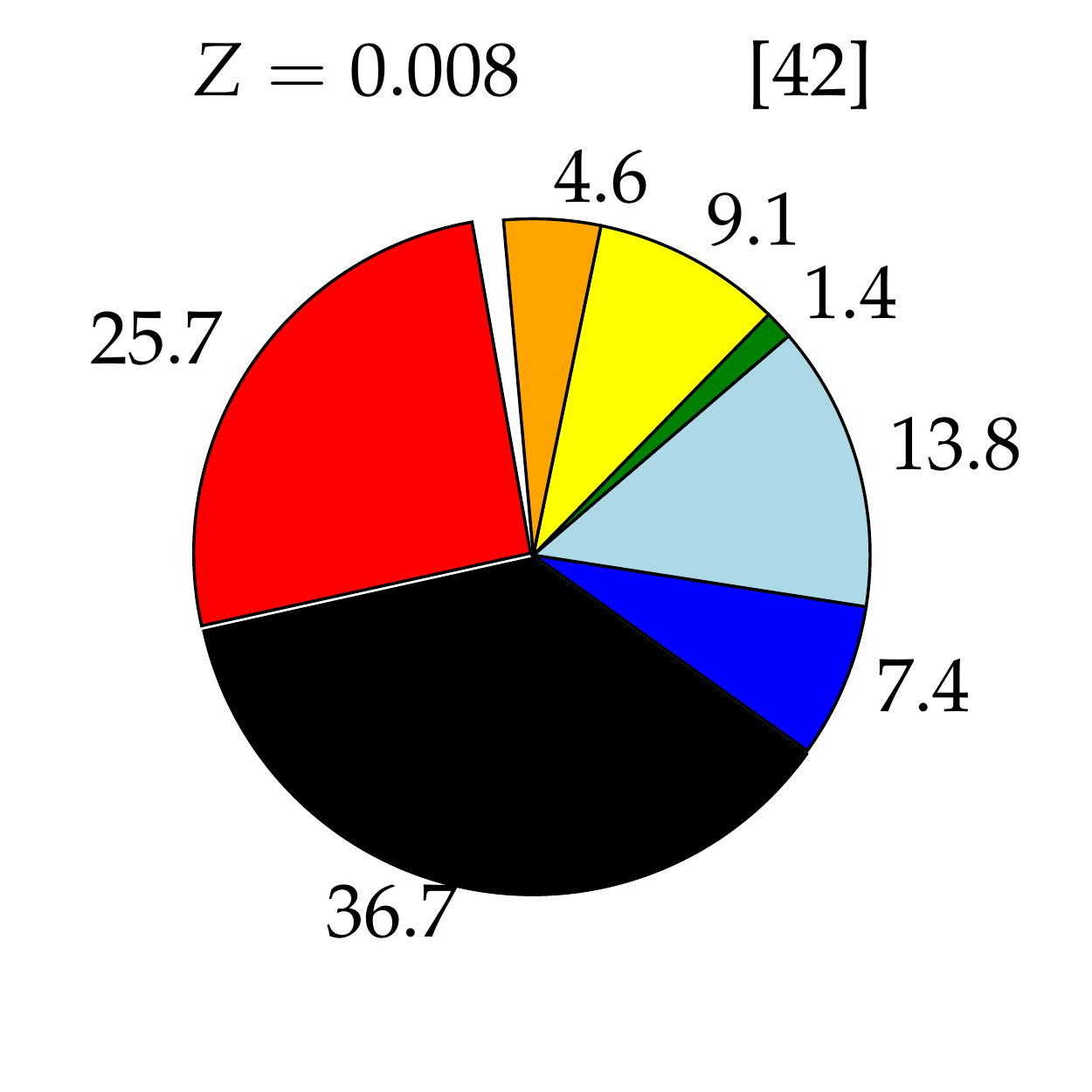}\\%
\includegraphics[width=\onedvariable\textwidth]{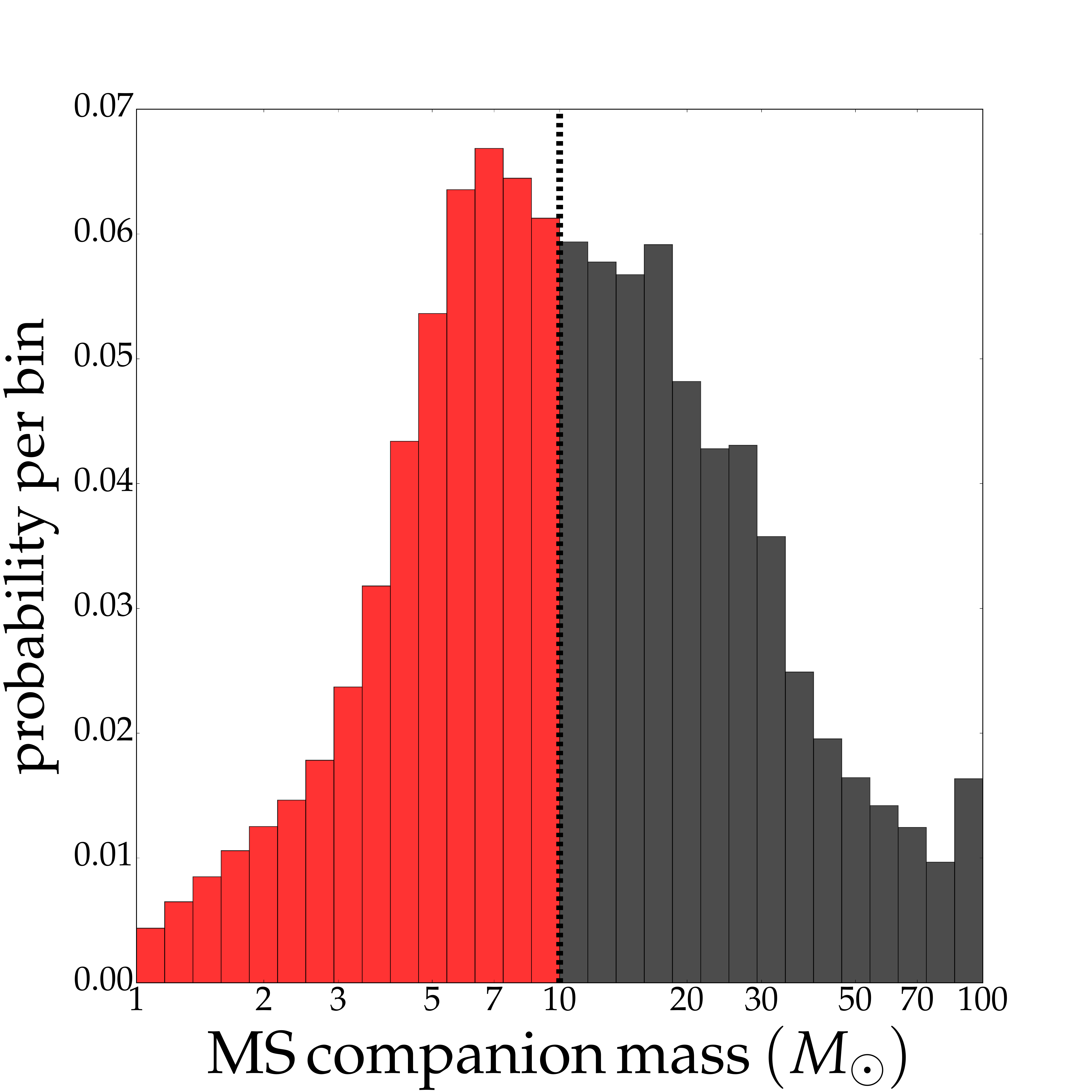}
\includegraphics[width=\onedvariable\textwidth]{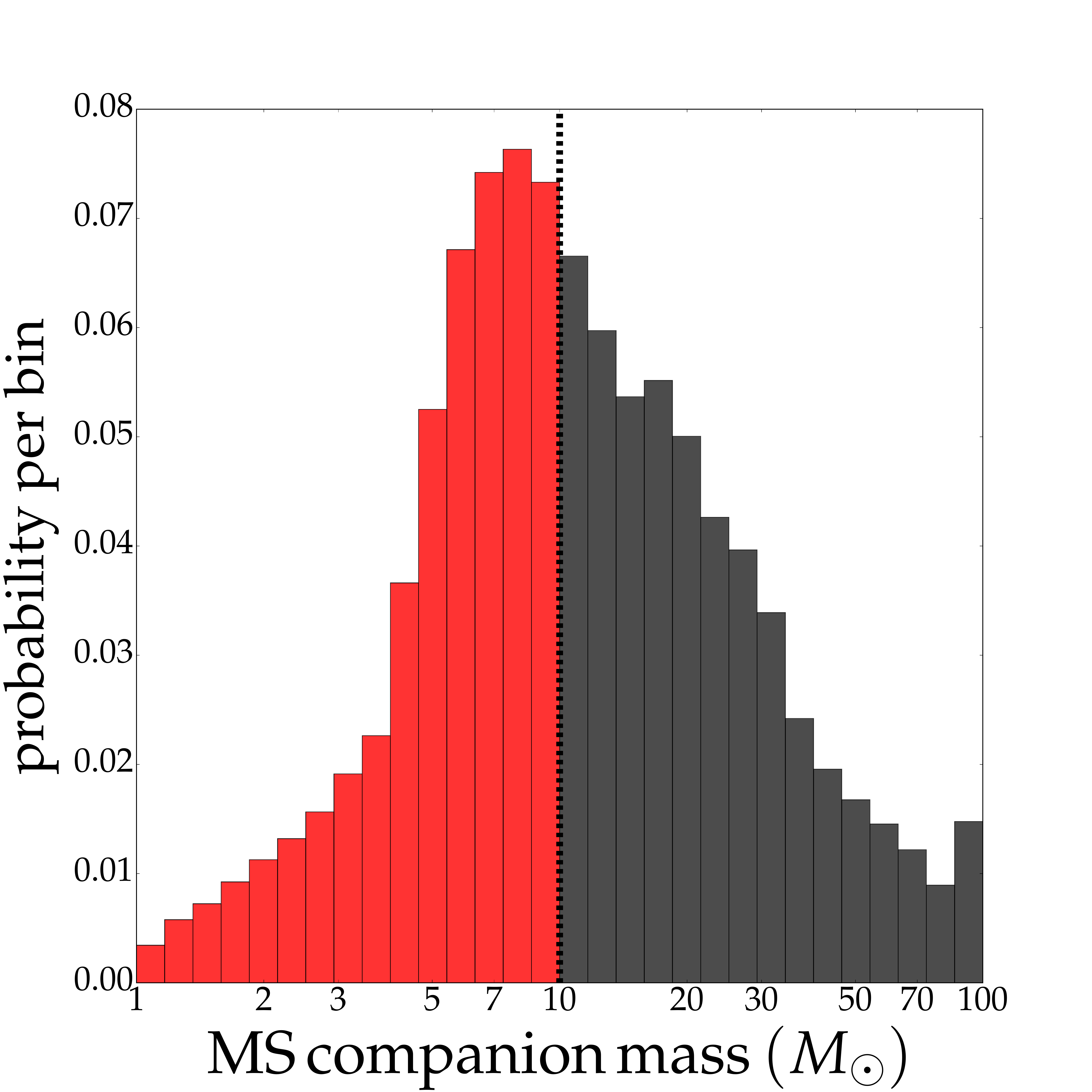}
\includegraphics[width=\onedvariable\textwidth]{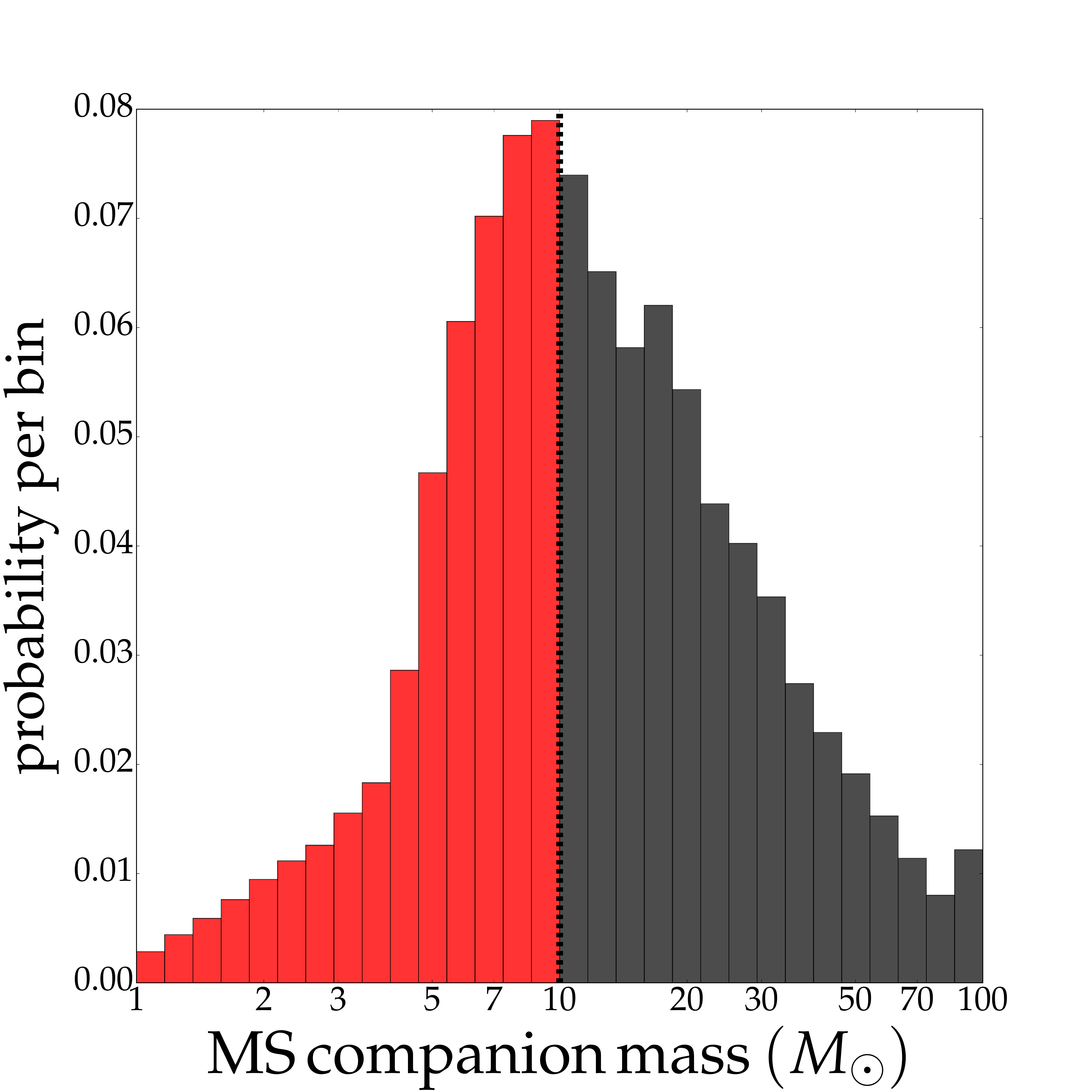}
\includegraphics[width=\onedvariable\textwidth]{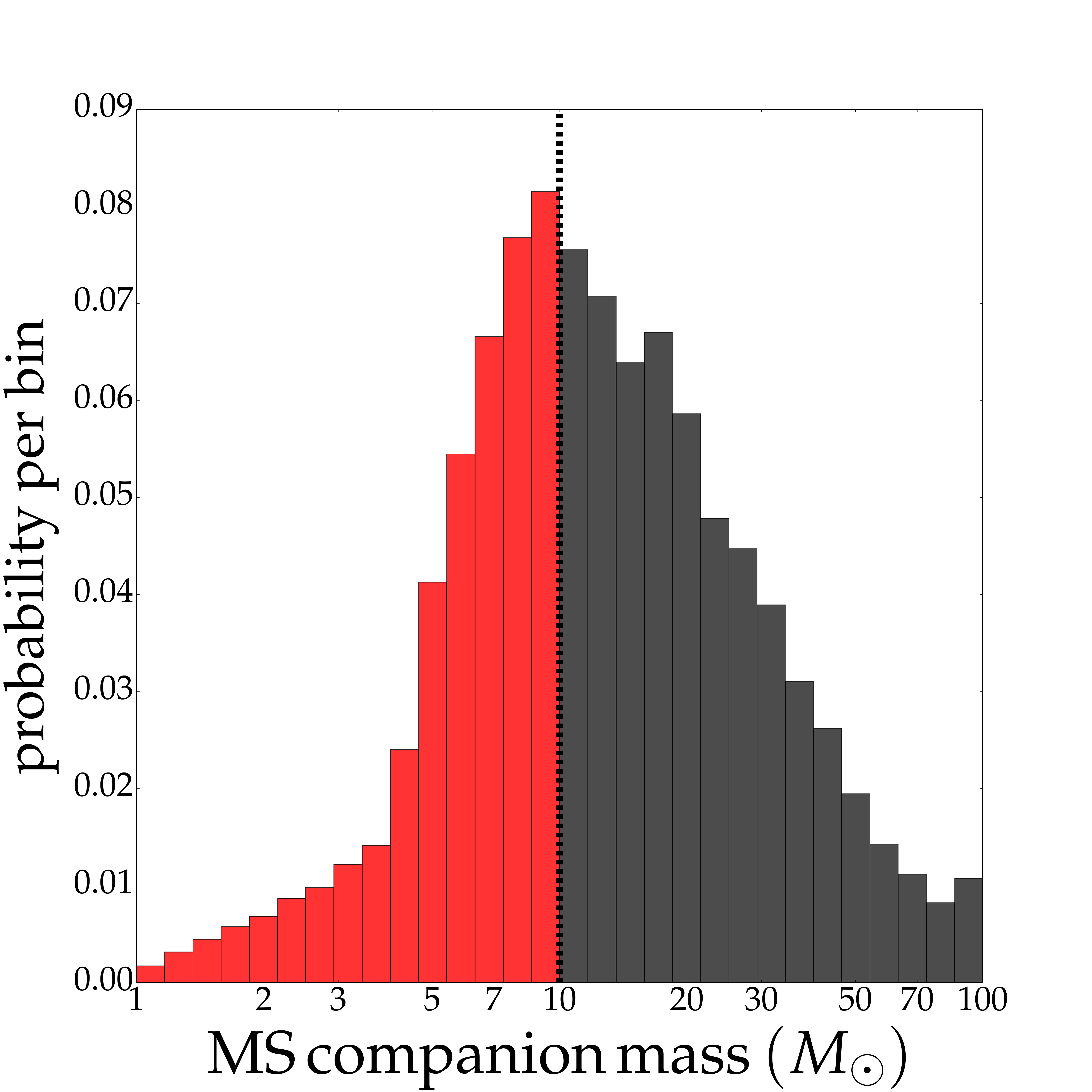}\\
\noindent\makebox[\linewidth]{\rule{0.8\paperwidth}{0.4pt}}\\
\includegraphics[width=\pievariable\textwidth]{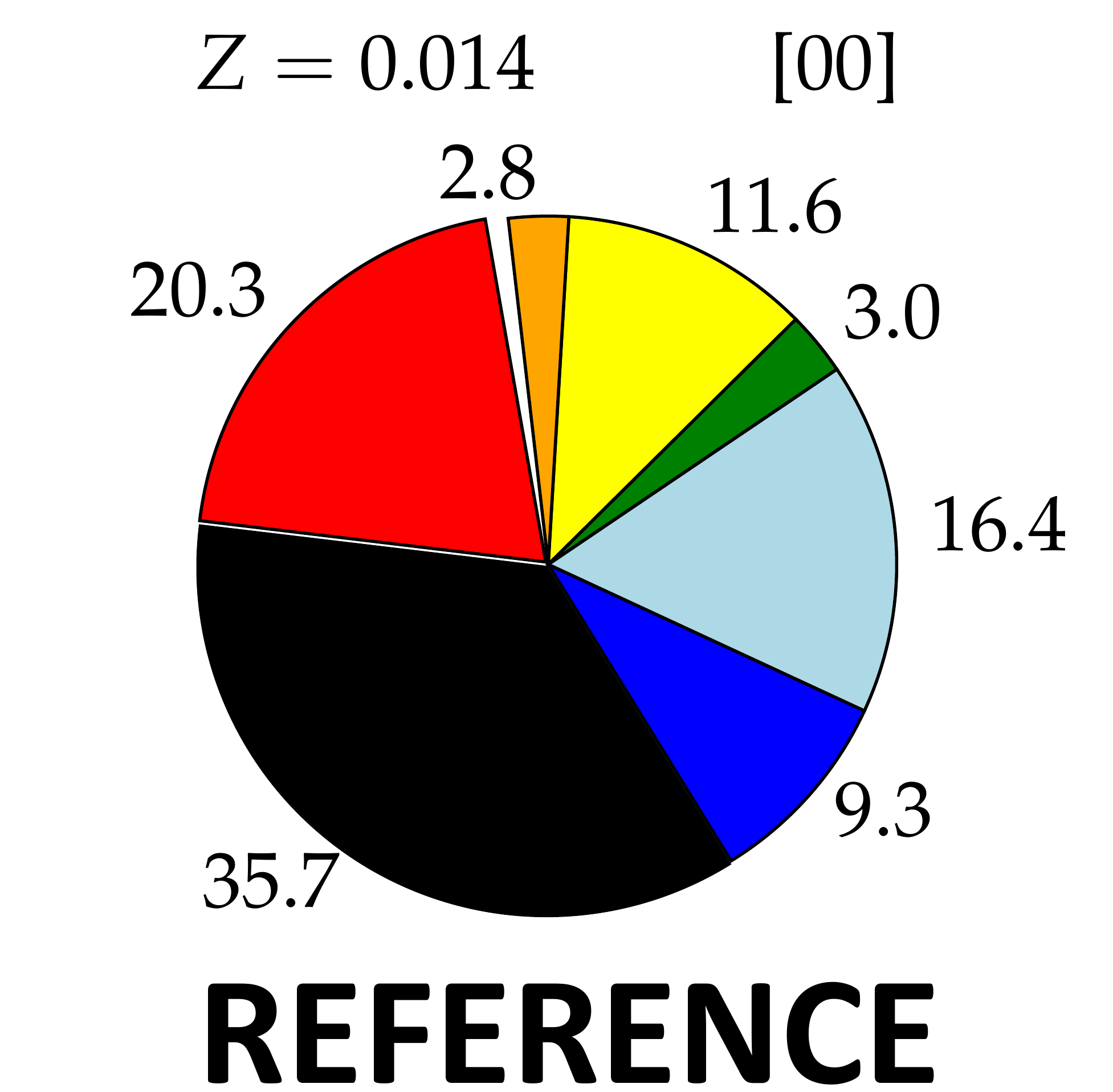}
\includegraphics[width=\pievariable\textwidth]{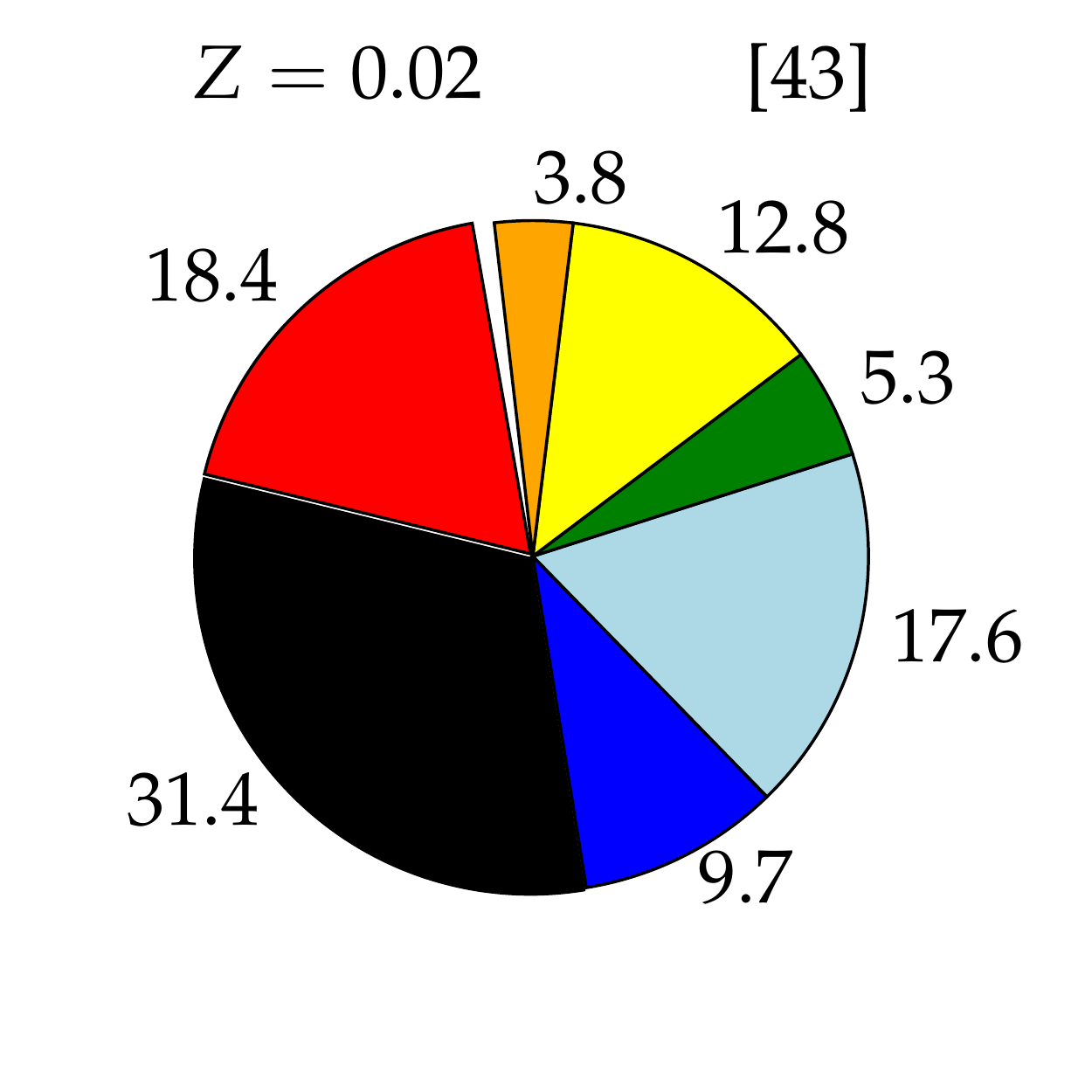}%
\includegraphics[width=\pievariable\textwidth]{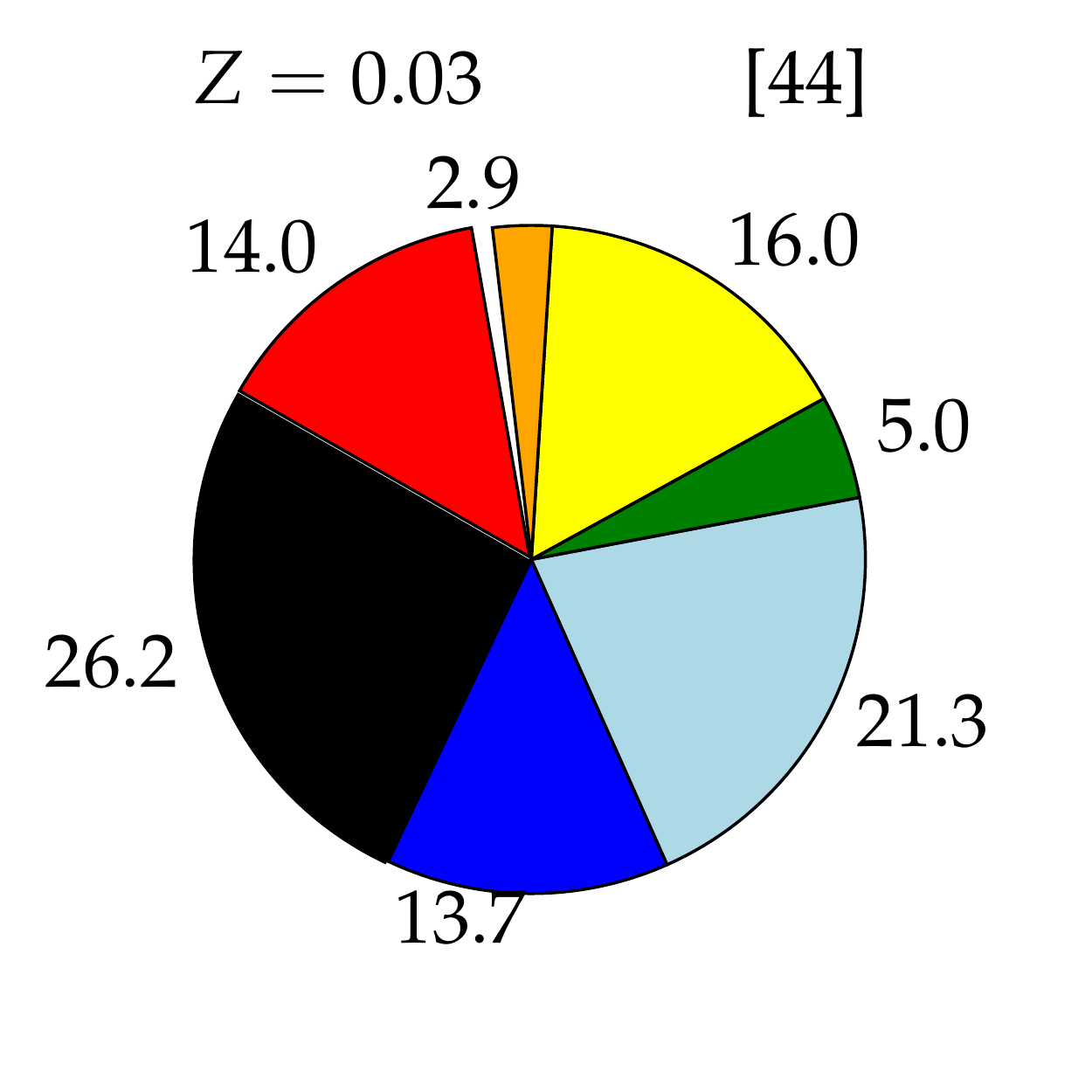}\\
\includegraphics[width=\onedvariable\textwidth]{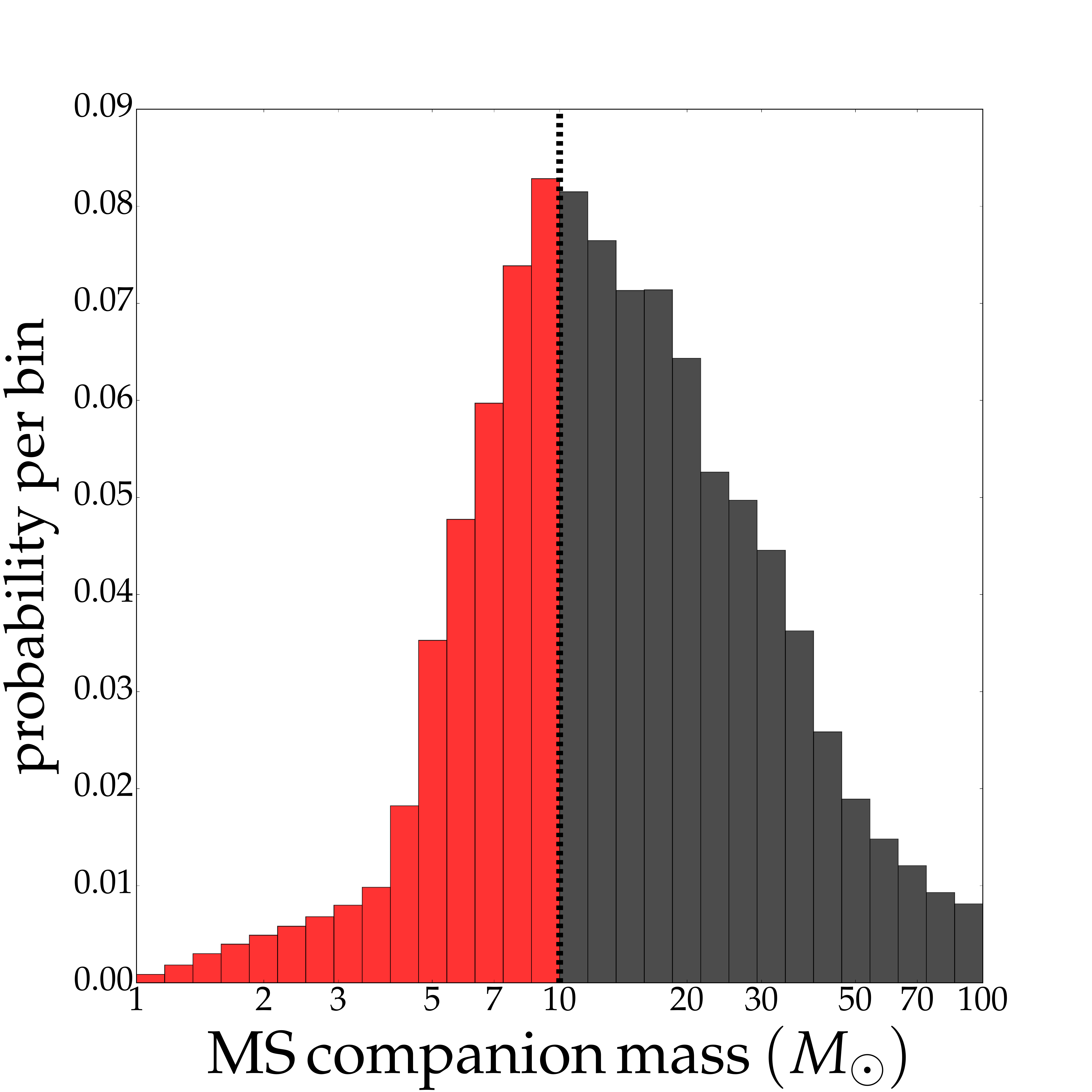}
\includegraphics[width=\onedvariable\textwidth]{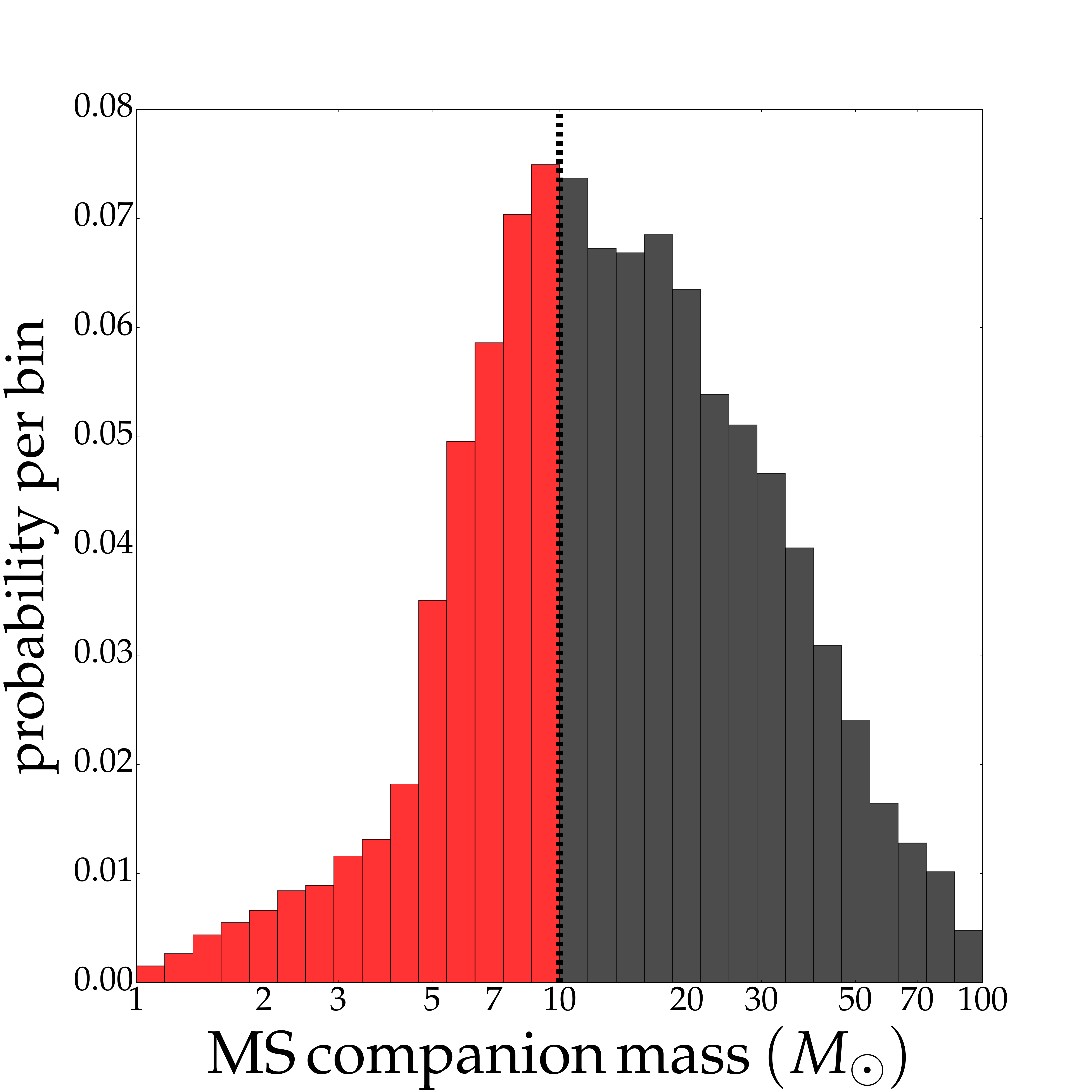}
\includegraphics[width=\onedvariable\textwidth]{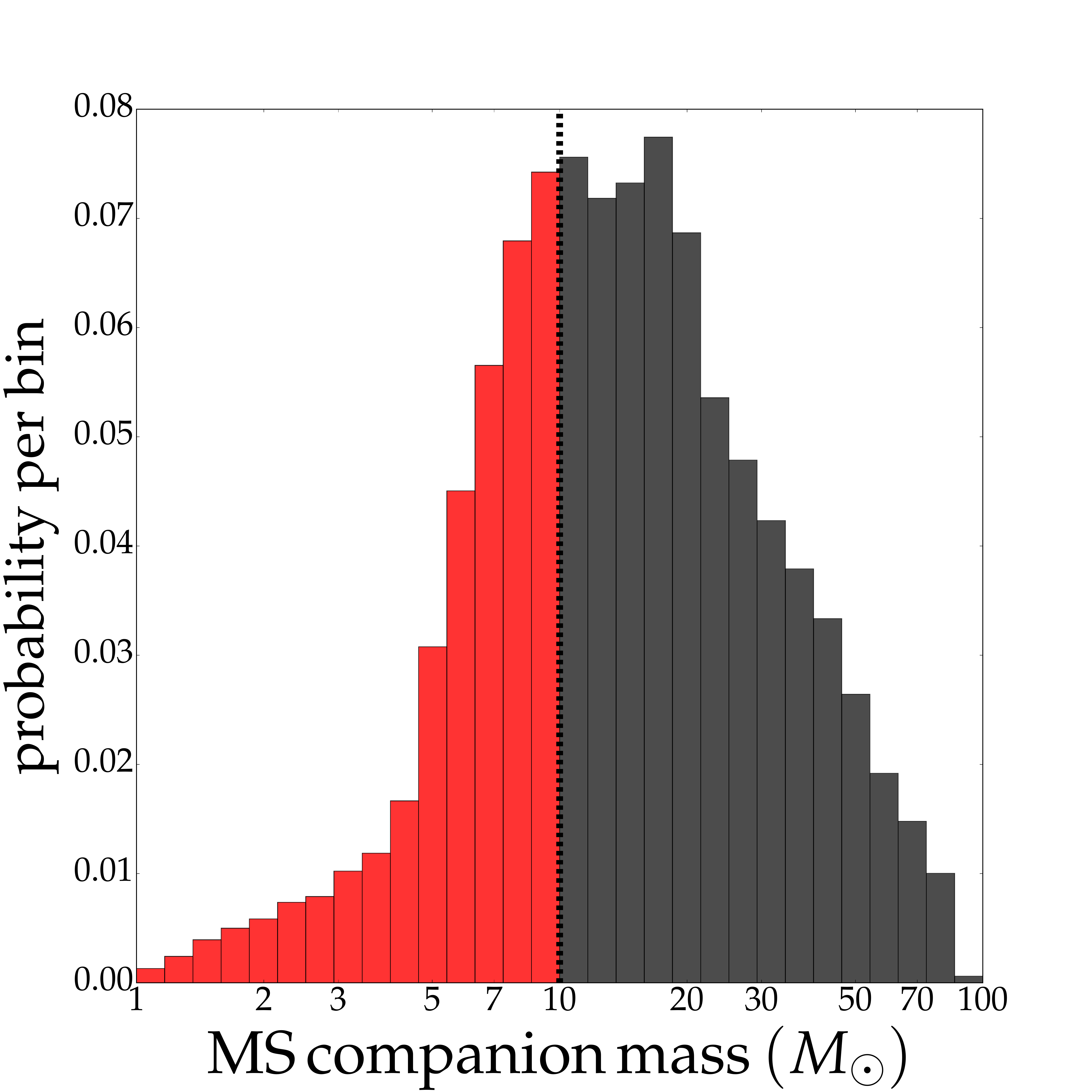}\\
\caption{As in Figure \ref{fig:pie} and \ref{fig:mass_dist}, at different metallicities. 
\emph{Pie chart:} Expected companions to stripped SNe. Colors are explained in the legend above and terms have the same meaning as in Figure \ref{fig:pie}. The assumed metallicity and the corresponding model number of the variation, as presented in Table 2 of \citetalias{Zapartas+2017}, are shown in the top-left and top-right corner of each pie chart, respectively. Minority channels with companions that are neither MS nor compact remnants (e.g., companions on the red supergiant phase) are omitted. \emph{Distribution:} Mass distribution of MS companions of stripped SNe, corresponding to the pie chart above it, for each model considered. In both type of figures, MS companions of mass $<10\Msun$ are shown in red whereas $>10 \Msun$ in black. The reference simulation for all the variations in the assumed physical parameters and initial conditions, which assumes solar metallicity ($Z=0.014$; Model 00 of \citetalias{Zapartas+2017}), is highlighted.
  }\label{fig:metallicity}
\end{center}
\end{figure*}

\begin{figure*}
\begin{center}
\includegraphics[width=\textwidth]{legend_piechart_compressed}
\noindent\makebox[\linewidth]{\rule{0.8\paperwidth}{0.4pt}}\\
\includegraphics[width=\pievariable\textwidth]{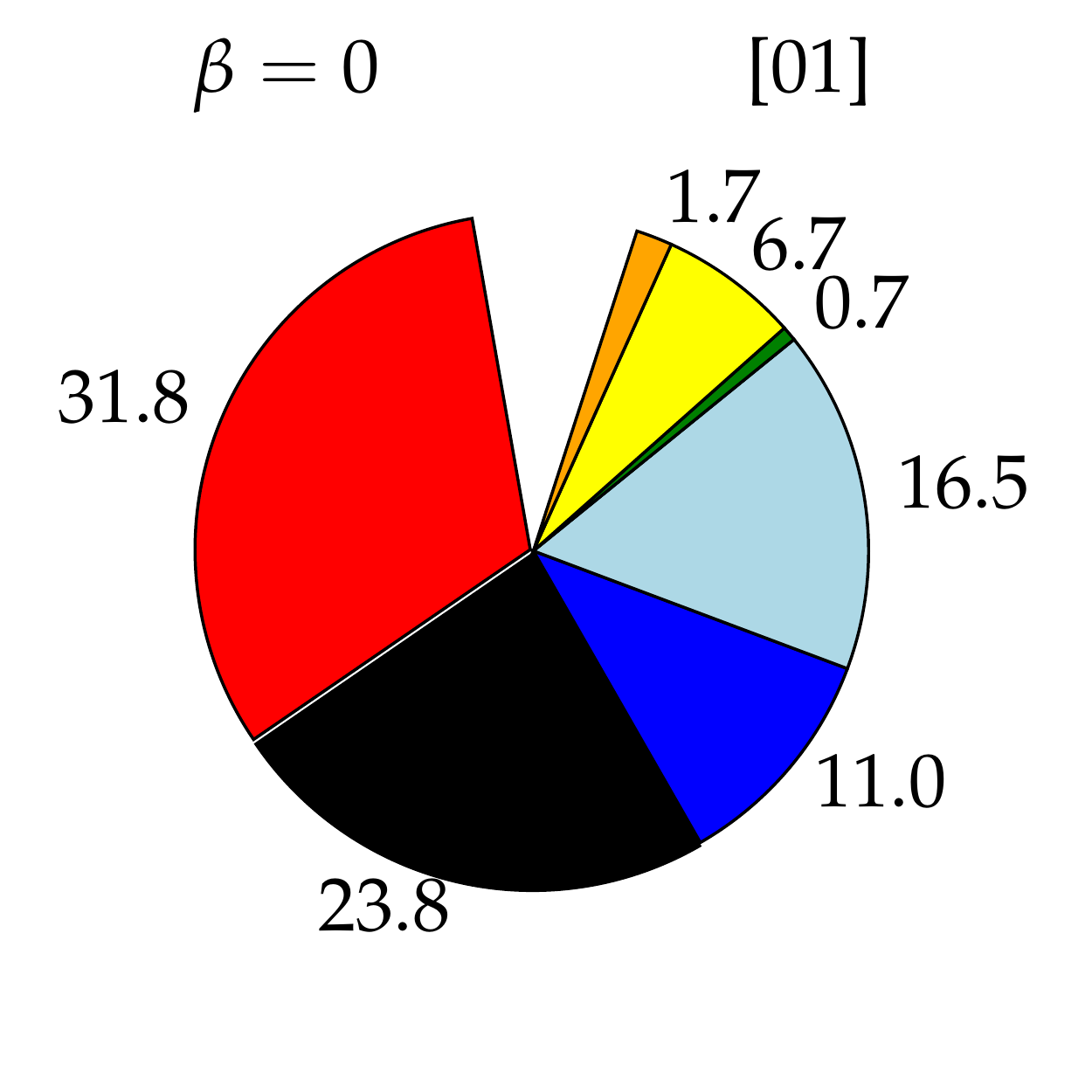}
\includegraphics[width=\pievariable\textwidth]{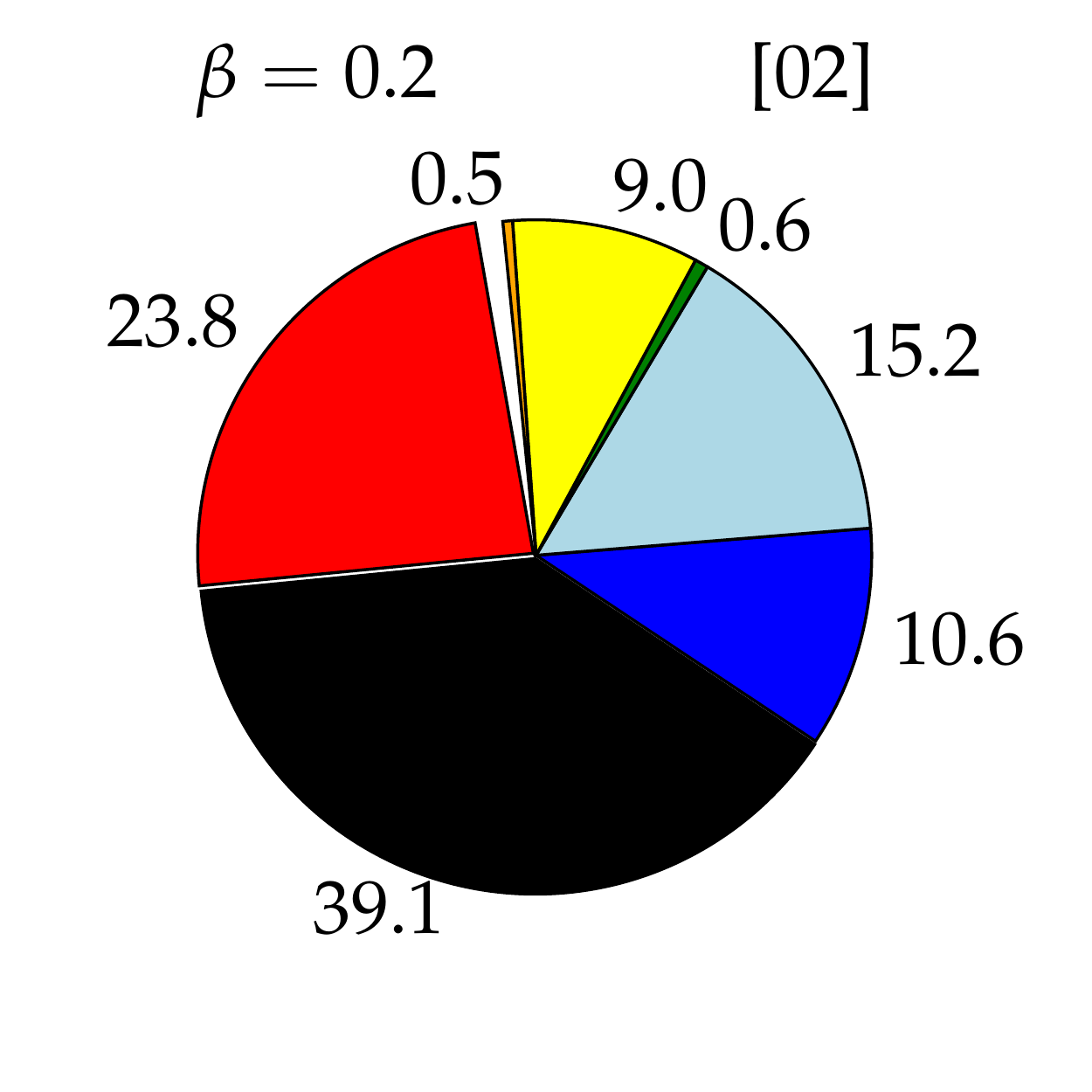}
\includegraphics[width=\pievariable\textwidth]{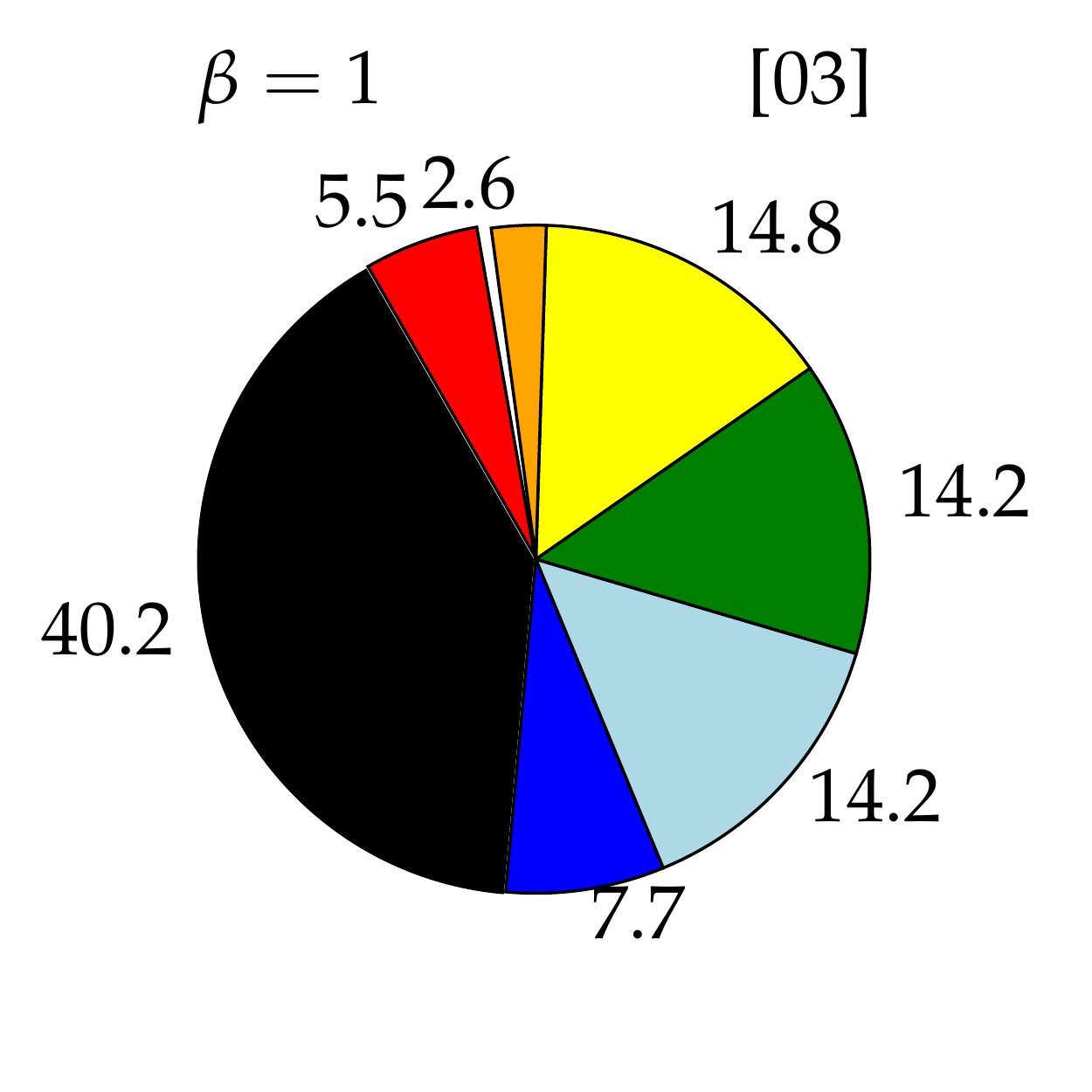}
\includegraphics[width=\pievariable\textwidth]{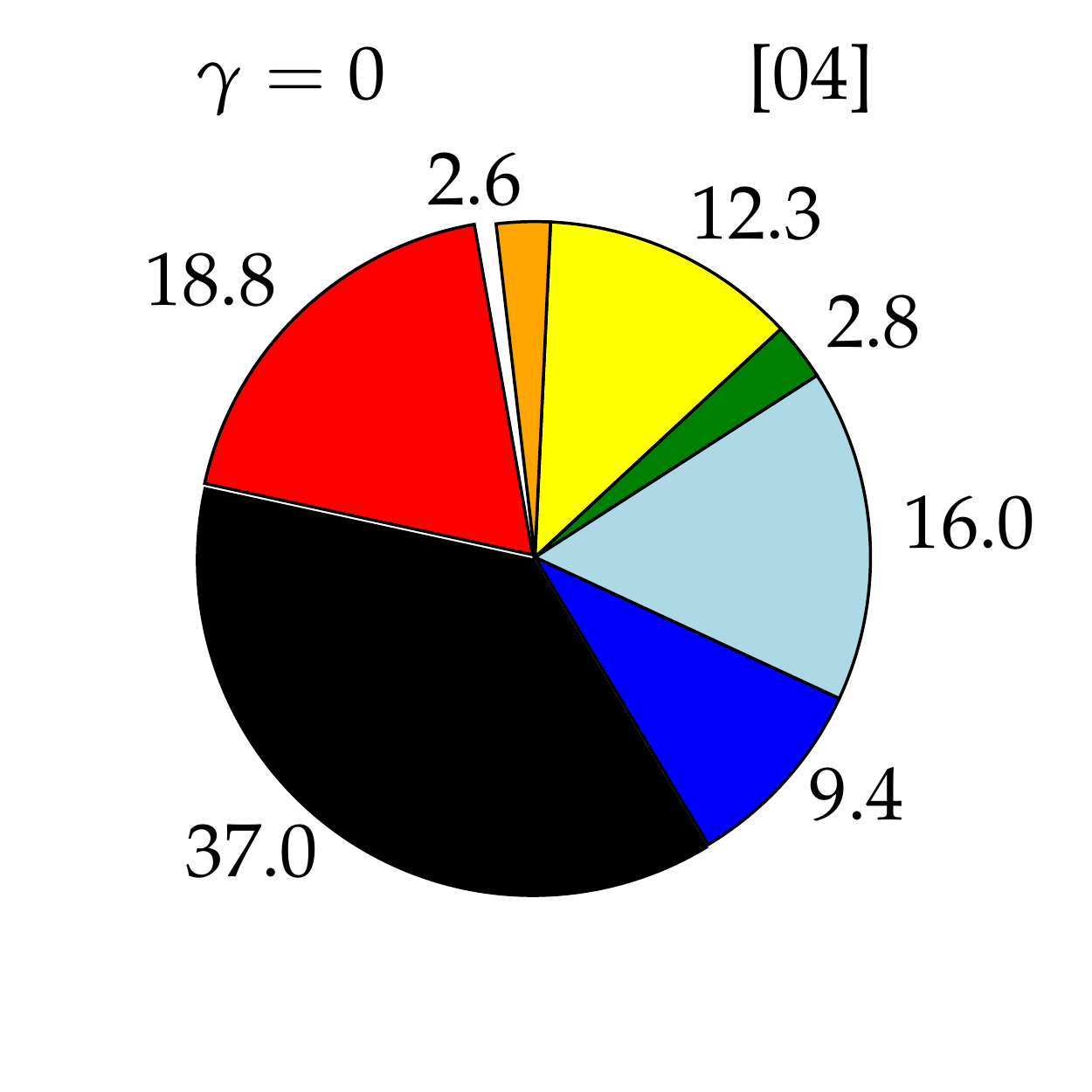}
\includegraphics[width=\pievariable\textwidth]{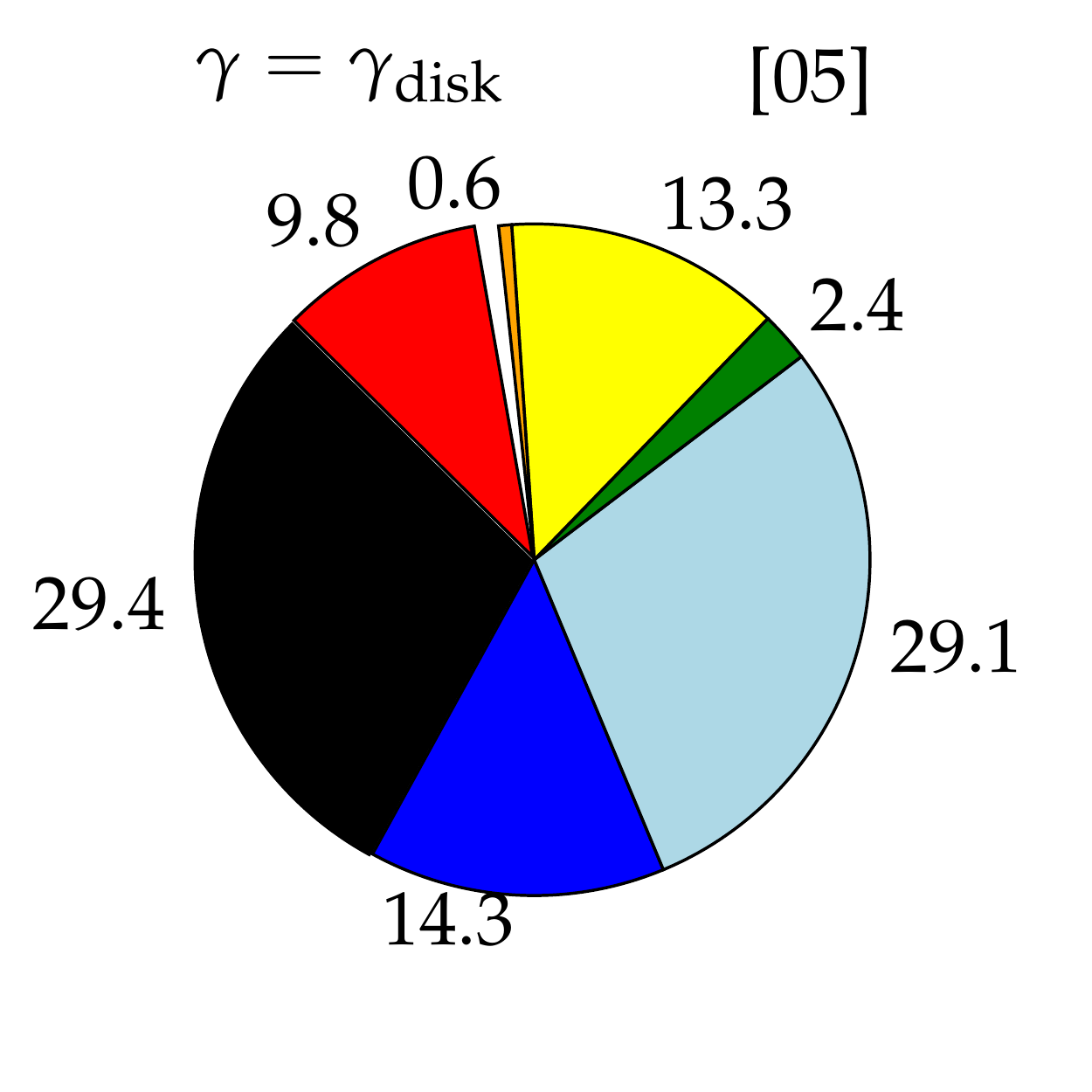}
\includegraphics[width=\pievariable\textwidth]{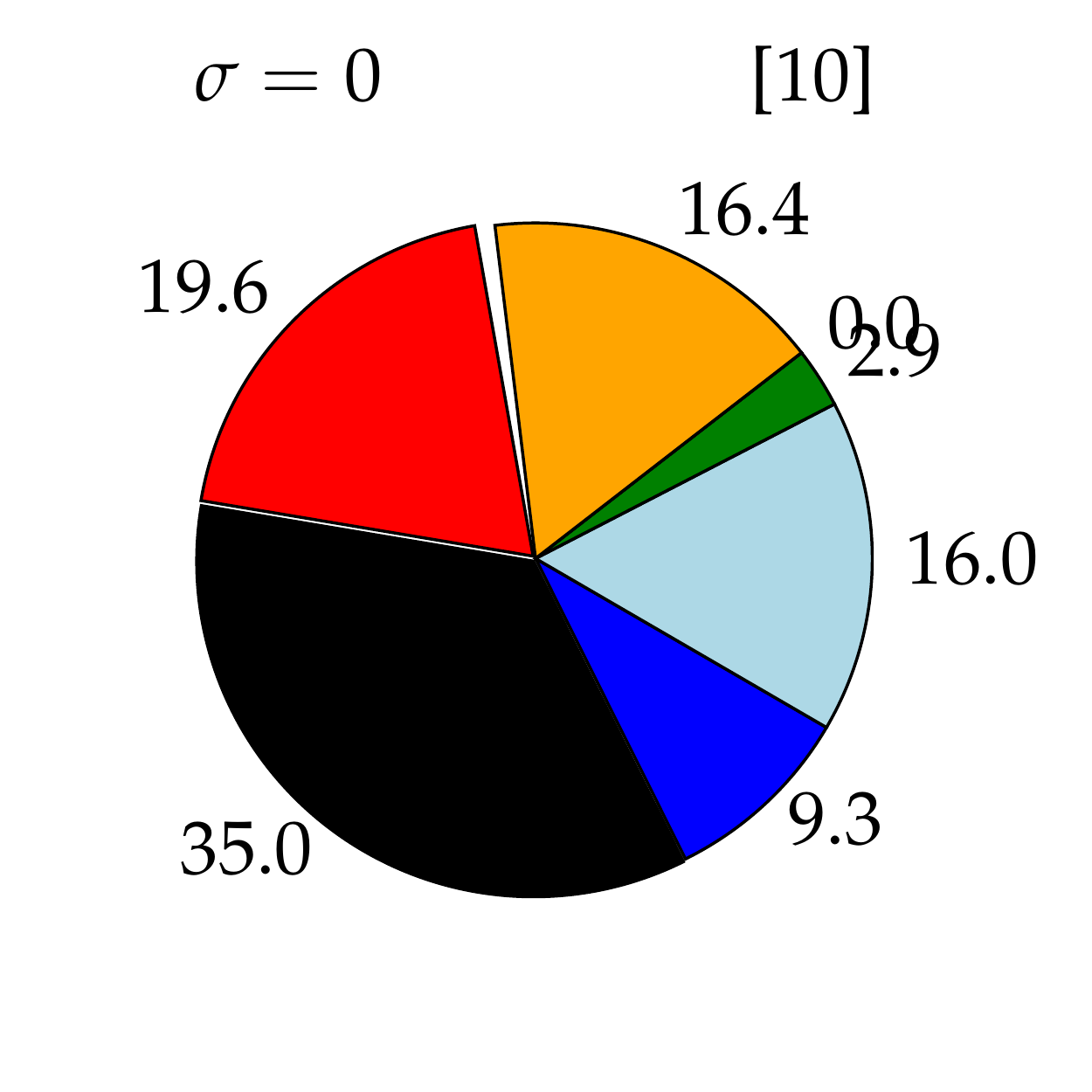}\\
\includegraphics[width=\onedvariable\textwidth]{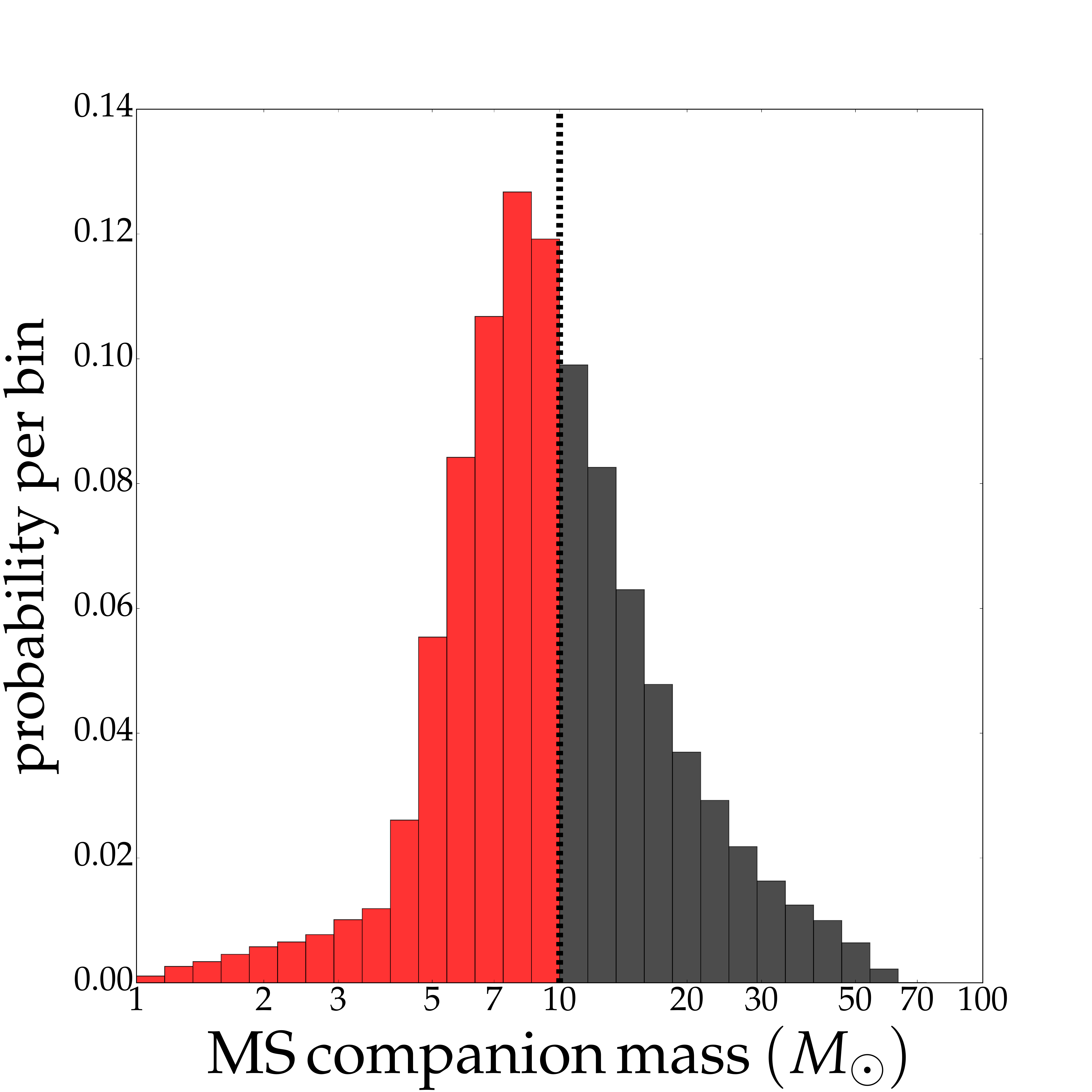}
\includegraphics[width=\onedvariable\textwidth]{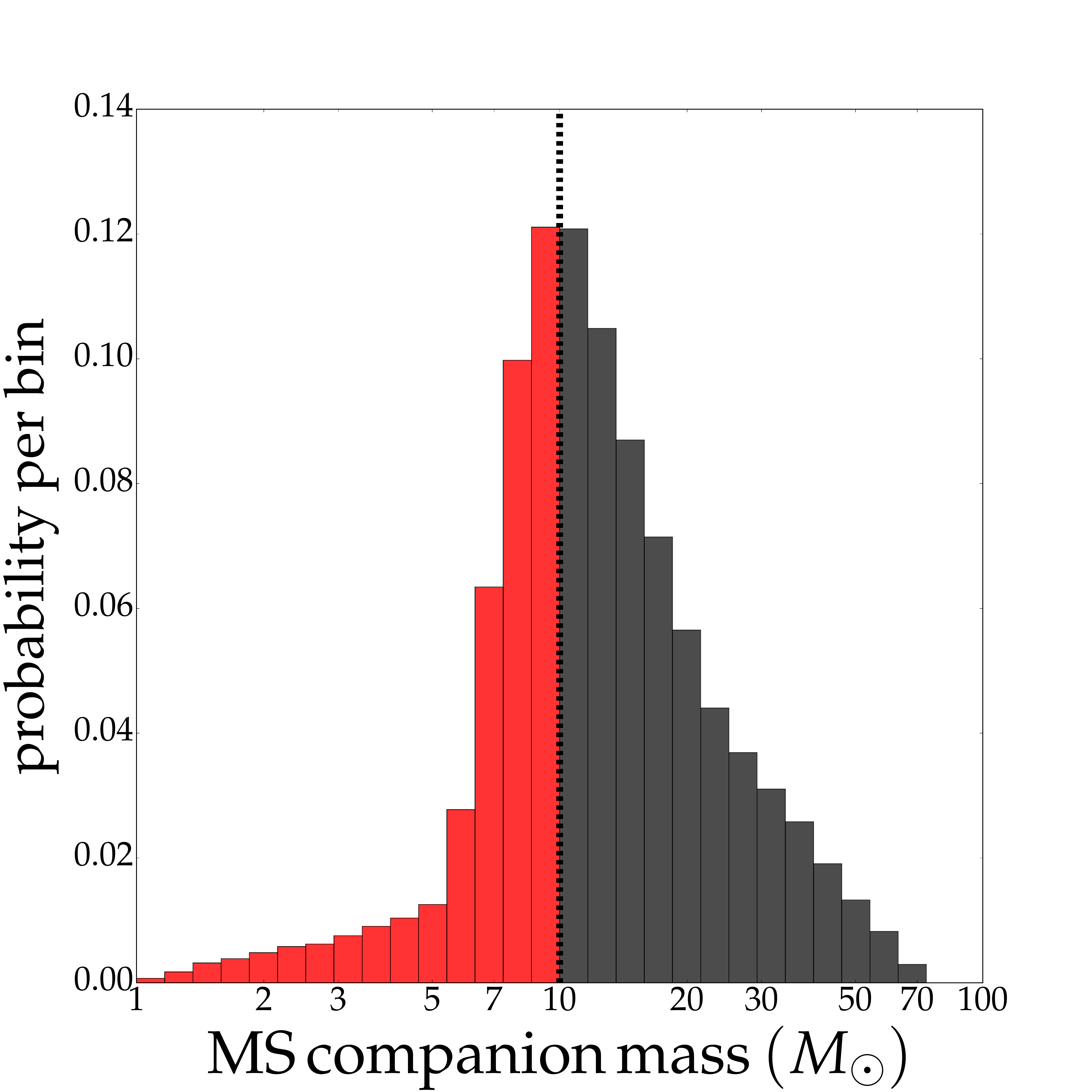}
\includegraphics[width=\onedvariable\textwidth]{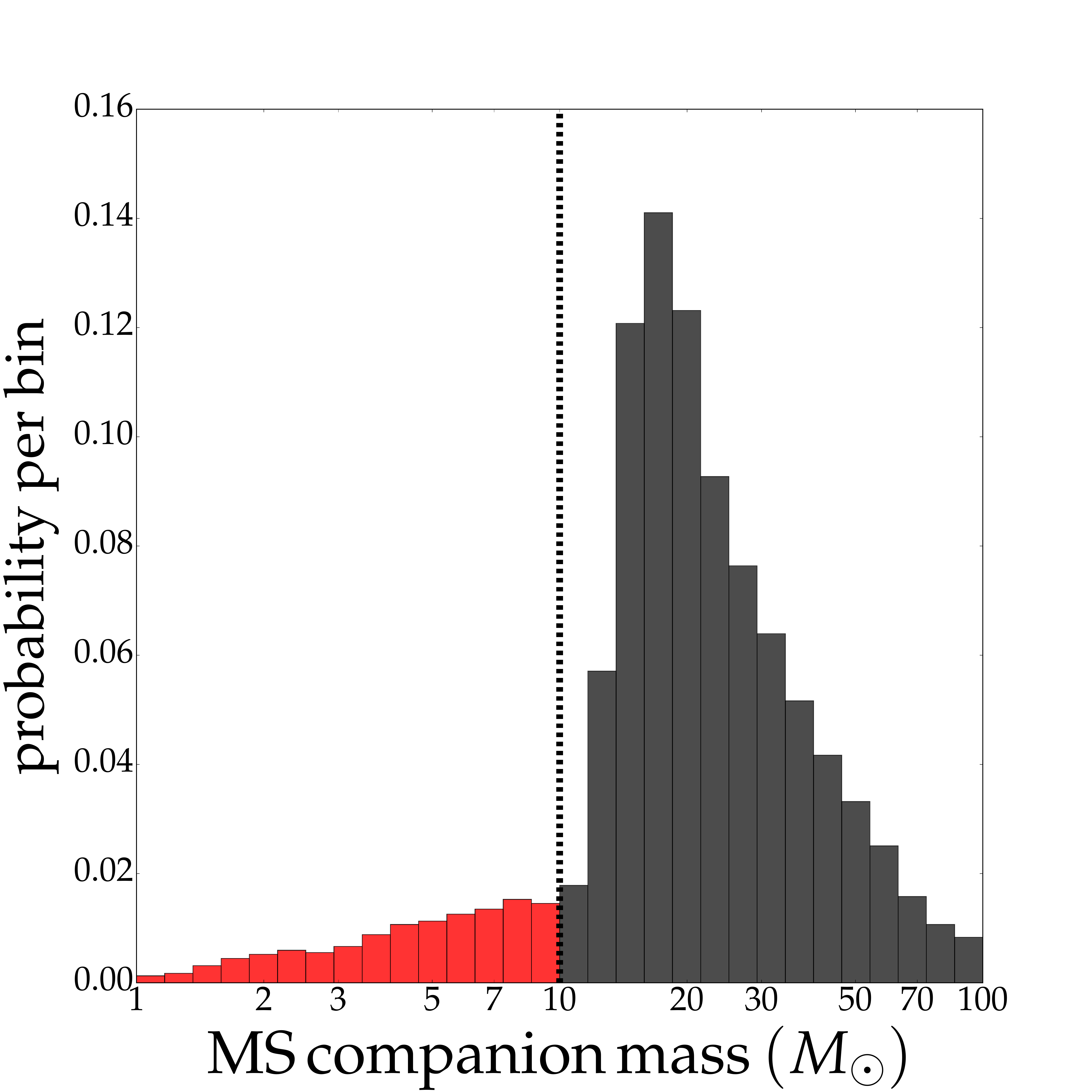}
\includegraphics[width=\onedvariable\textwidth]{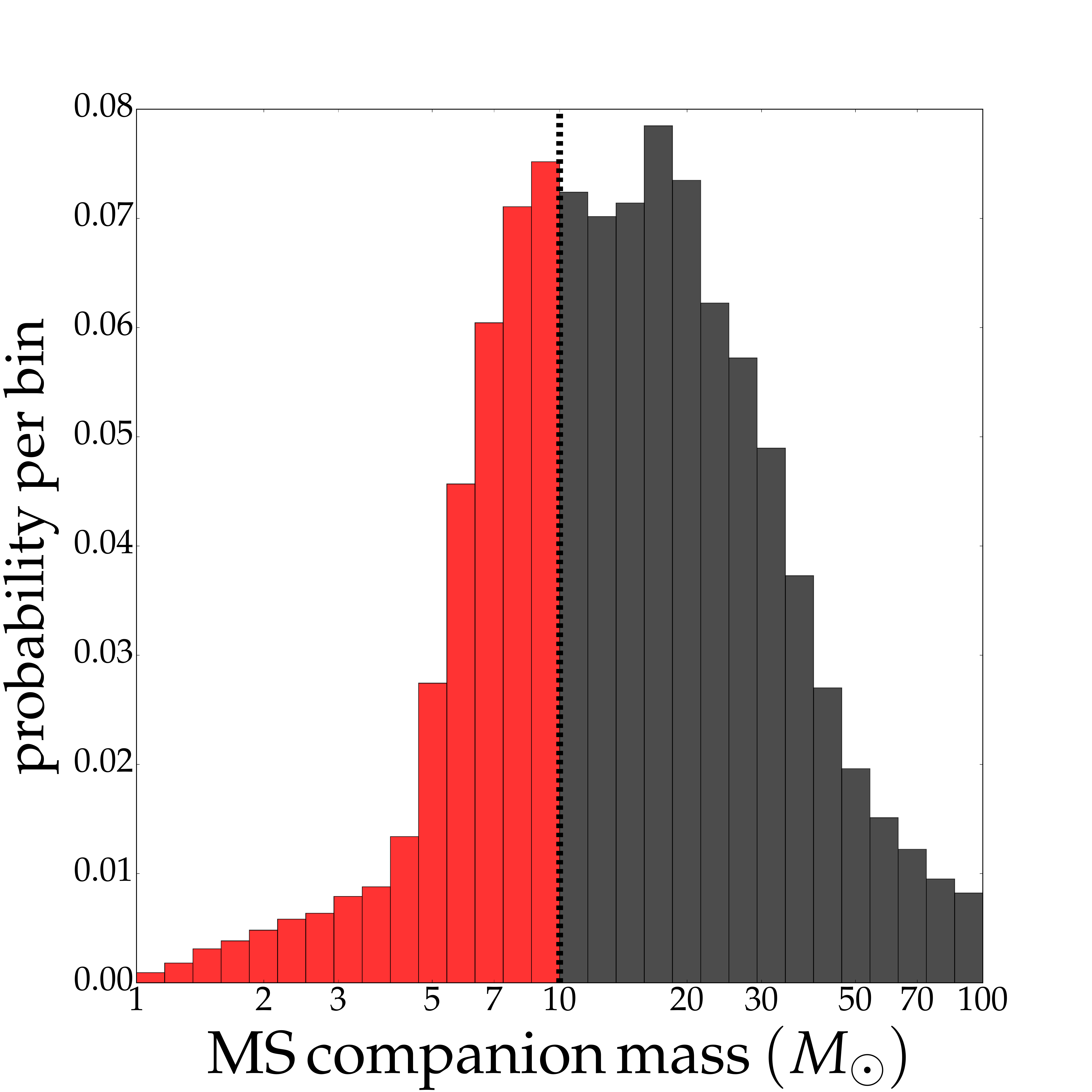}
\includegraphics[width=\onedvariable\textwidth]{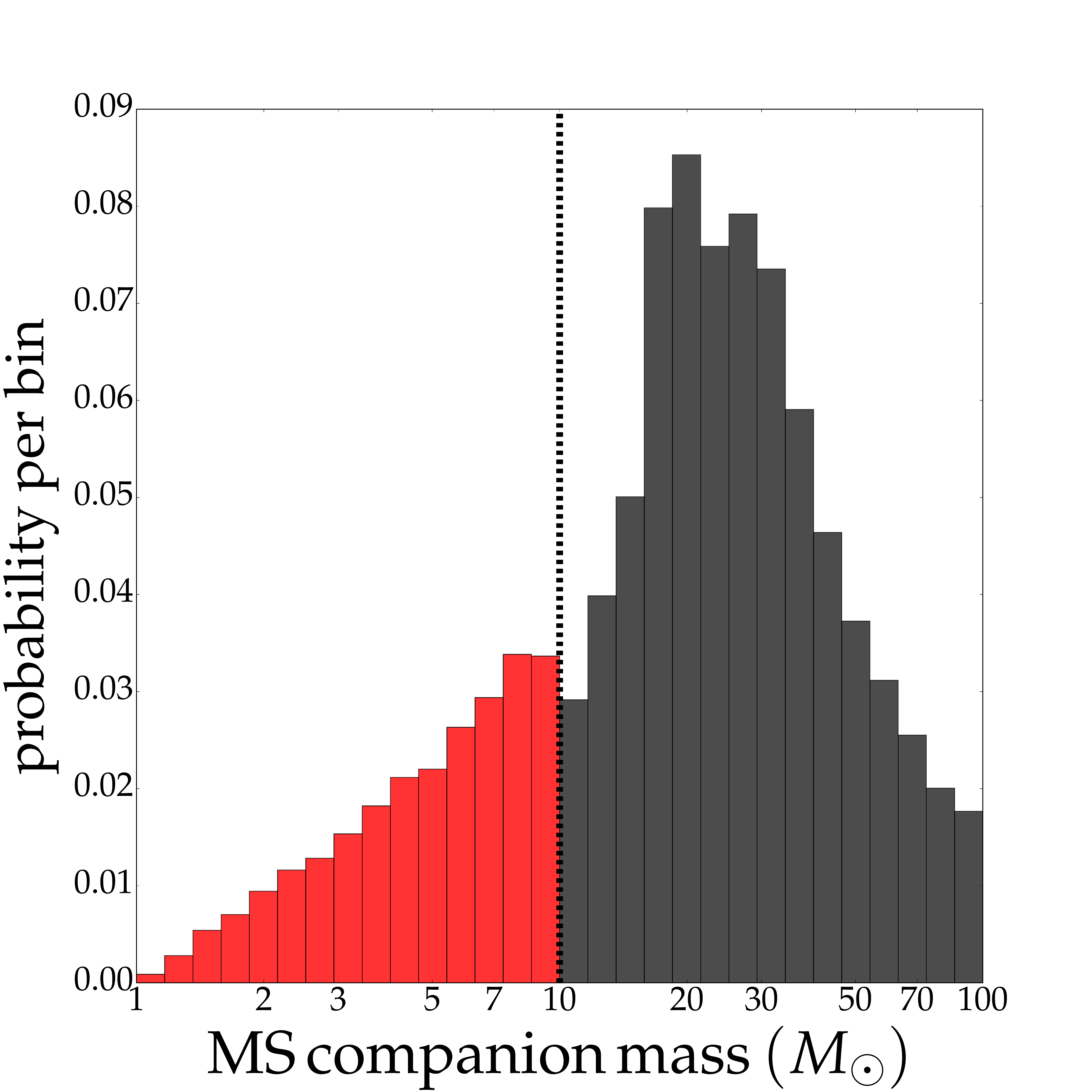}
\includegraphics[width=\onedvariable\textwidth]{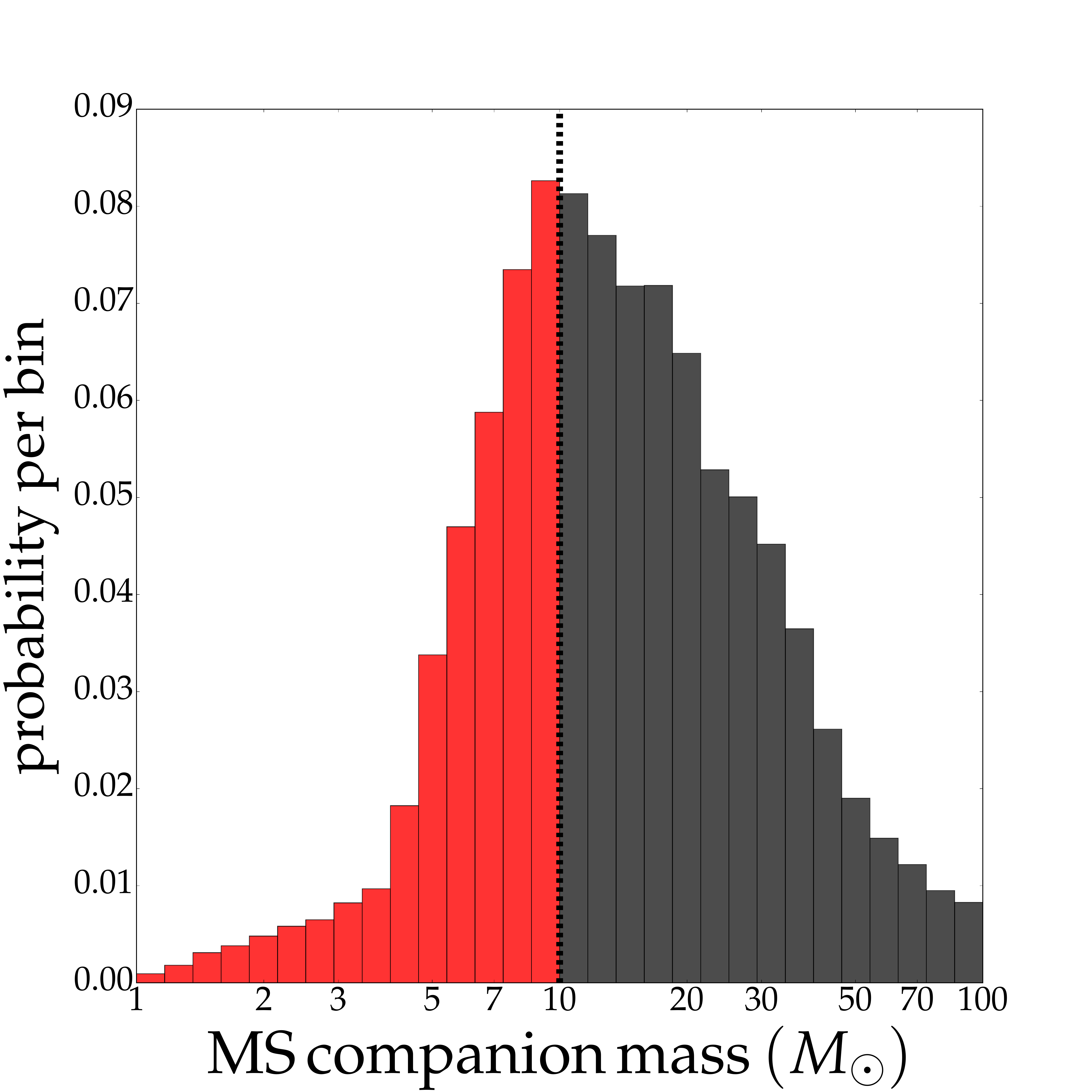}\\
\noindent\makebox[\linewidth]{\rule{0.8\paperwidth}{0.4pt}}\\
\includegraphics[width=\pievariable\textwidth]{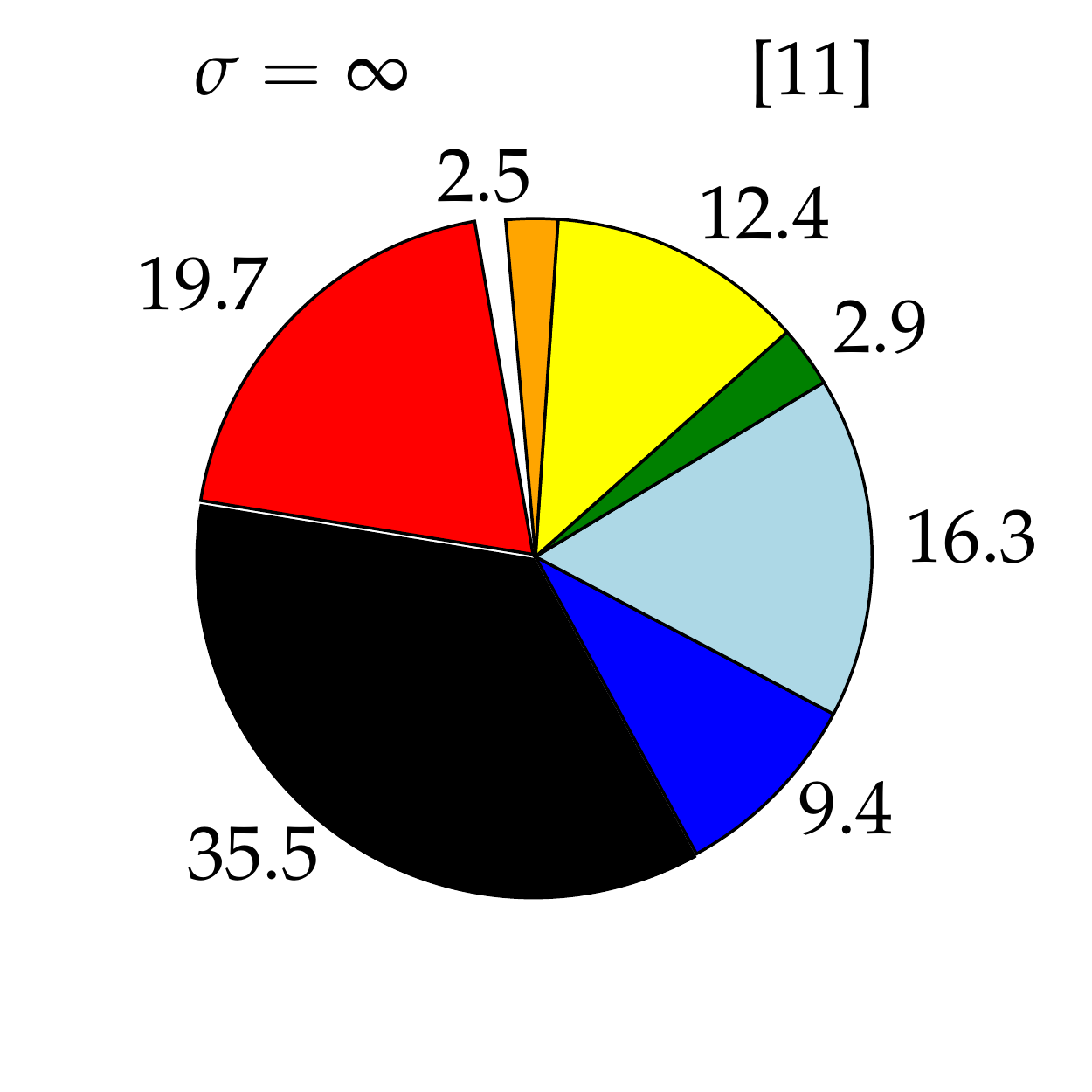}
\includegraphics[width=\pievariable\textwidth]{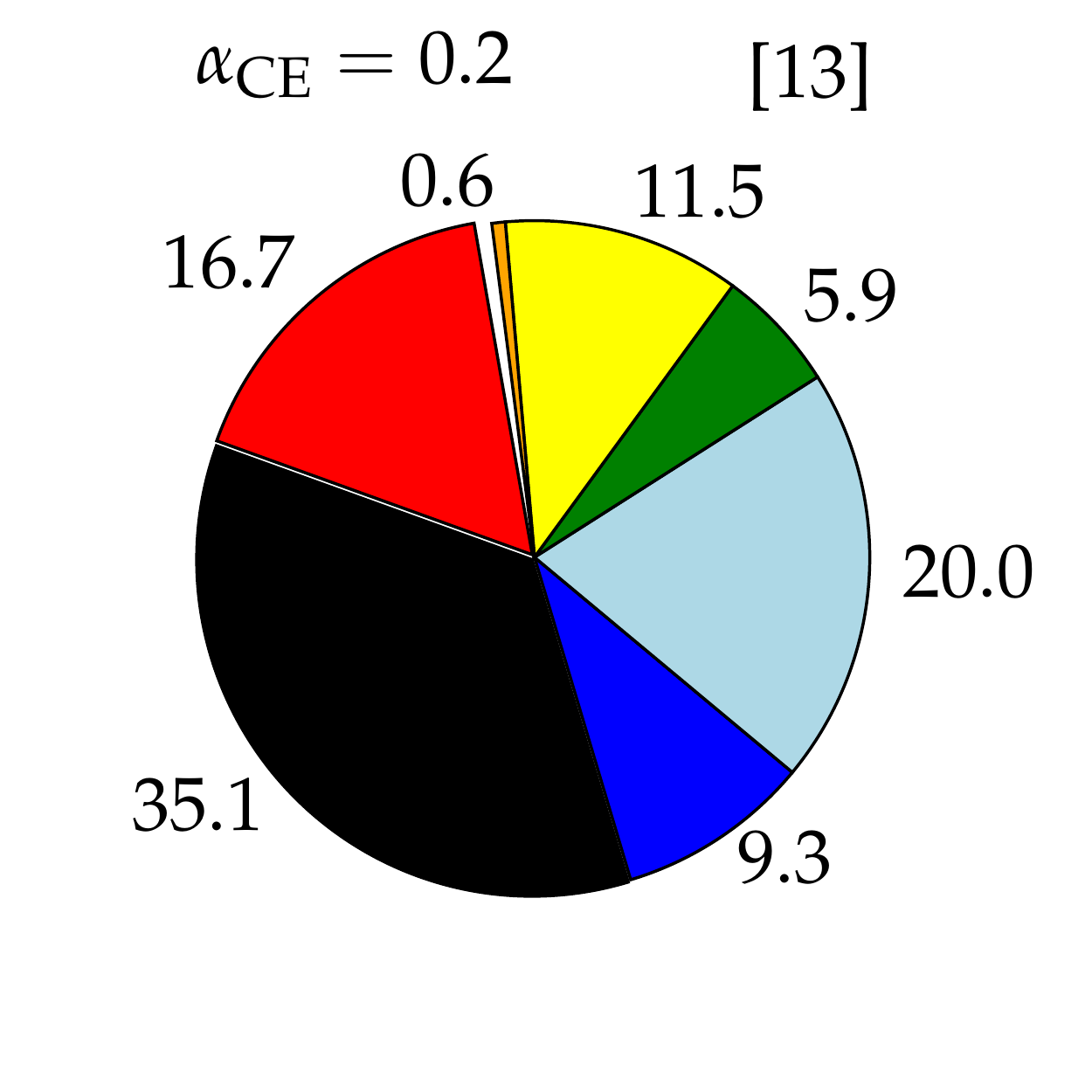}
\includegraphics[width=\pievariable\textwidth]{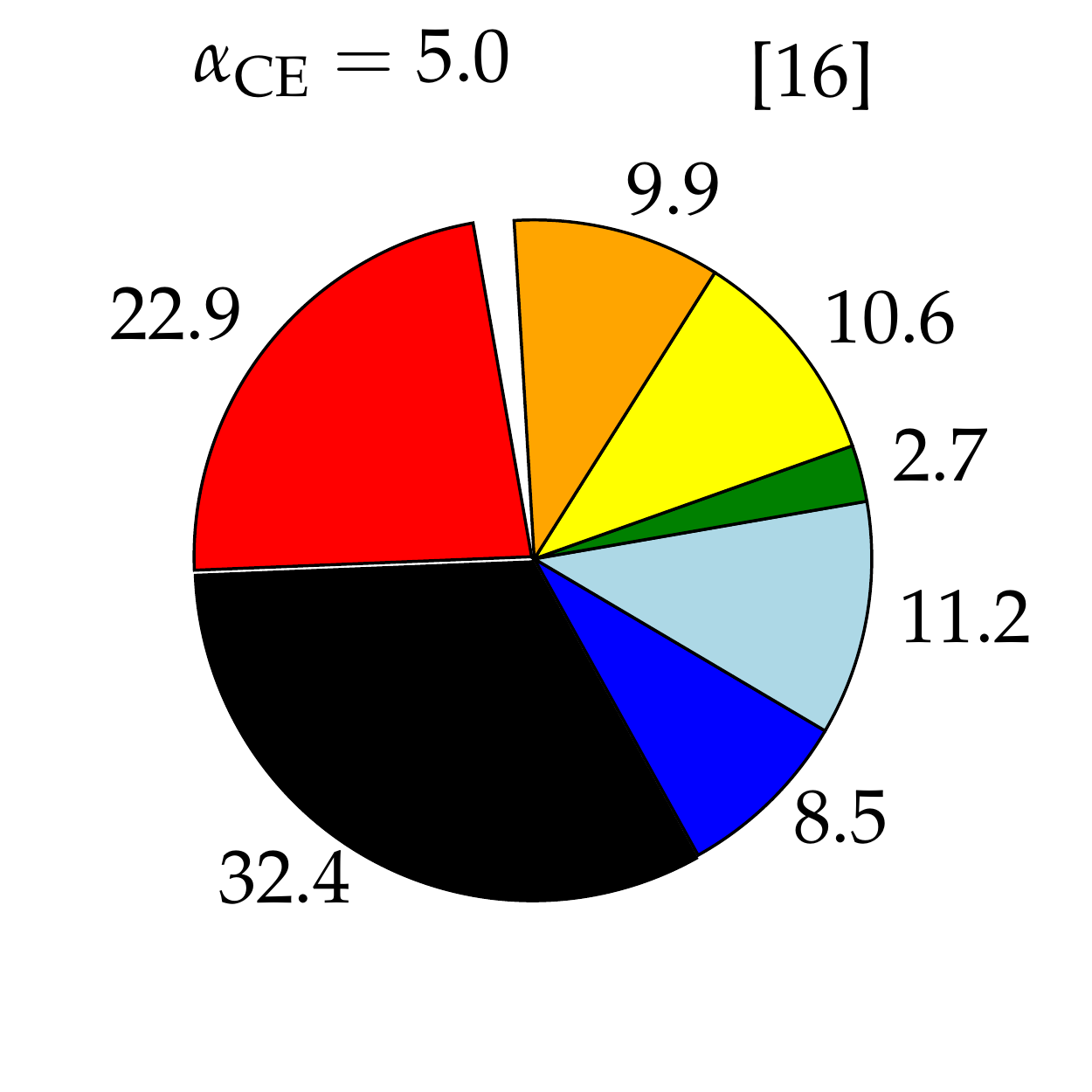}
\includegraphics[width=\pievariable\textwidth]{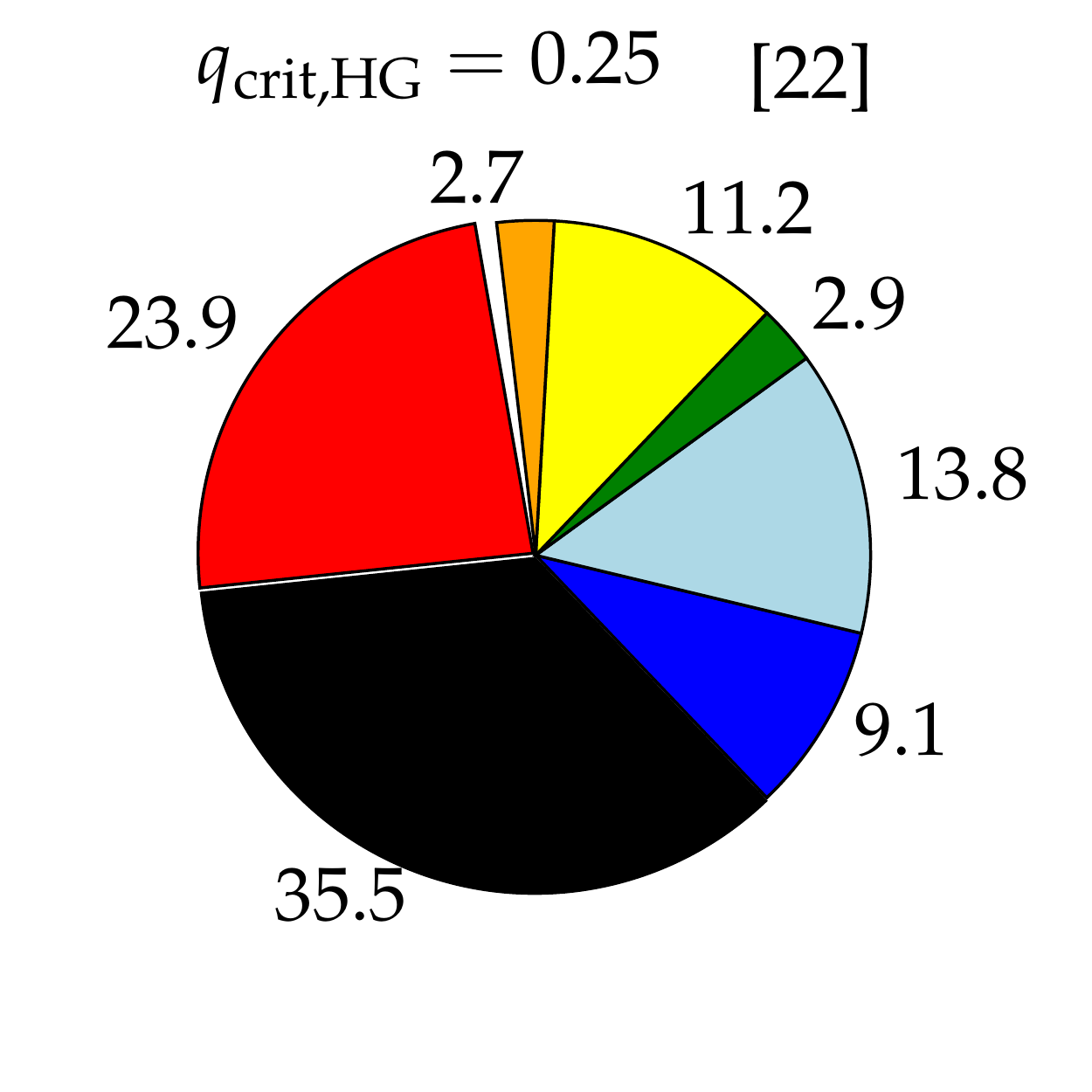}
\includegraphics[width=\pievariable\textwidth]{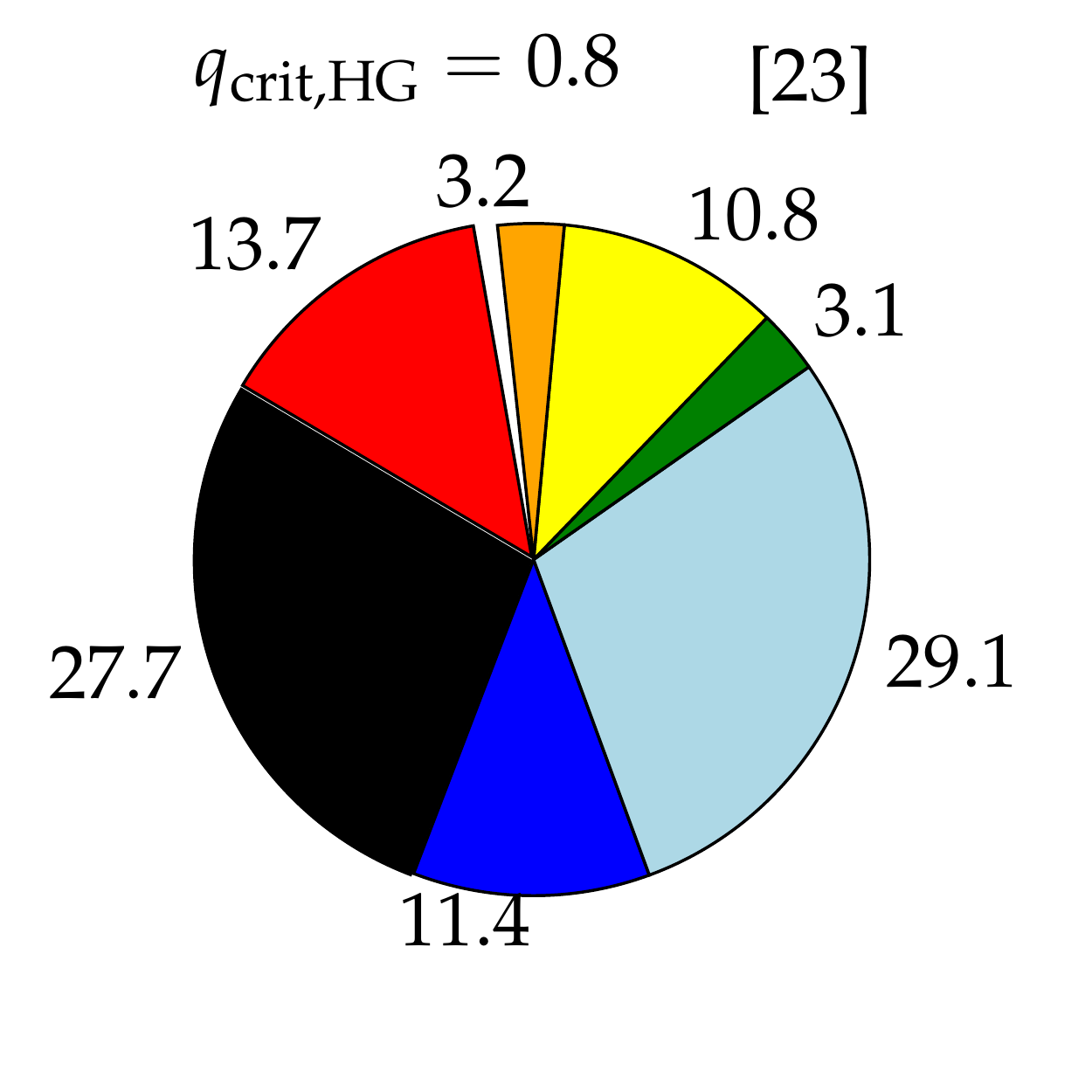}
\includegraphics[width=\pievariable\textwidth]{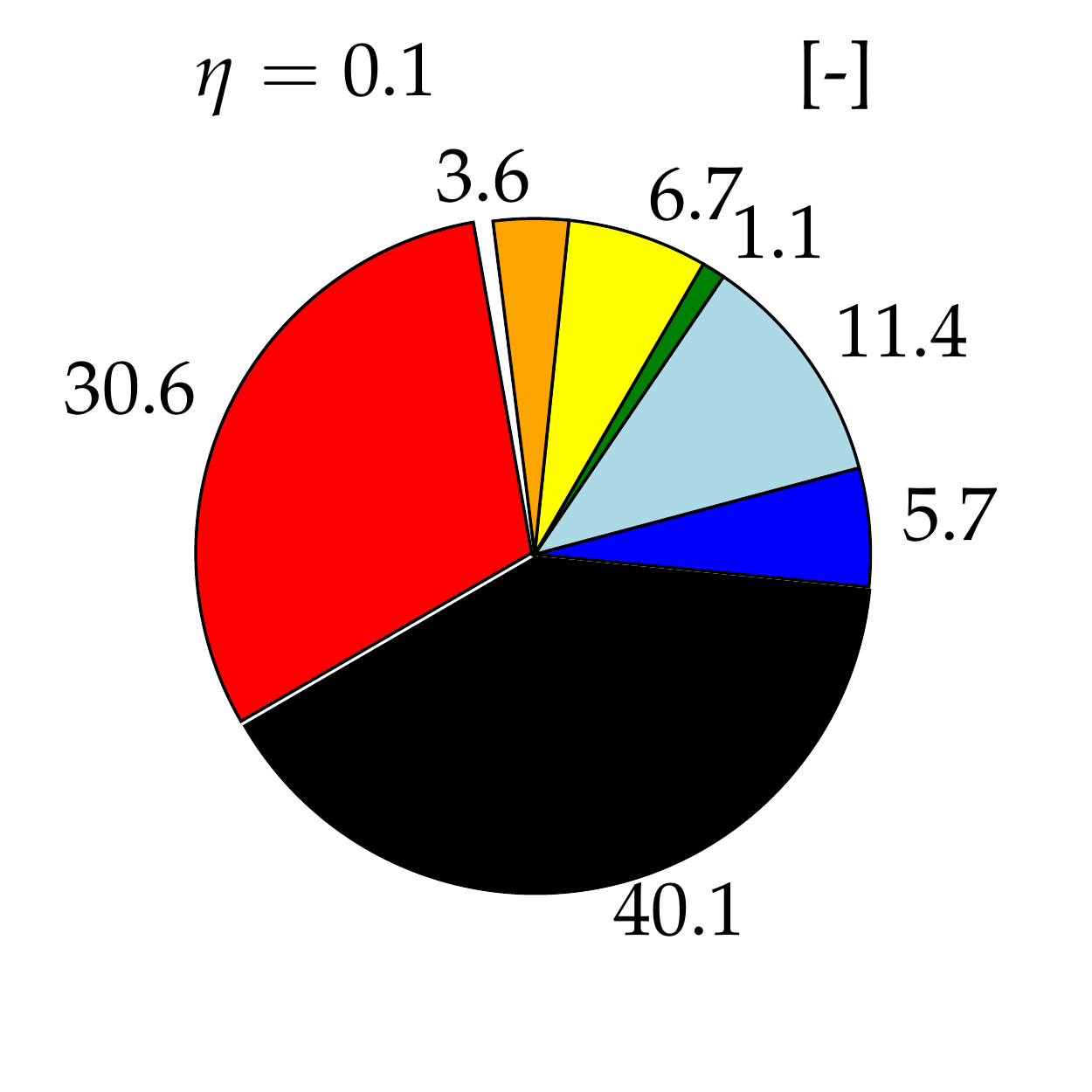}\\
\includegraphics[width=\onedvariable\textwidth]{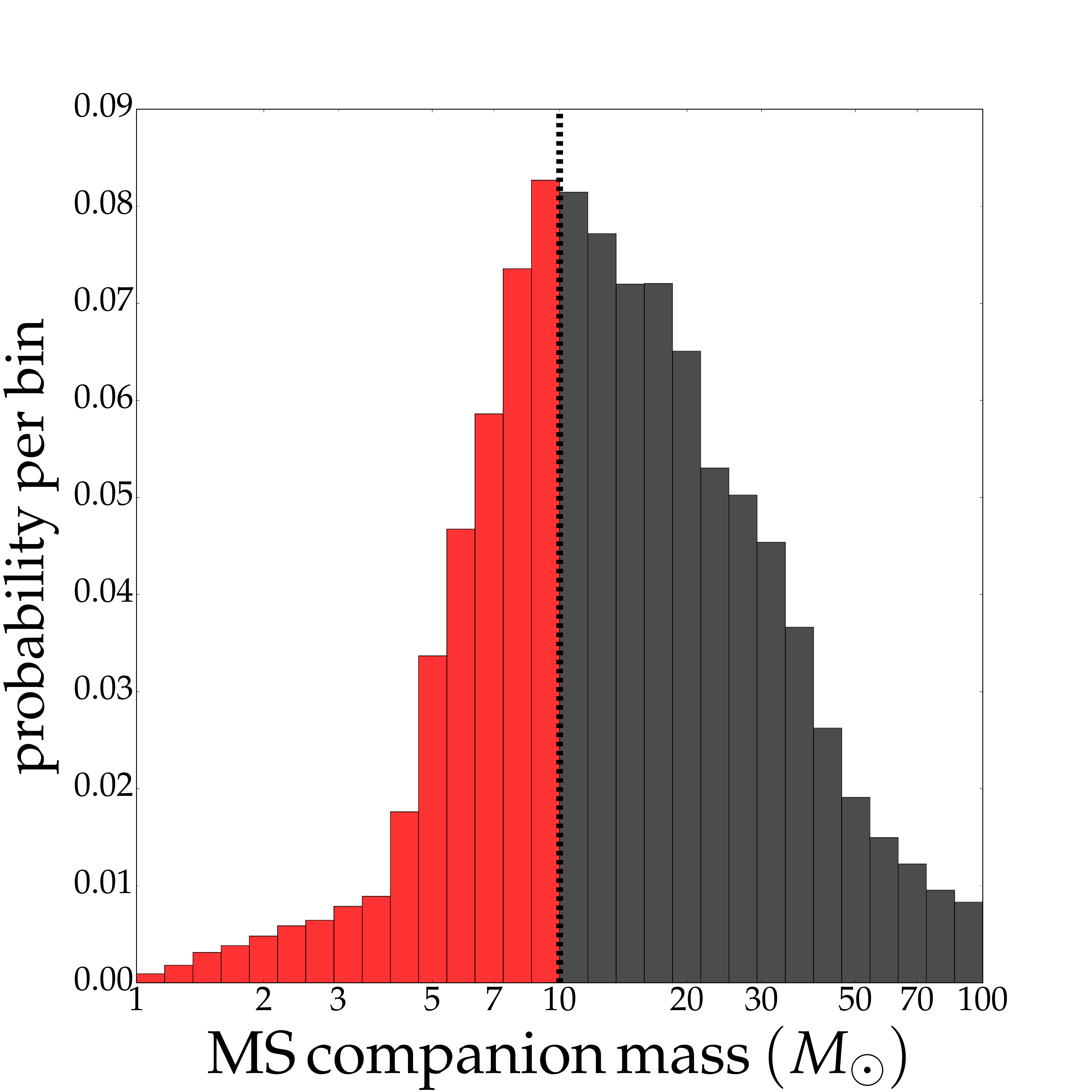}
\includegraphics[width=\onedvariable\textwidth]{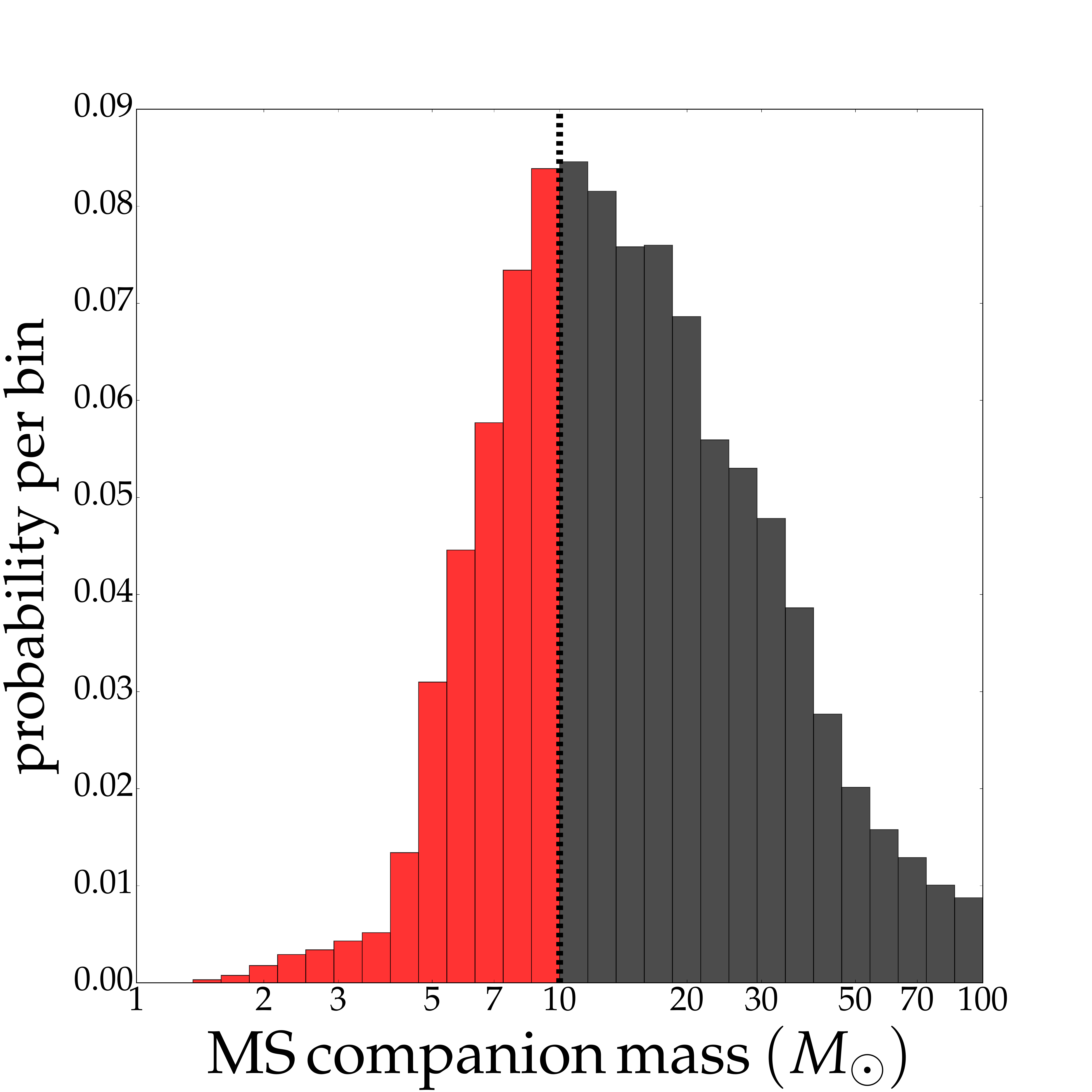}
\includegraphics[width=\onedvariable\textwidth]{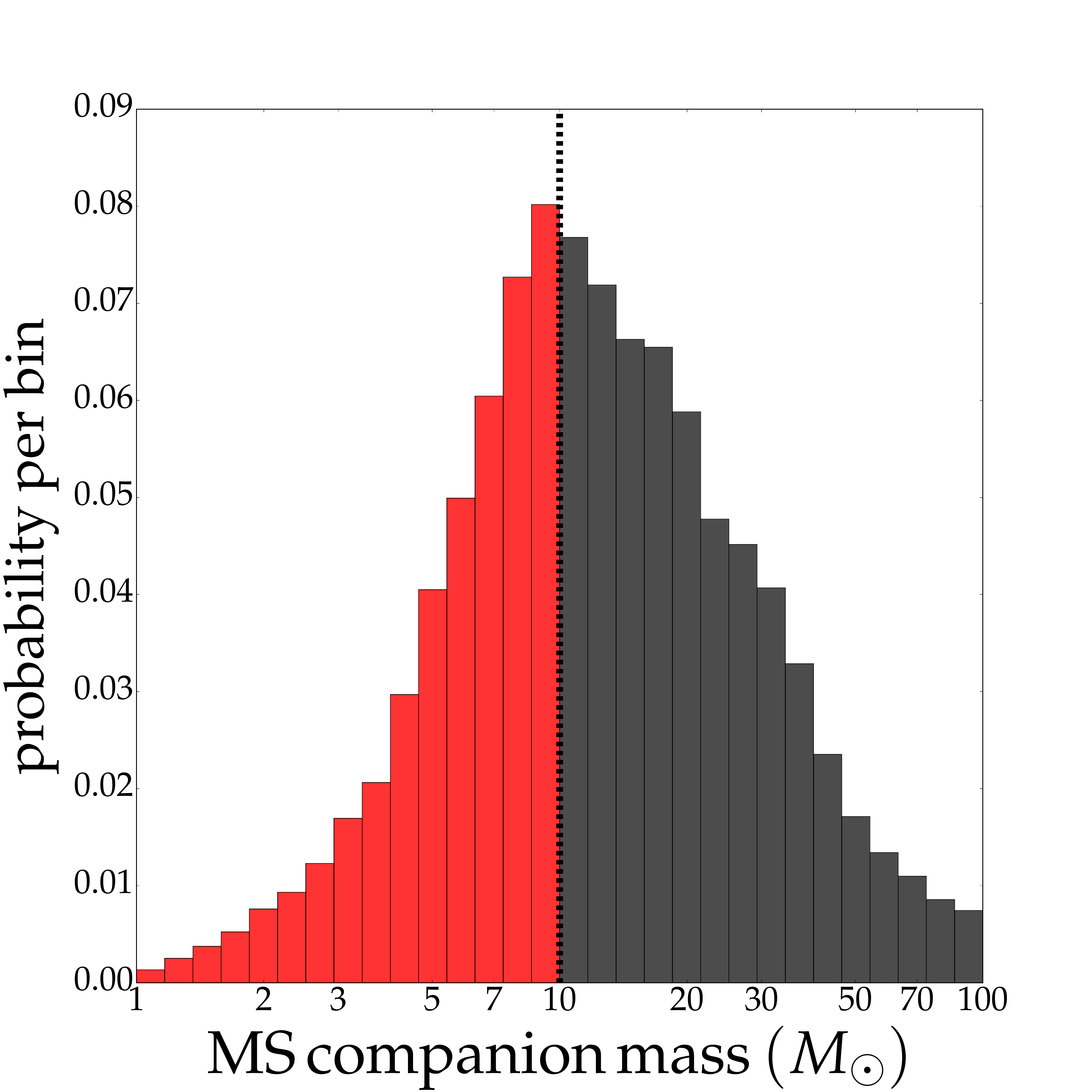}
\includegraphics[width=\onedvariable\textwidth]{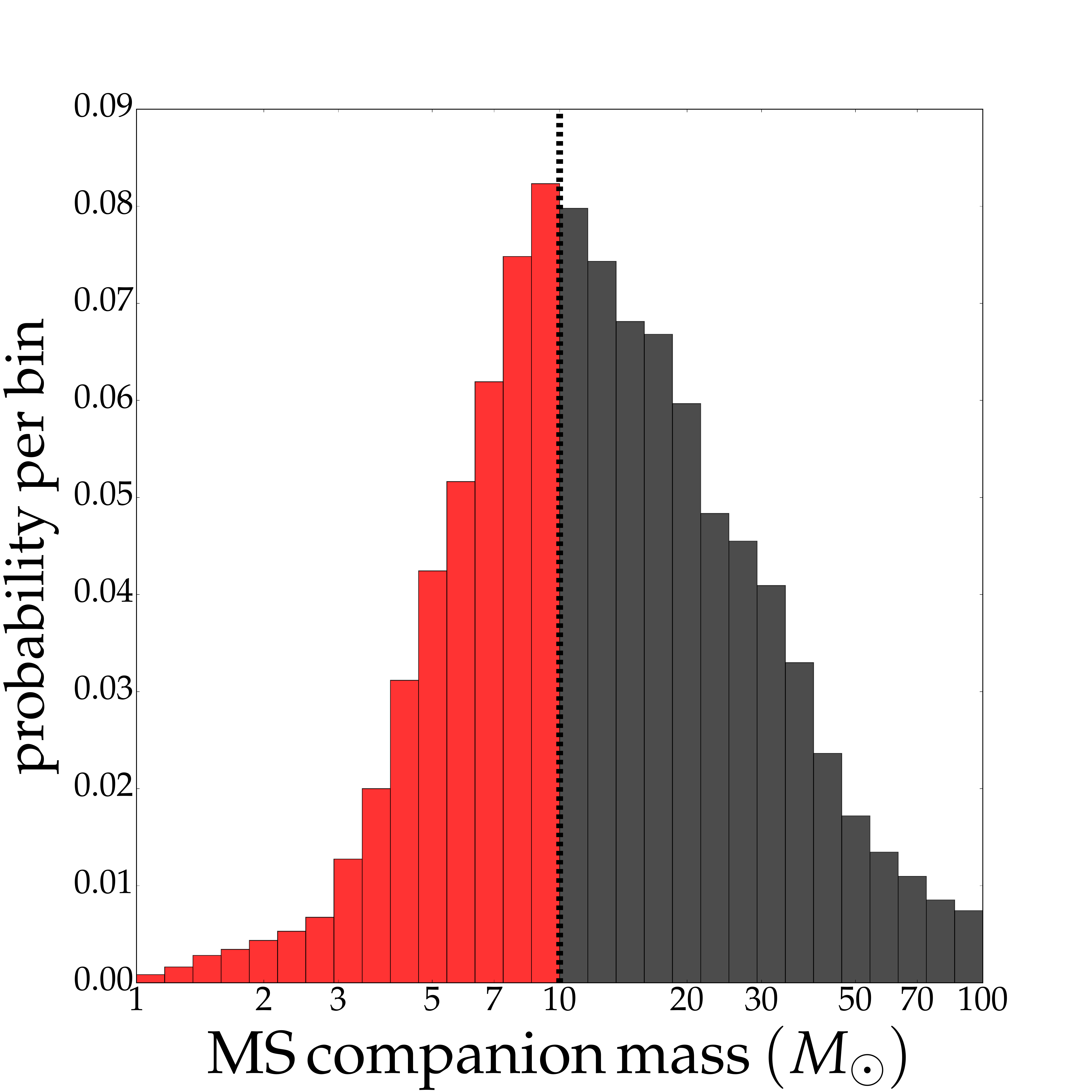}
\includegraphics[width=\onedvariable\textwidth]{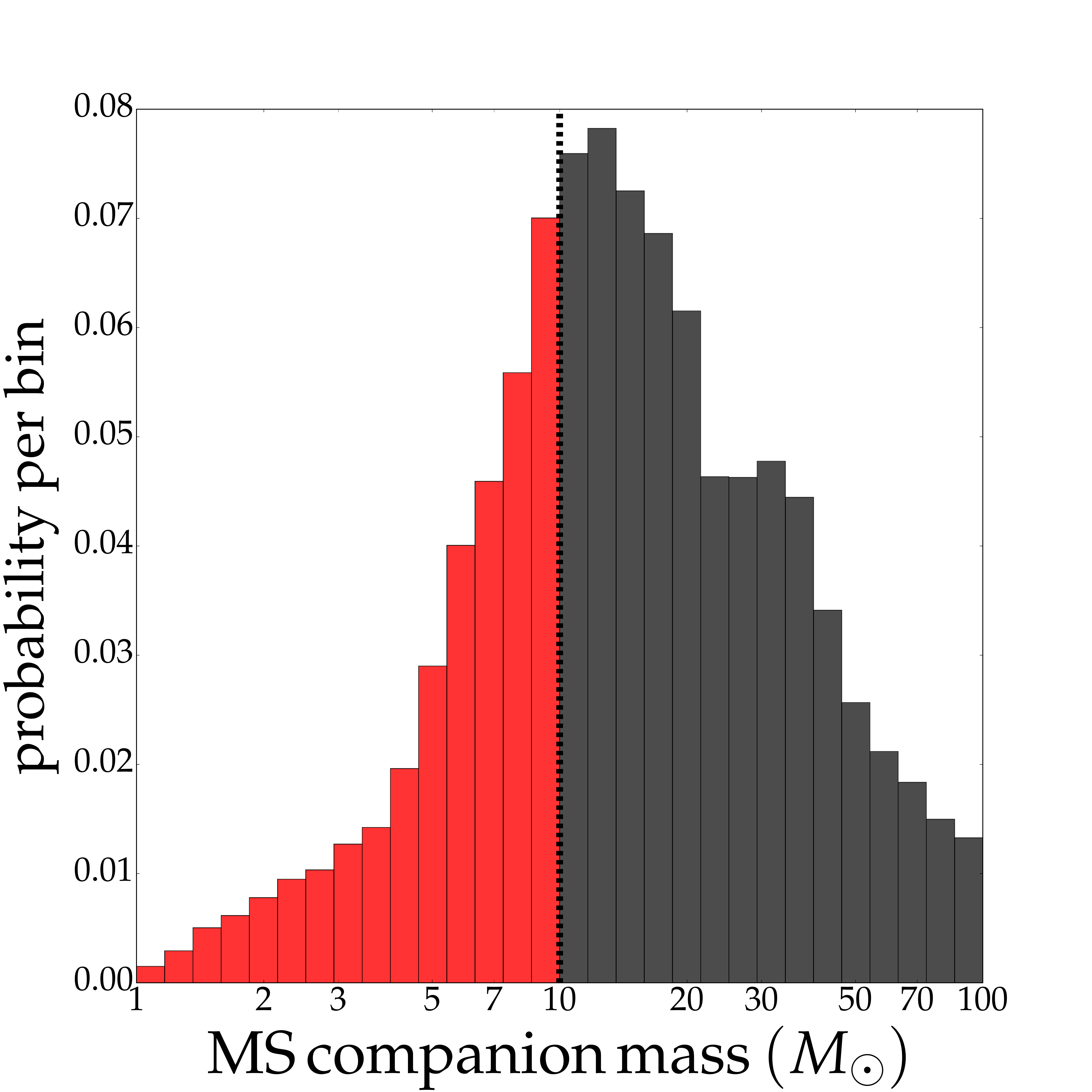}
\includegraphics[width=\onedvariable\textwidth]{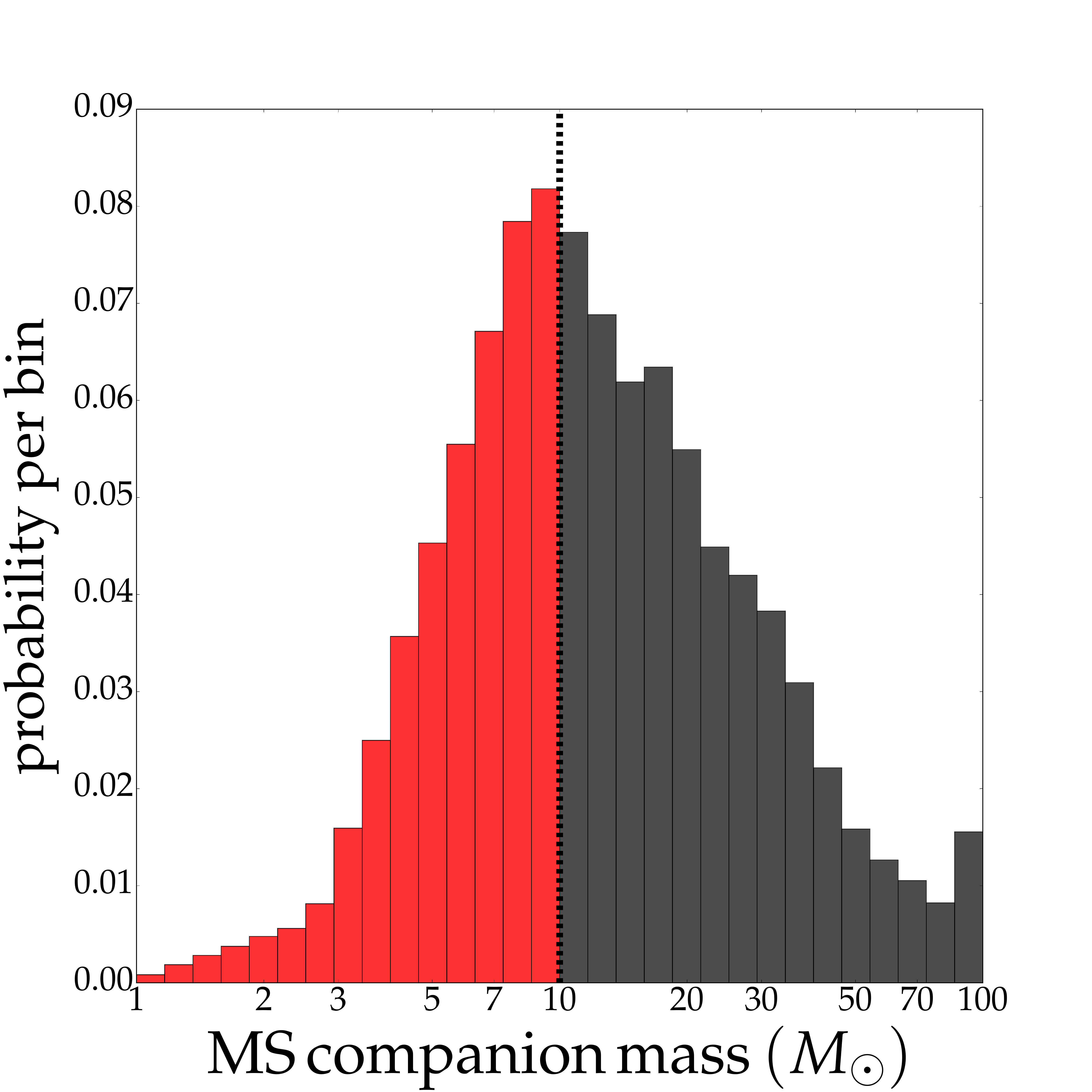}\\
\noindent\makebox[\linewidth]{\rule{0.8\paperwidth}{0.4pt}}\\
\includegraphics[width=\pievariable\textwidth]{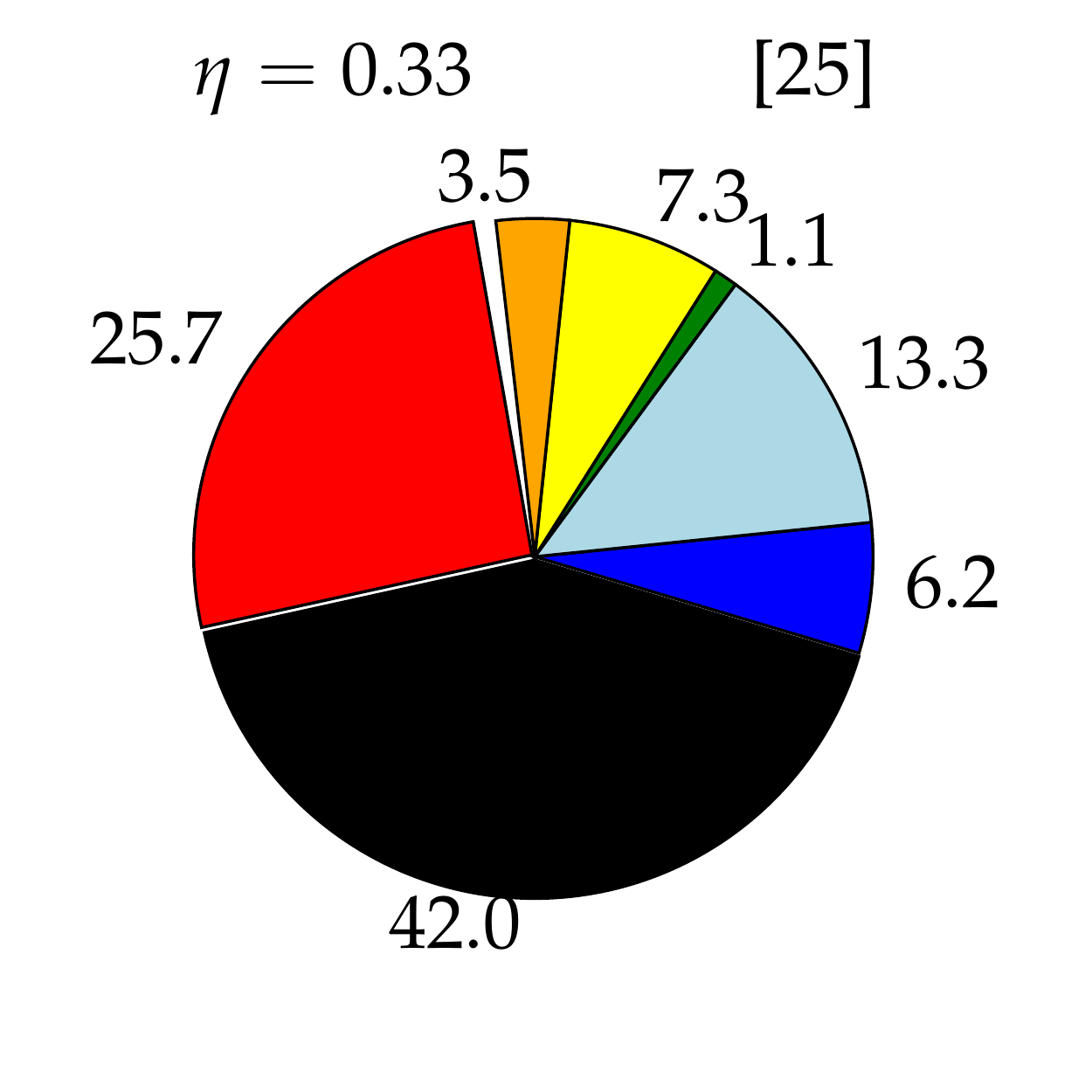}
\includegraphics[width=\pievariable\textwidth]{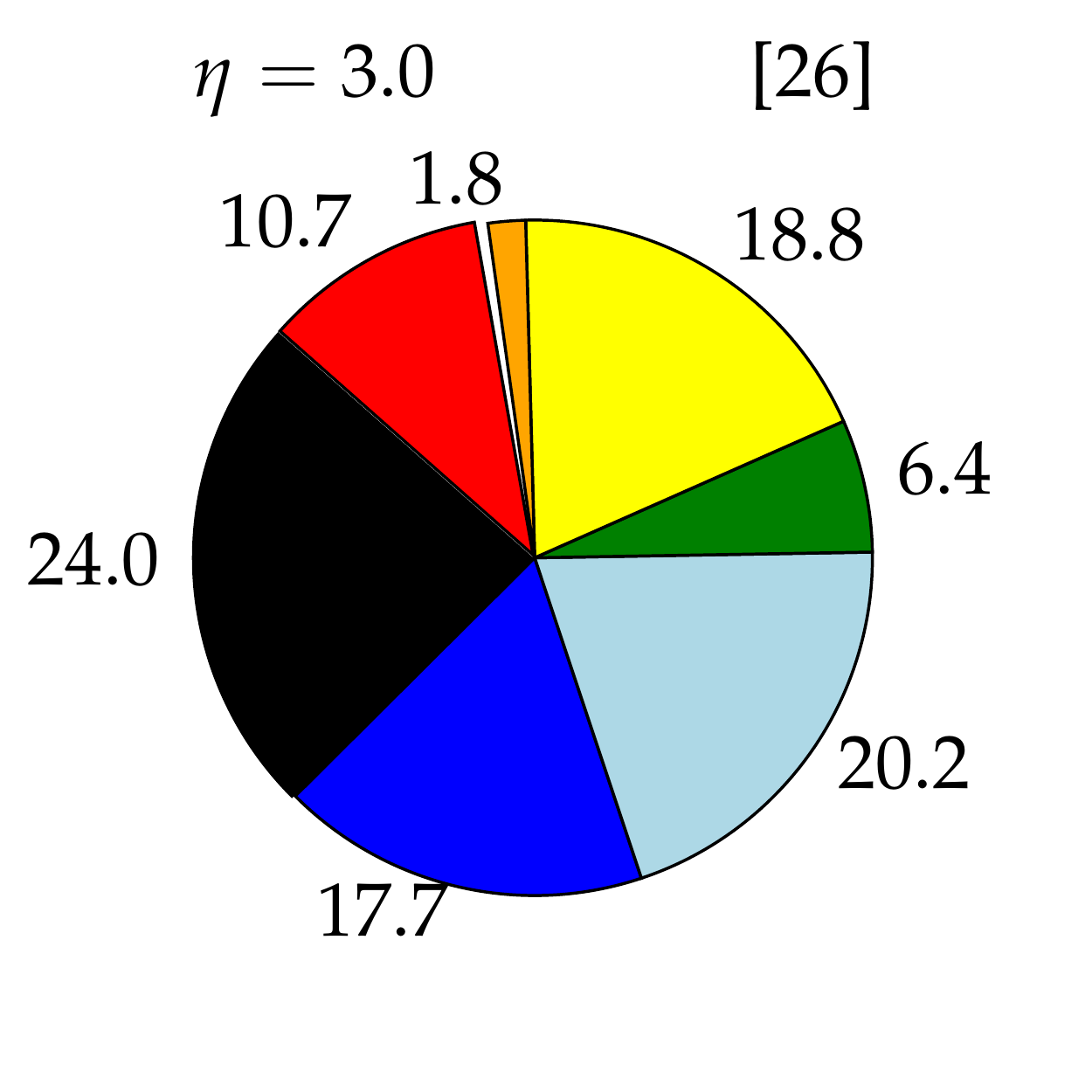}
\includegraphics[width=\pievariable\textwidth]{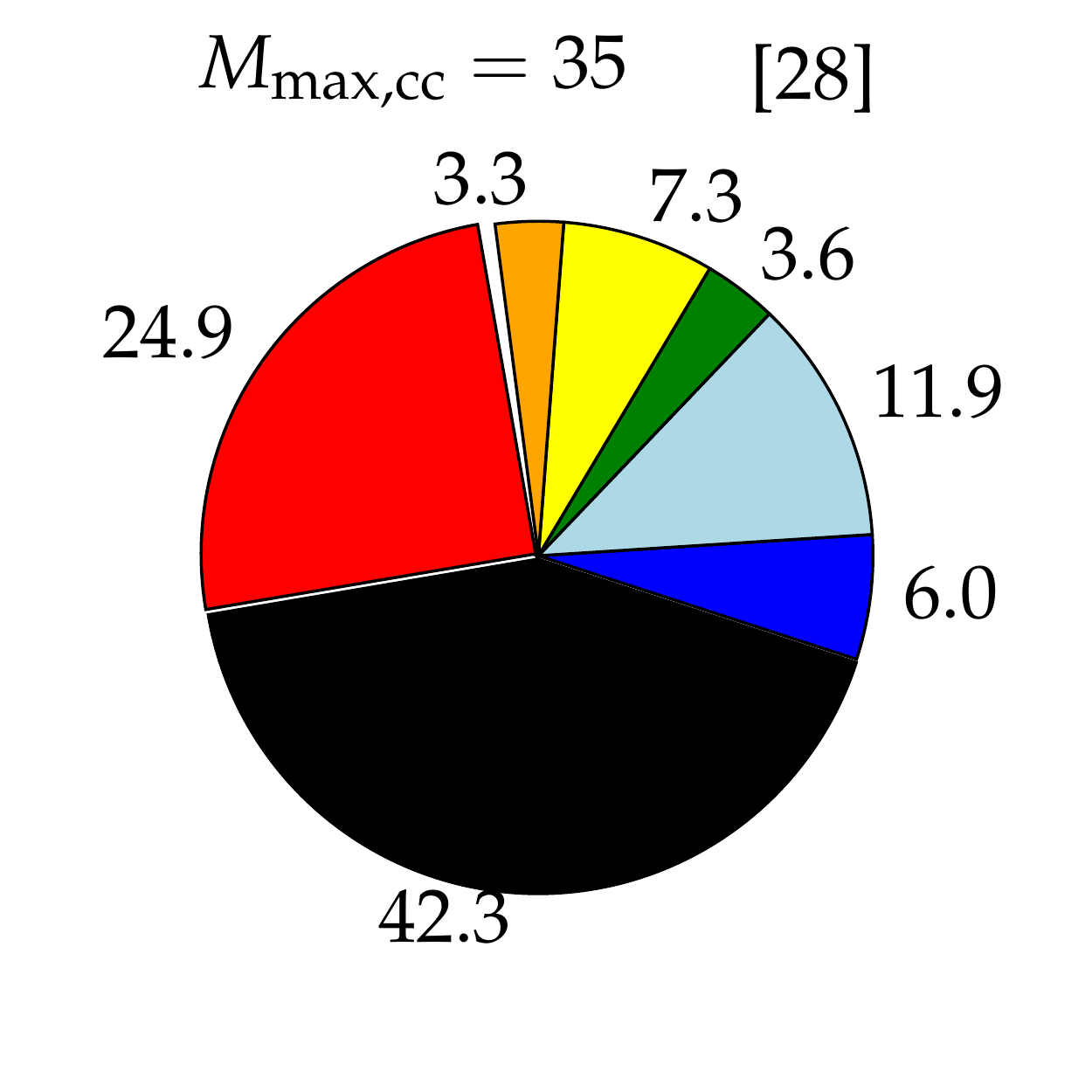}
\includegraphics[width=\pievariable\textwidth]{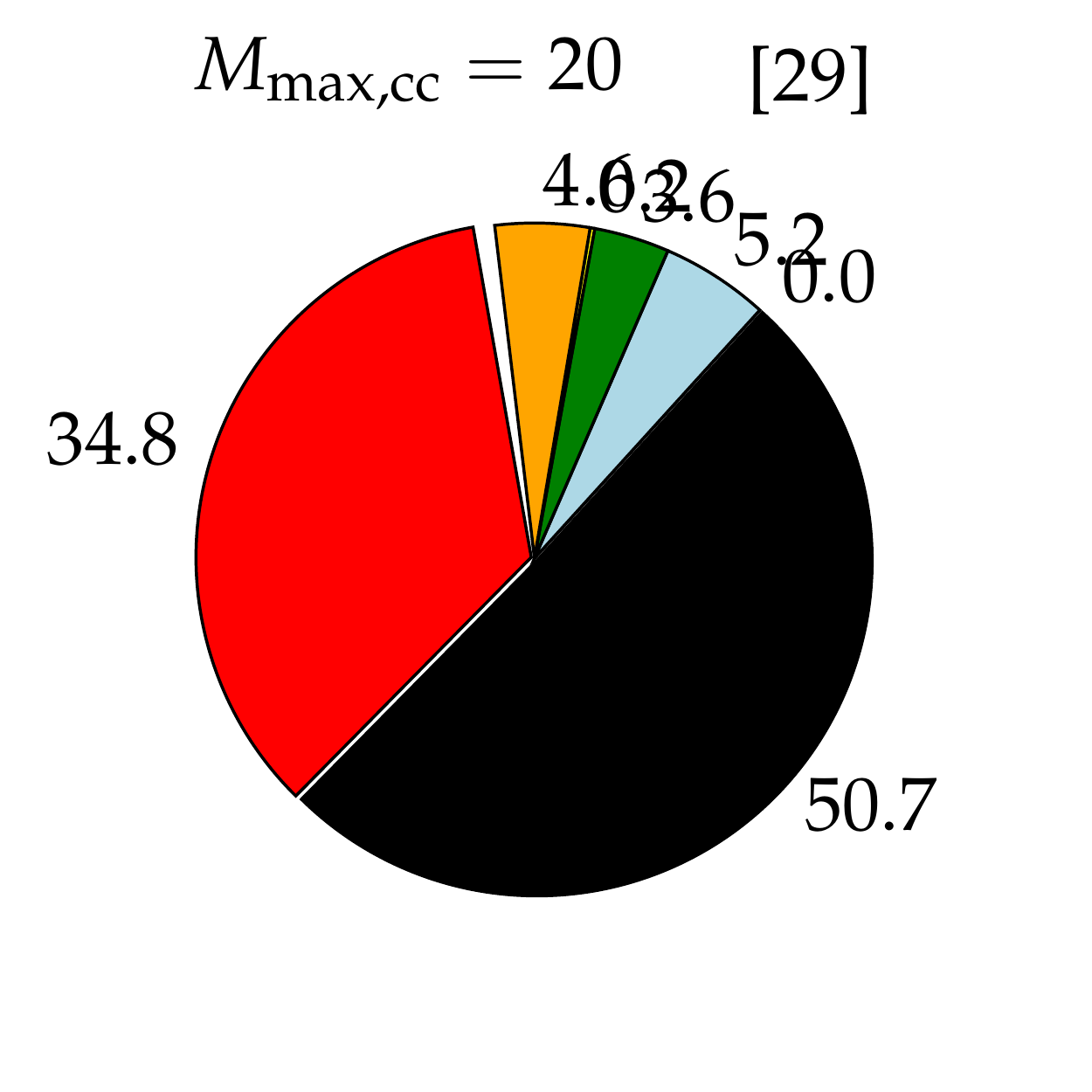}\\
\includegraphics[width=\onedvariable\textwidth]{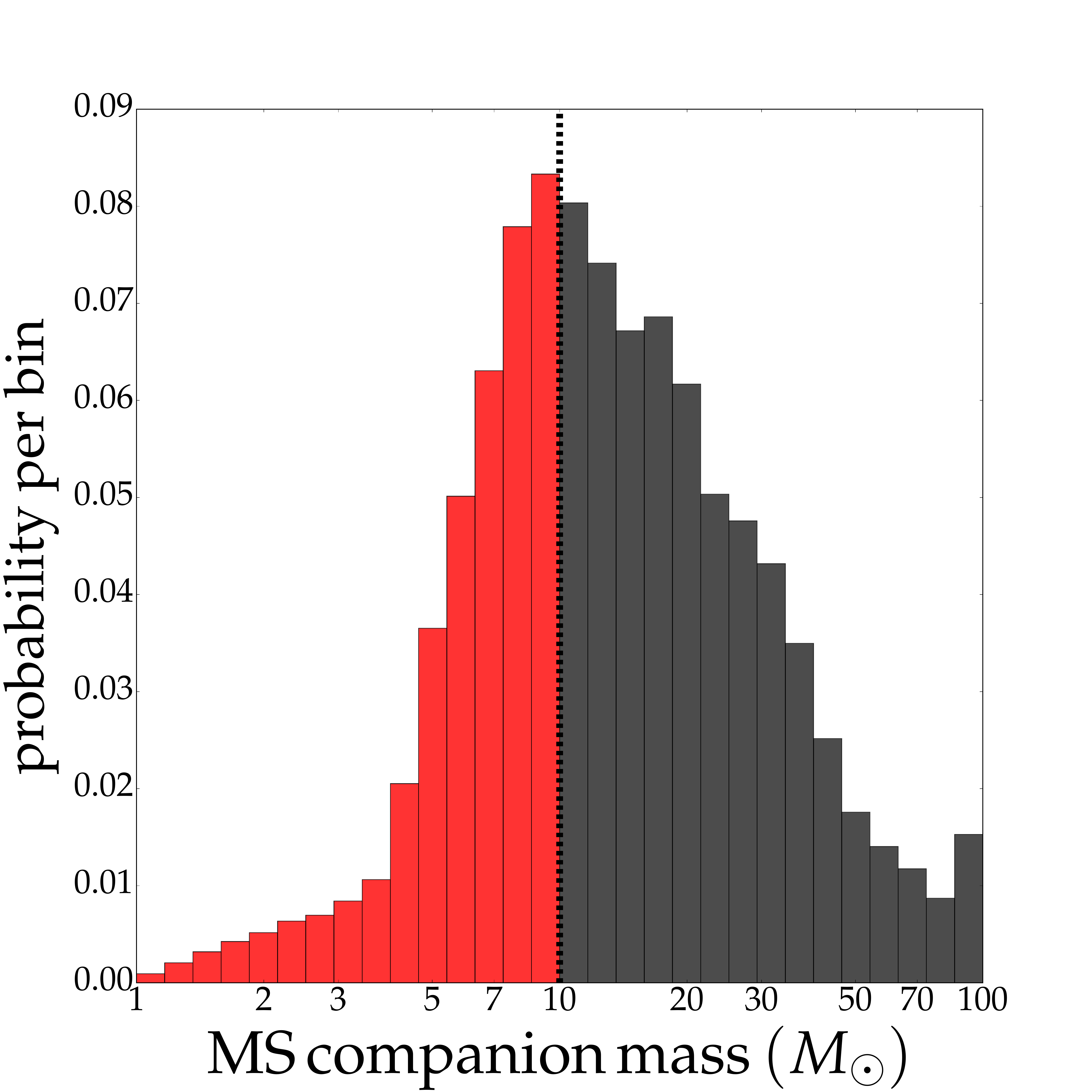}
\includegraphics[width=\onedvariable\textwidth]{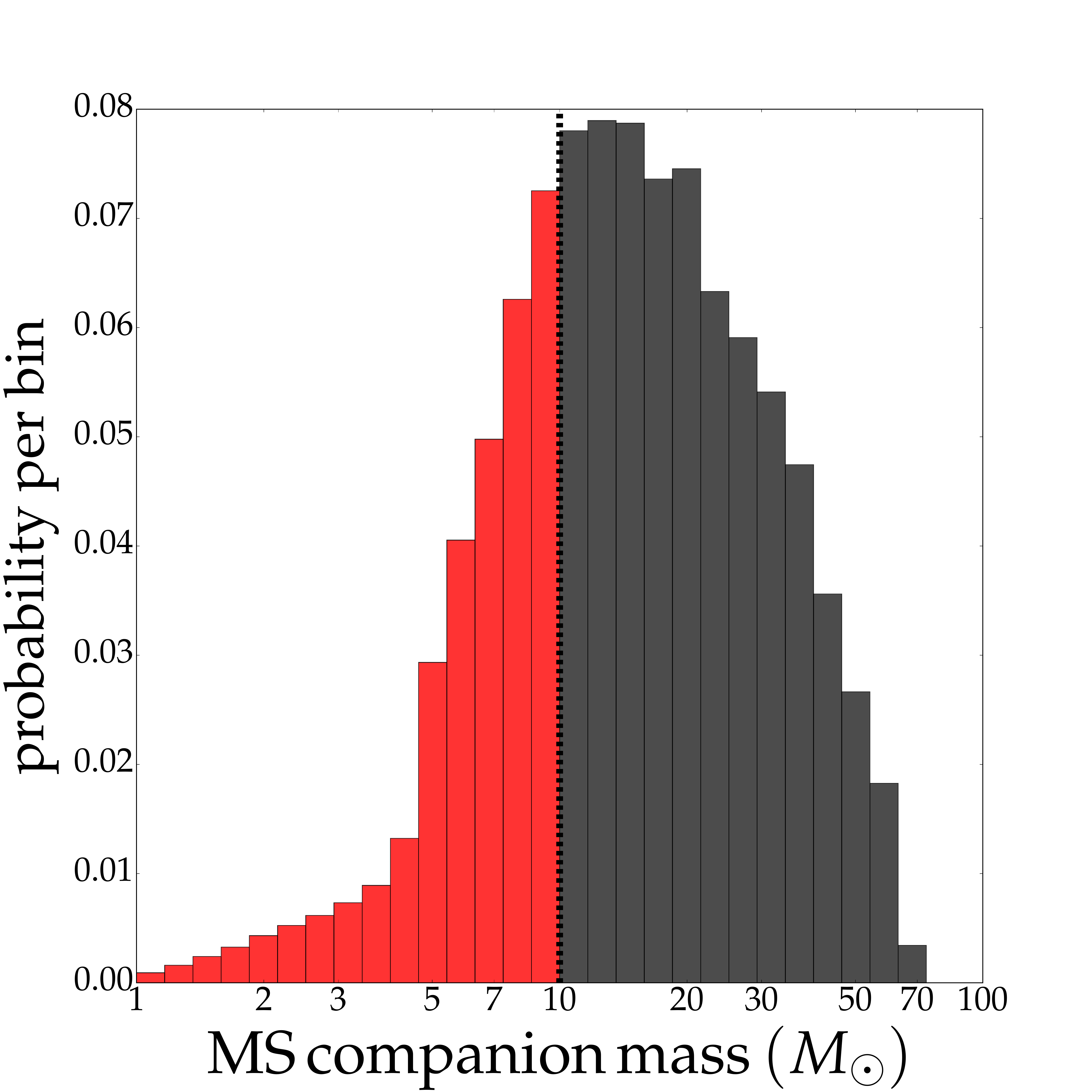}
\includegraphics[width=\onedvariable\textwidth]{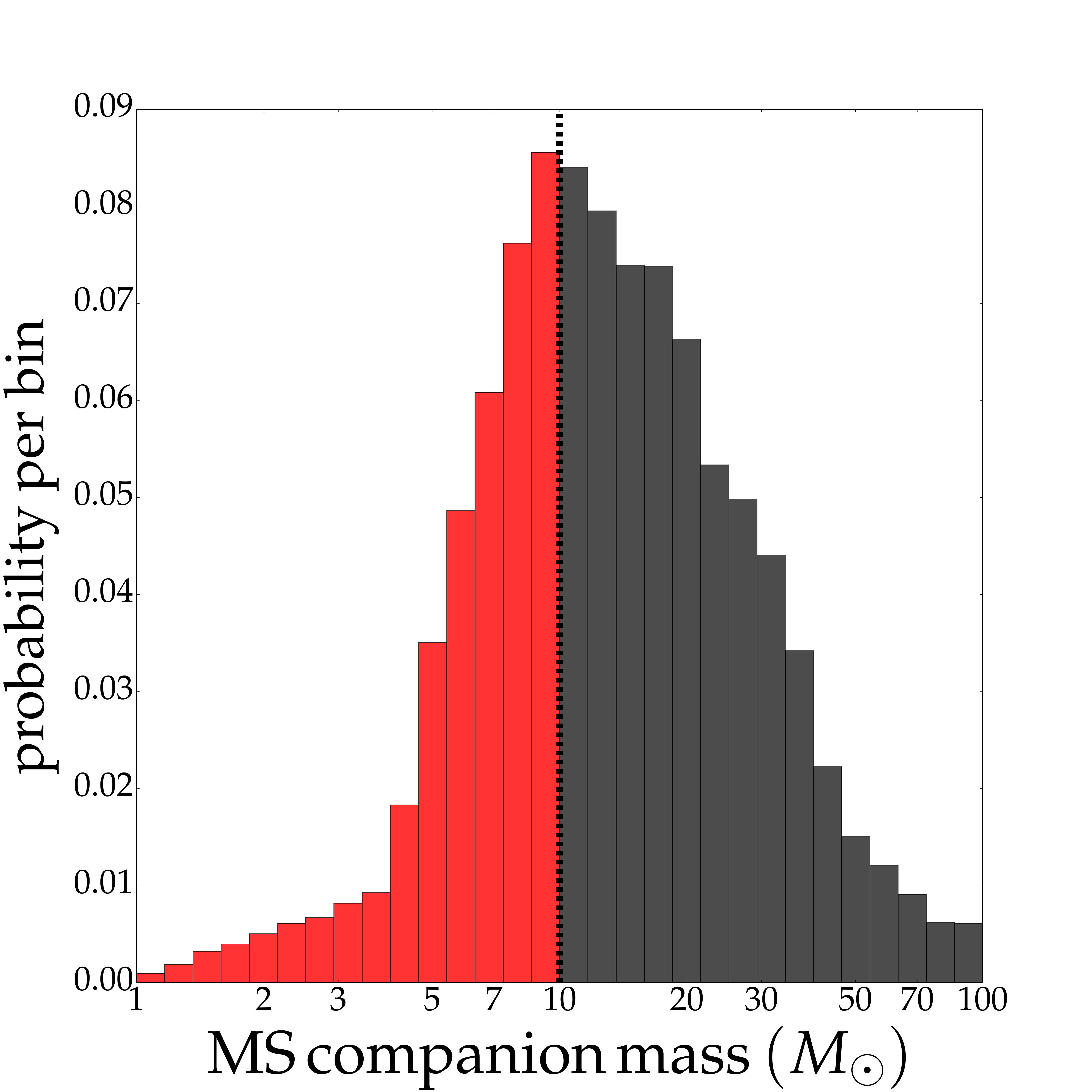}
\includegraphics[width=\onedvariable\textwidth]{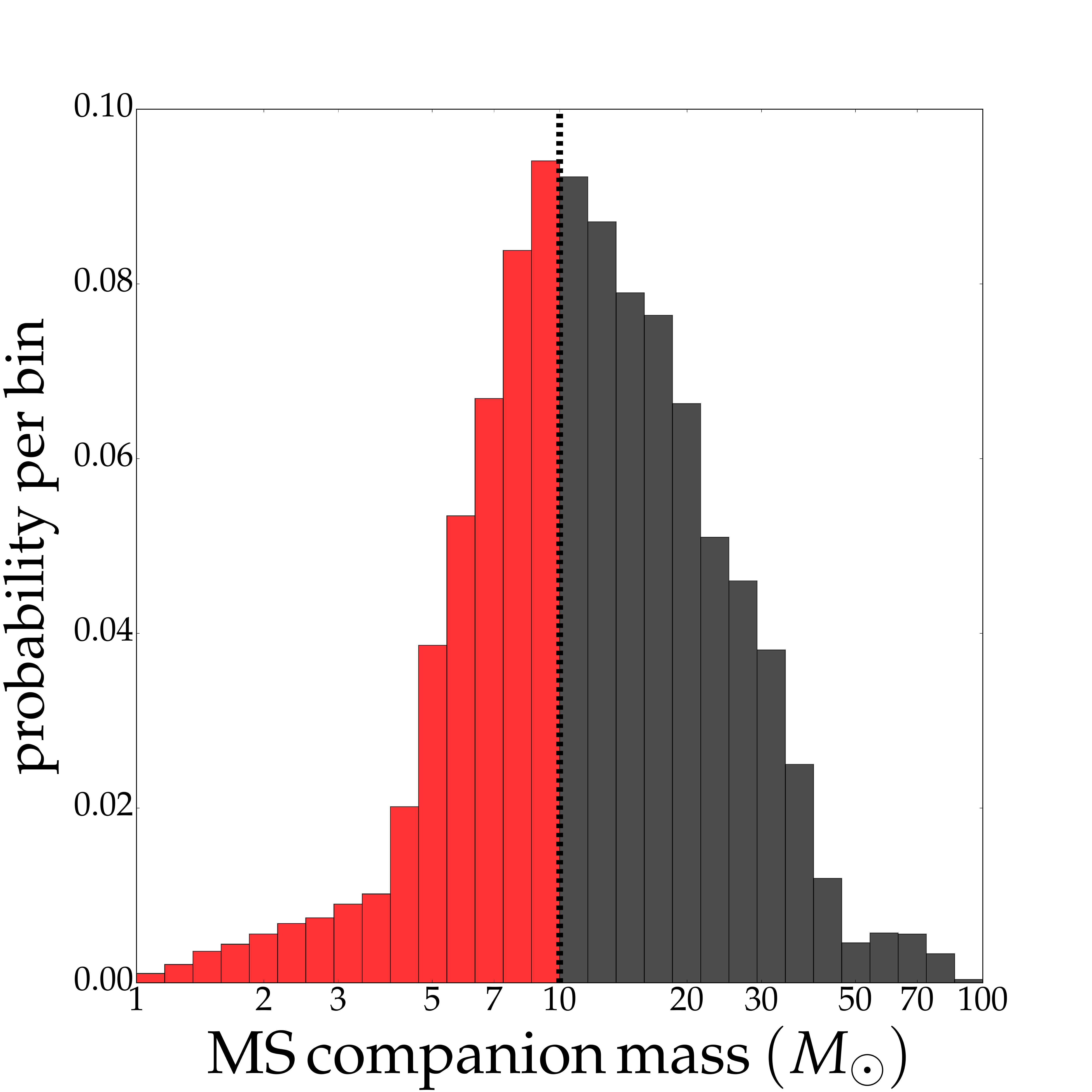}\\
\caption{Same as the reference simulation in Figure \ref{fig:metallicity} ($Z=0.014$; Model 00 of \citetalias{Zapartas+2017}) 
but with different physical assumptions. A dash in model number means that this assumed variation is not considered by \citetalias{Zapartas+2017}.
  }\label{fig:physical_assumptions}
\end{center}
\end{figure*}

\begin{figure*}
\begin{center}
\includegraphics[width=\textwidth]{legend_piechart_compressed}
\noindent\makebox[\linewidth]{\rule{0.8\paperwidth}{0.4pt}}\\
\includegraphics[width=\pievariable\textwidth]{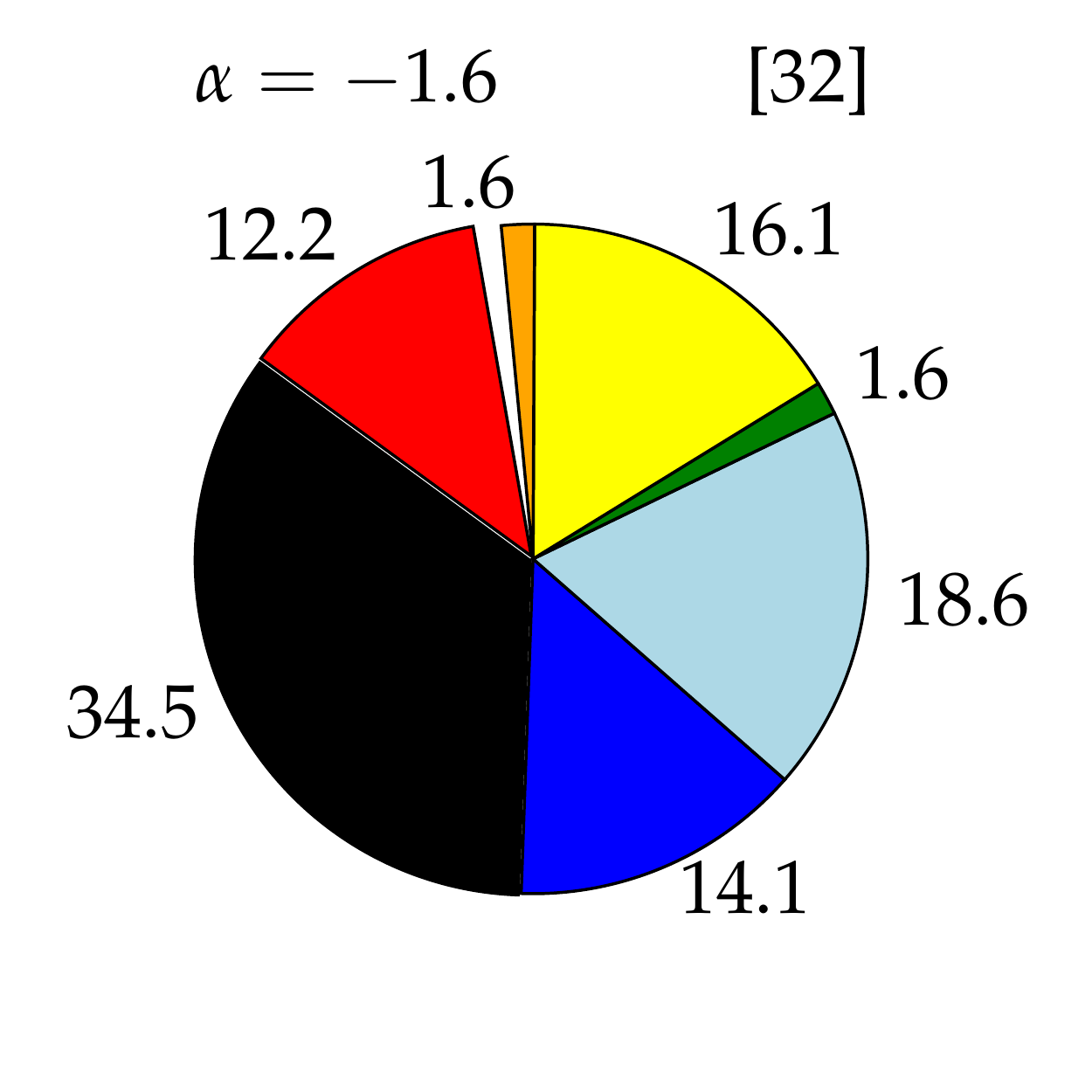}
\includegraphics[width=\pievariable\textwidth]{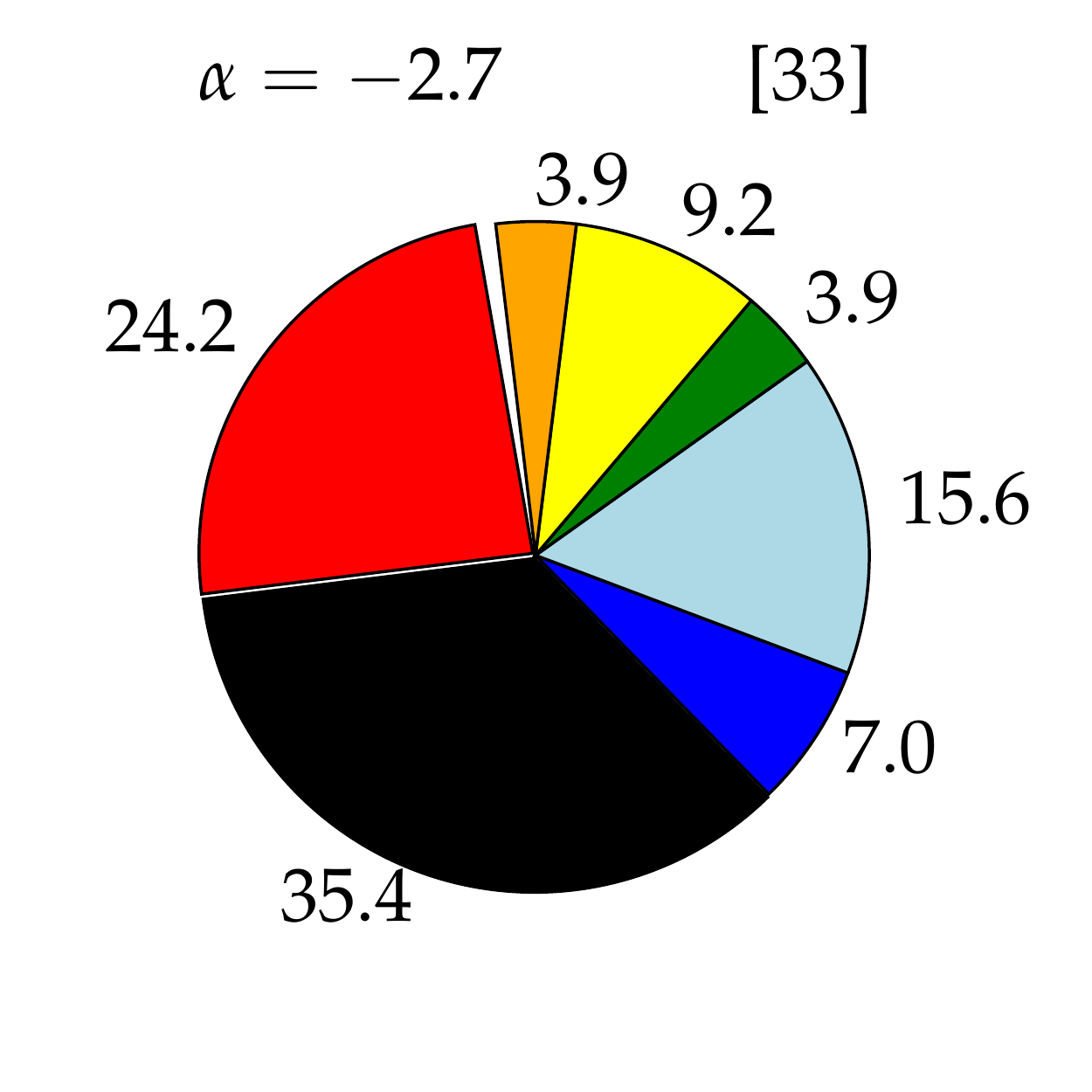}
\includegraphics[width=\pievariable\textwidth]{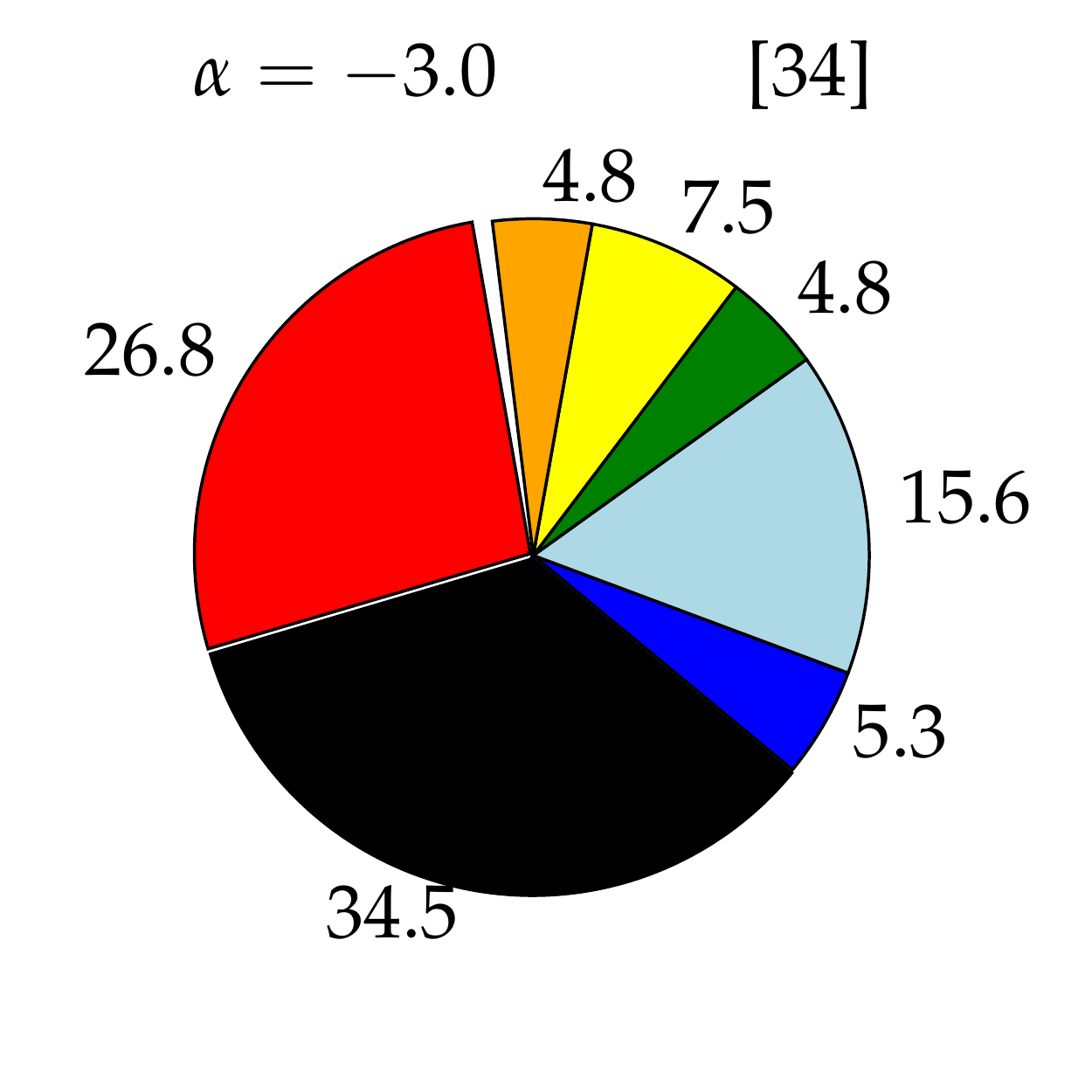}
\includegraphics[width=\pievariable\textwidth]{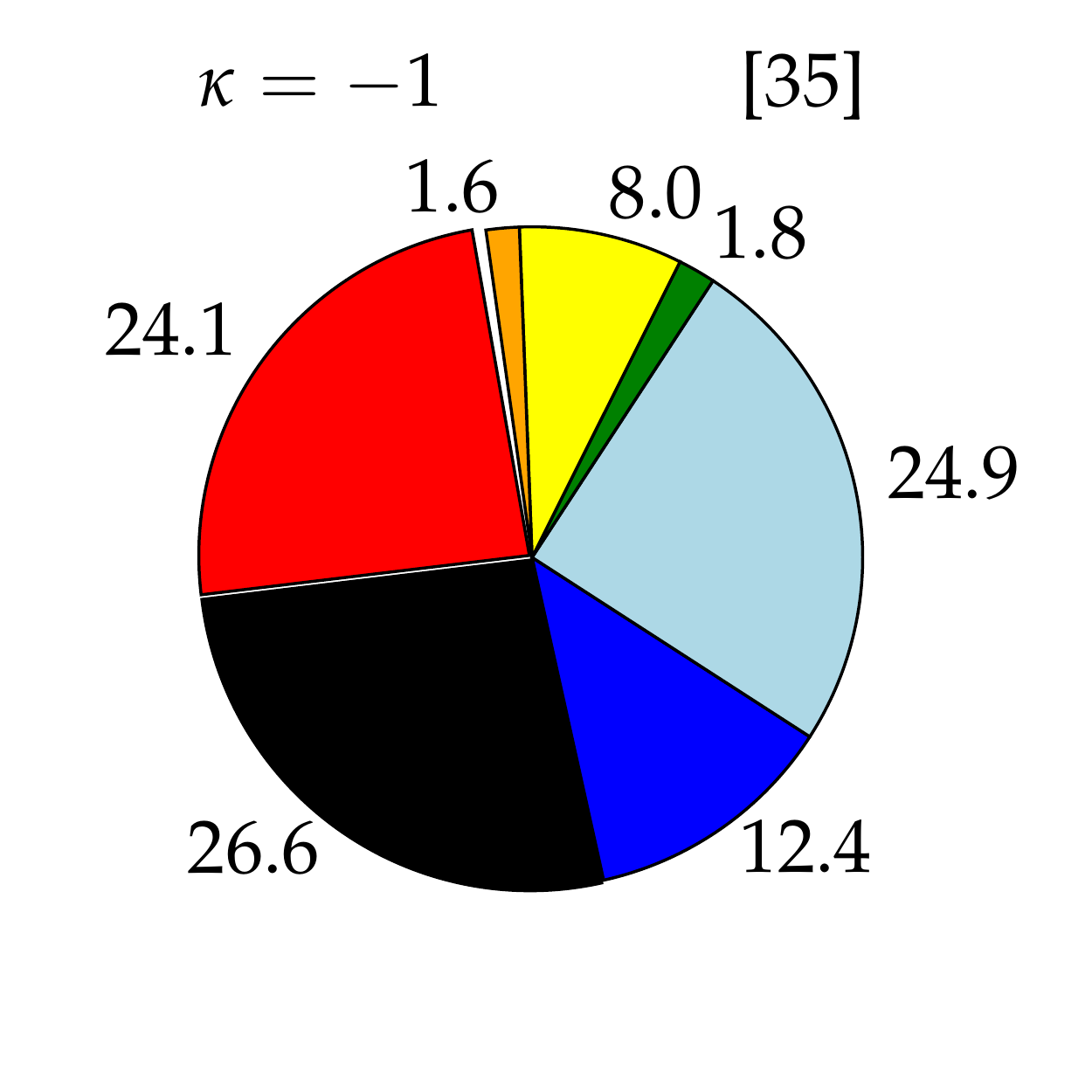}
\includegraphics[width=\pievariable\textwidth]{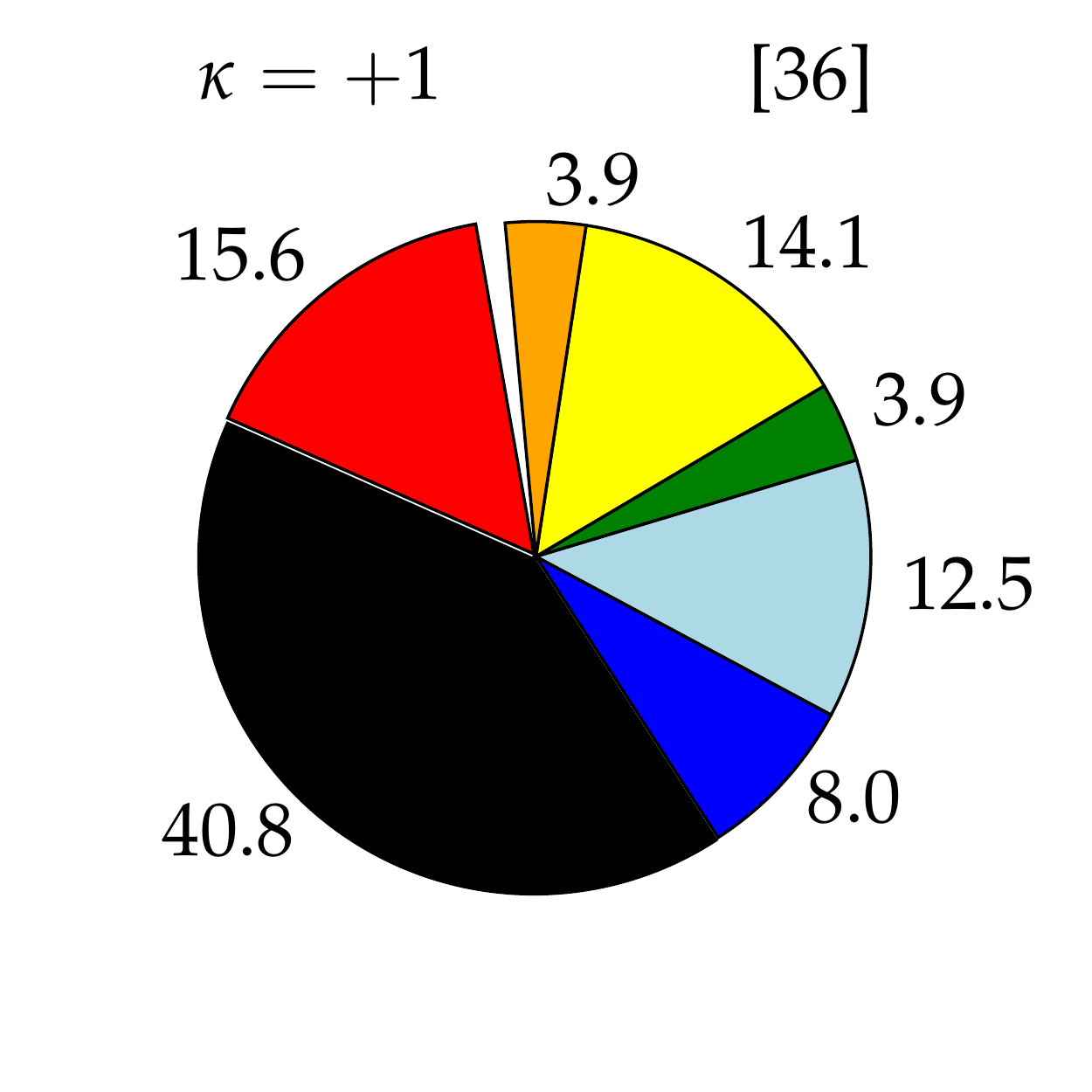}\\
\includegraphics[width=\onedvariable\textwidth]{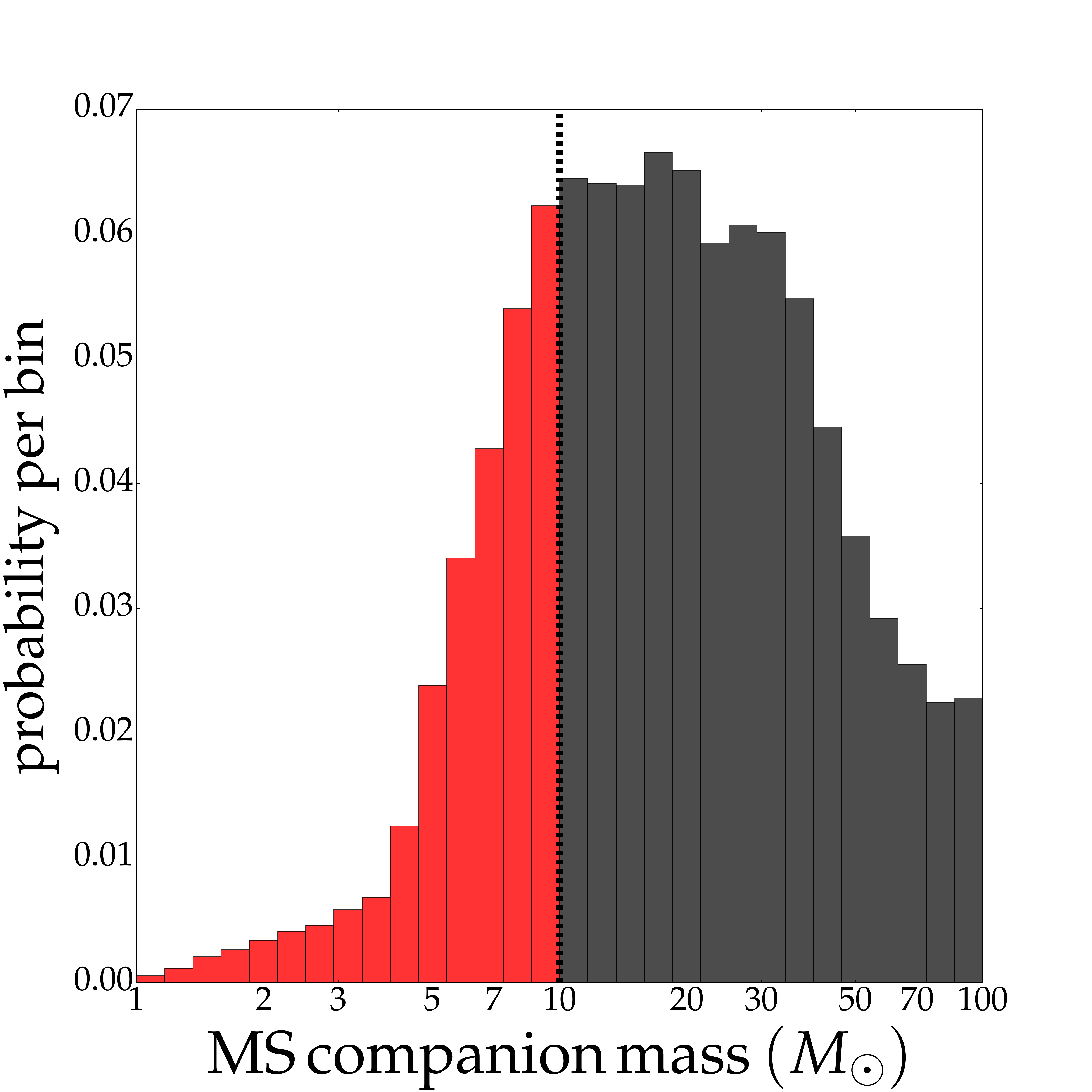}
\includegraphics[width=\onedvariable\textwidth]{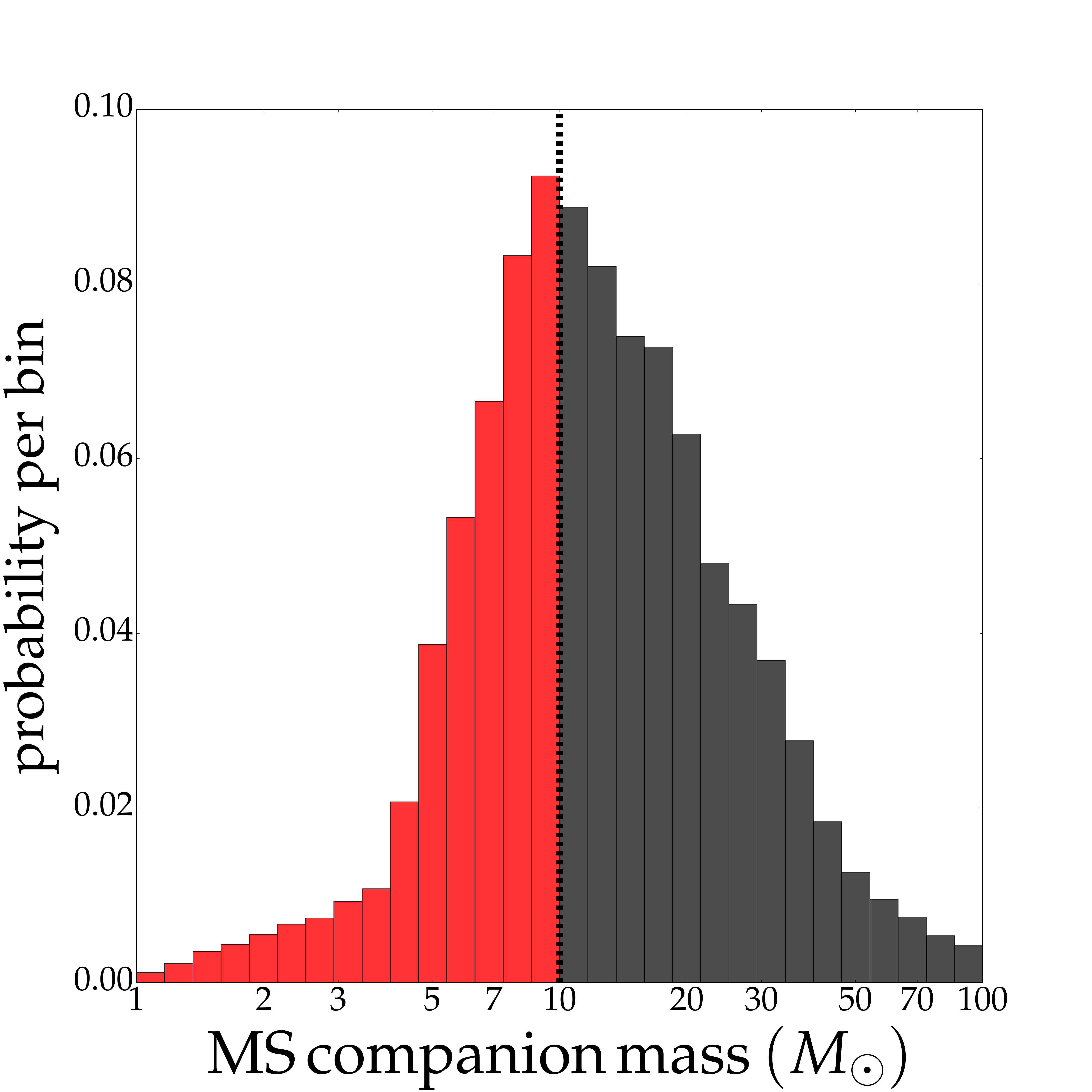}
\includegraphics[width=\onedvariable\textwidth]{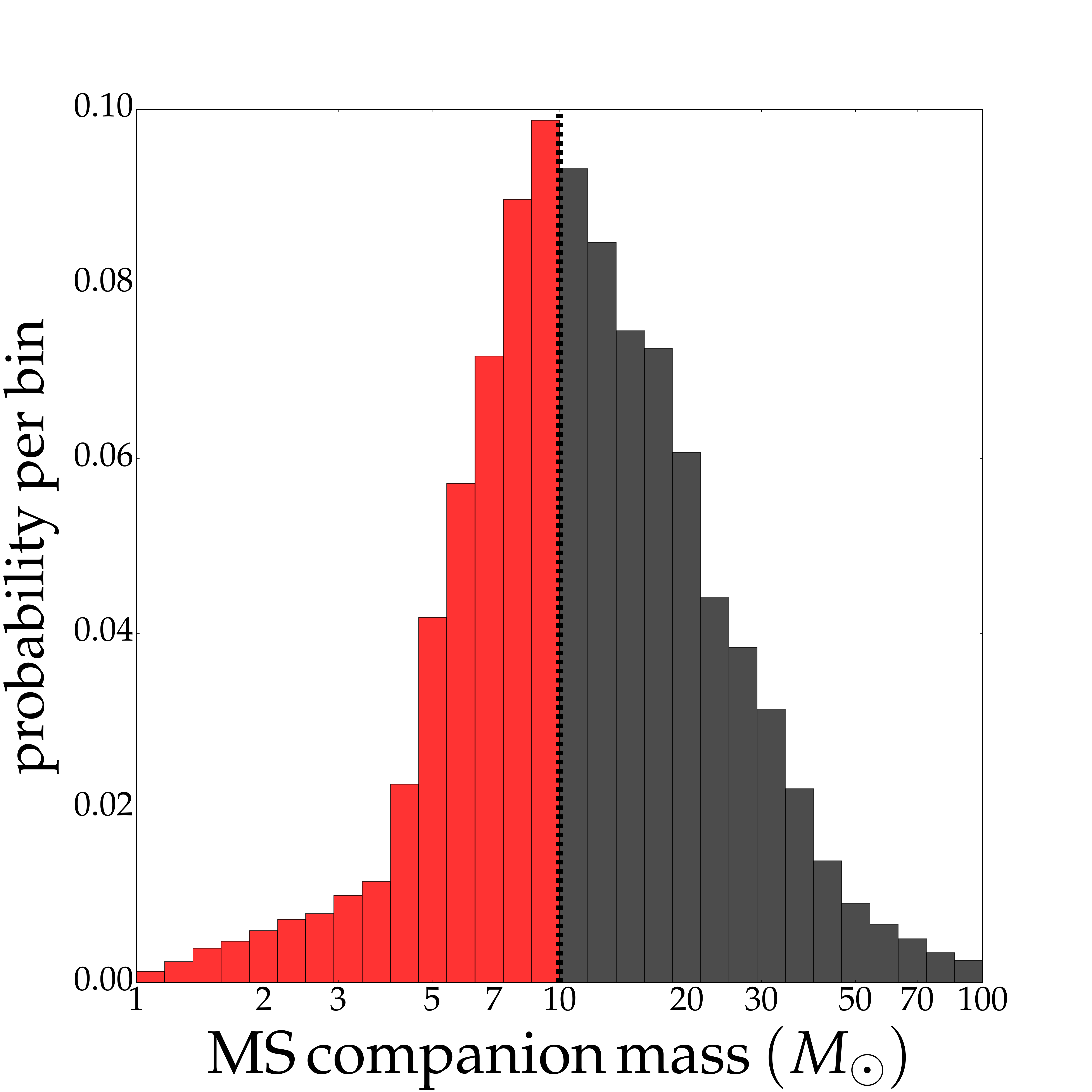}
\includegraphics[width=\onedvariable\textwidth]{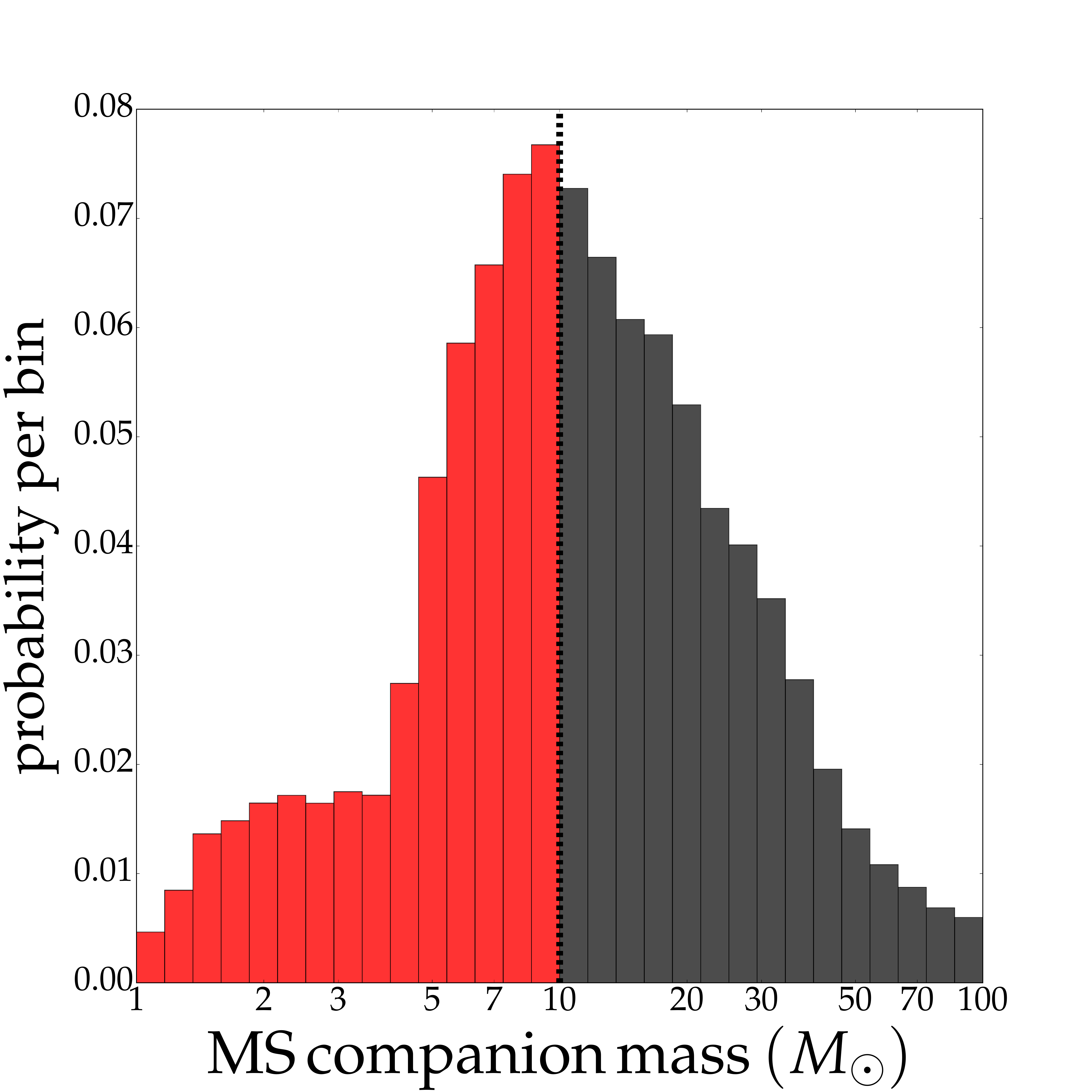}
\includegraphics[width=\onedvariable\textwidth]{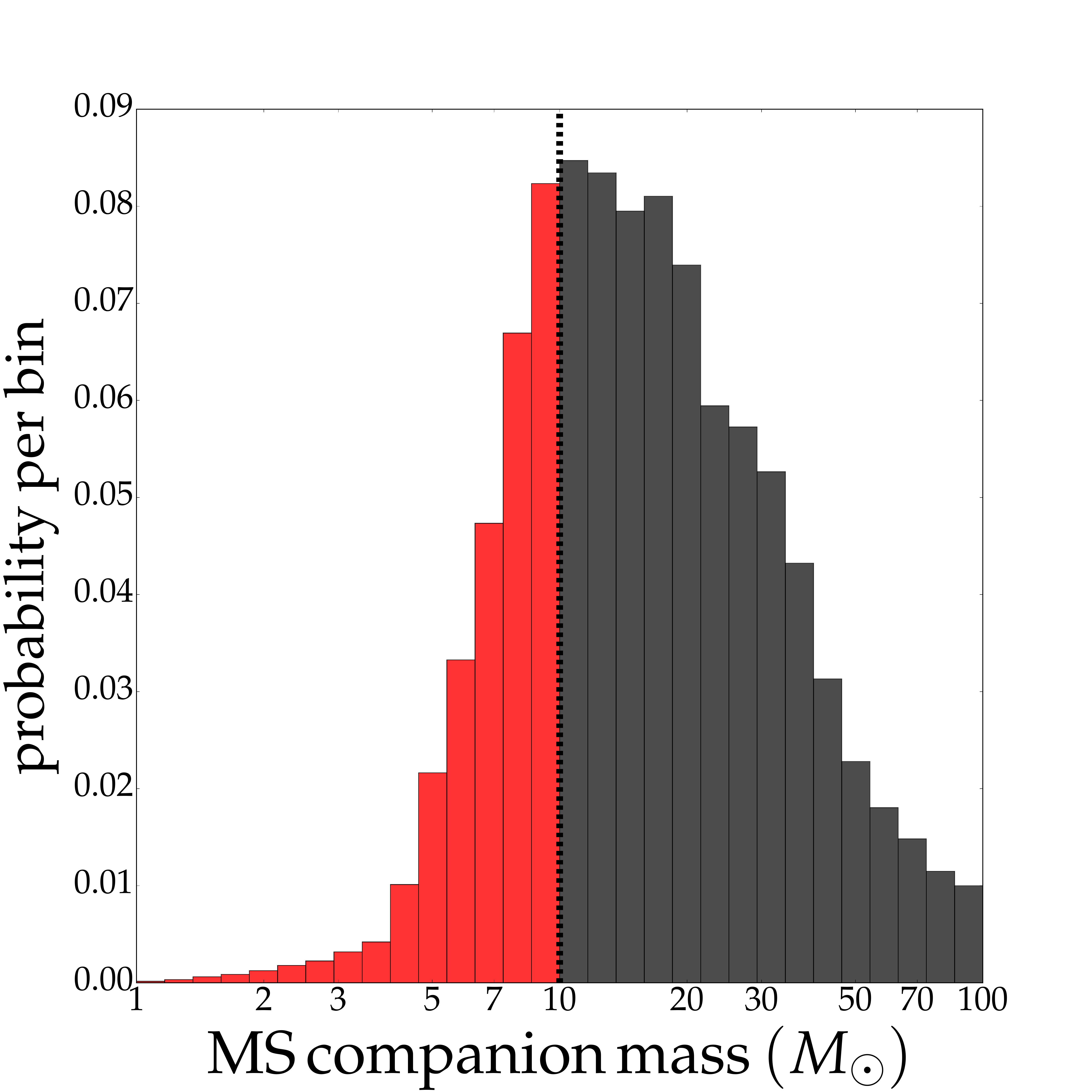}\\
\noindent\makebox[\linewidth]{\rule{0.8\paperwidth}{0.4pt}}\\
\includegraphics[width=\pievariable\textwidth]{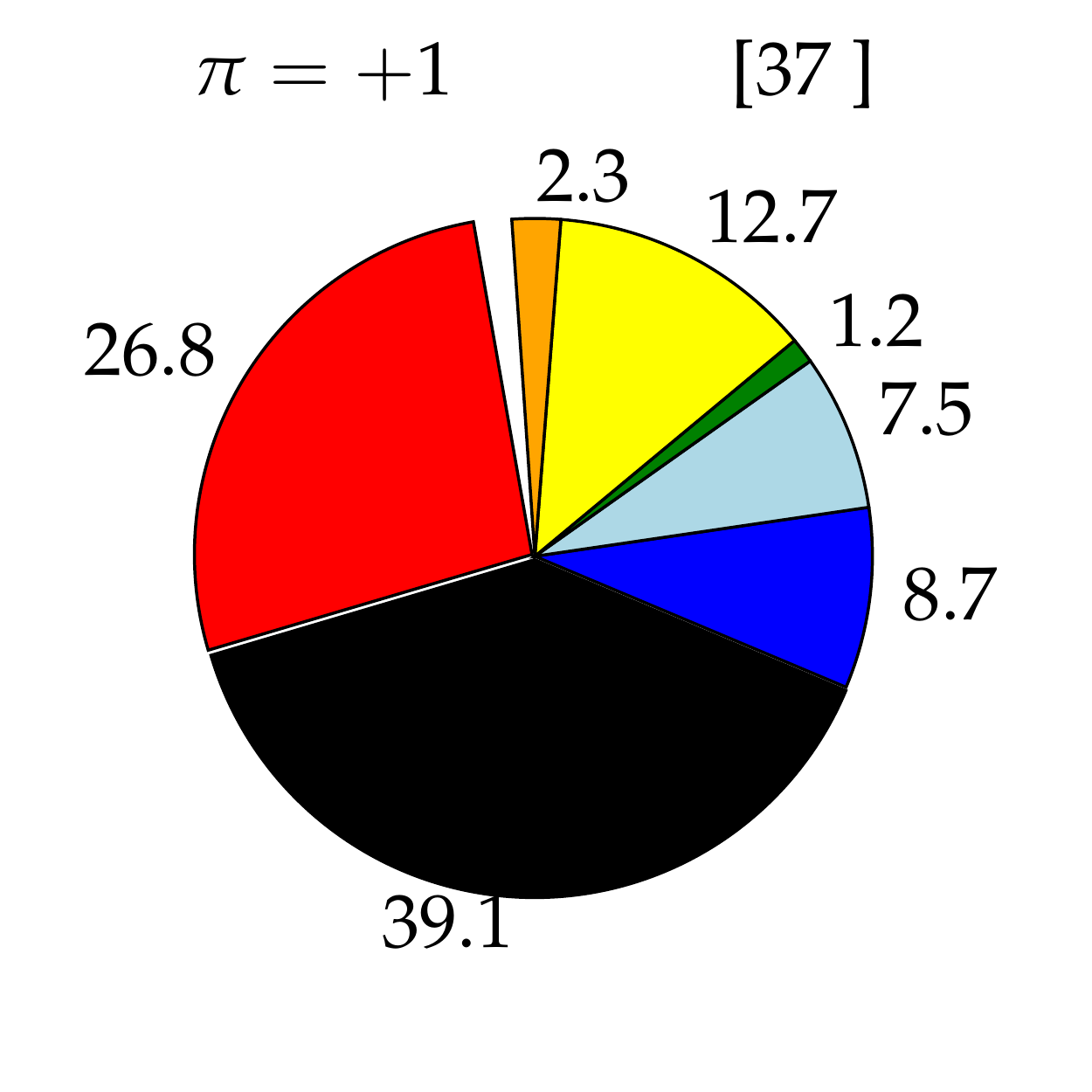}
\includegraphics[width=\pievariable\textwidth]{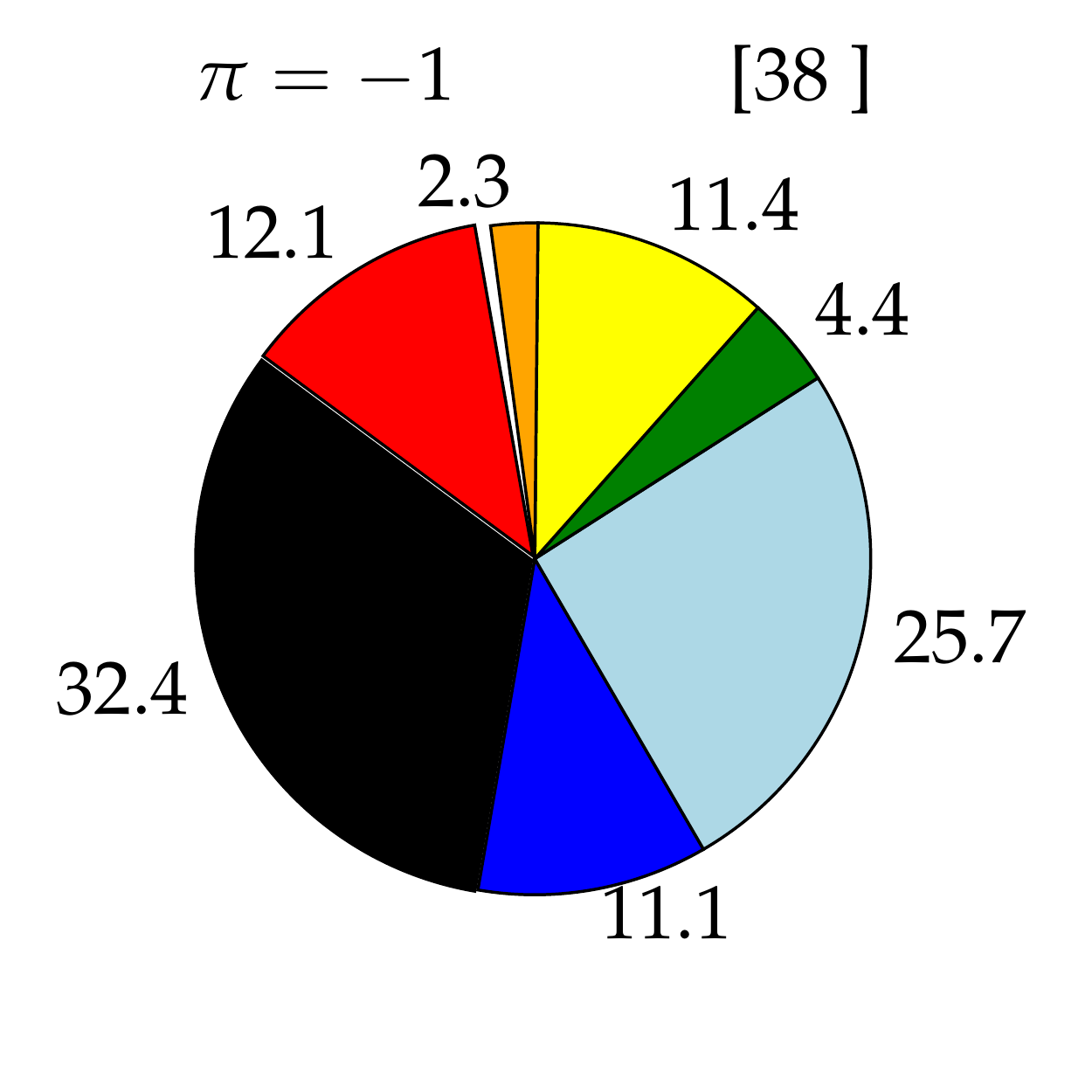}
\includegraphics[width=\pievariable\textwidth]{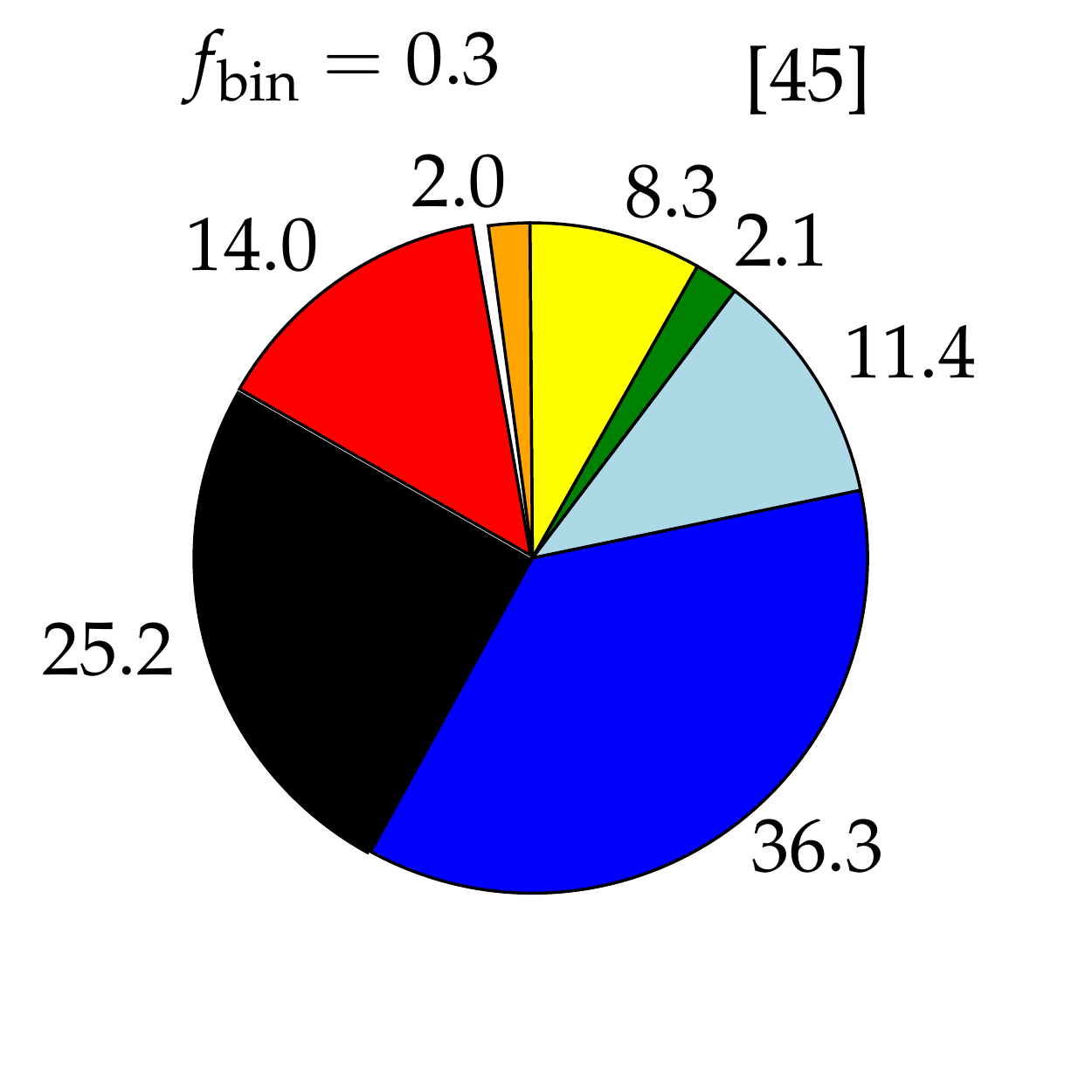}
\includegraphics[width=\pievariable\textwidth]{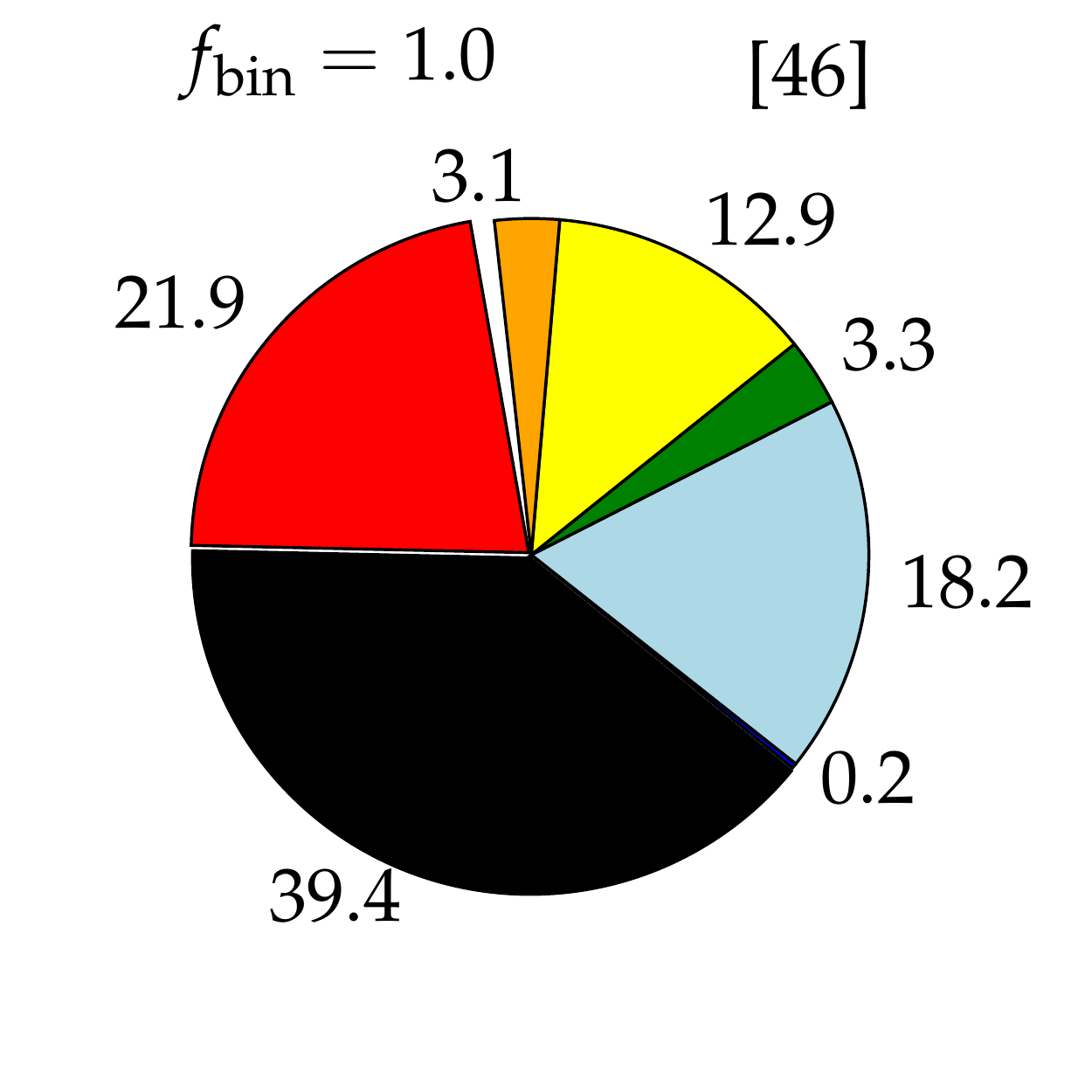}
\includegraphics[width=\pievariable\textwidth]{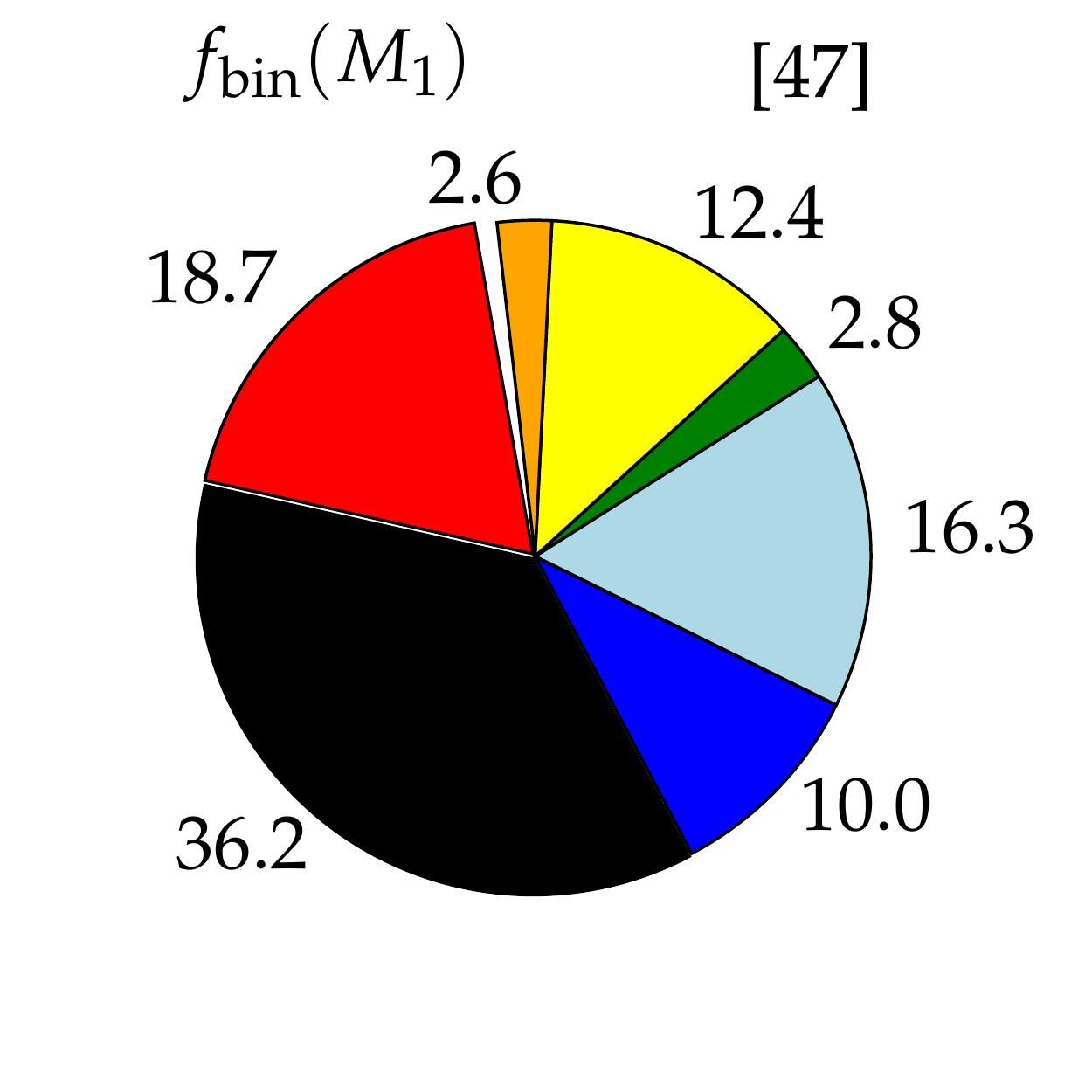}\\
\includegraphics[width=\onedvariable\textwidth]{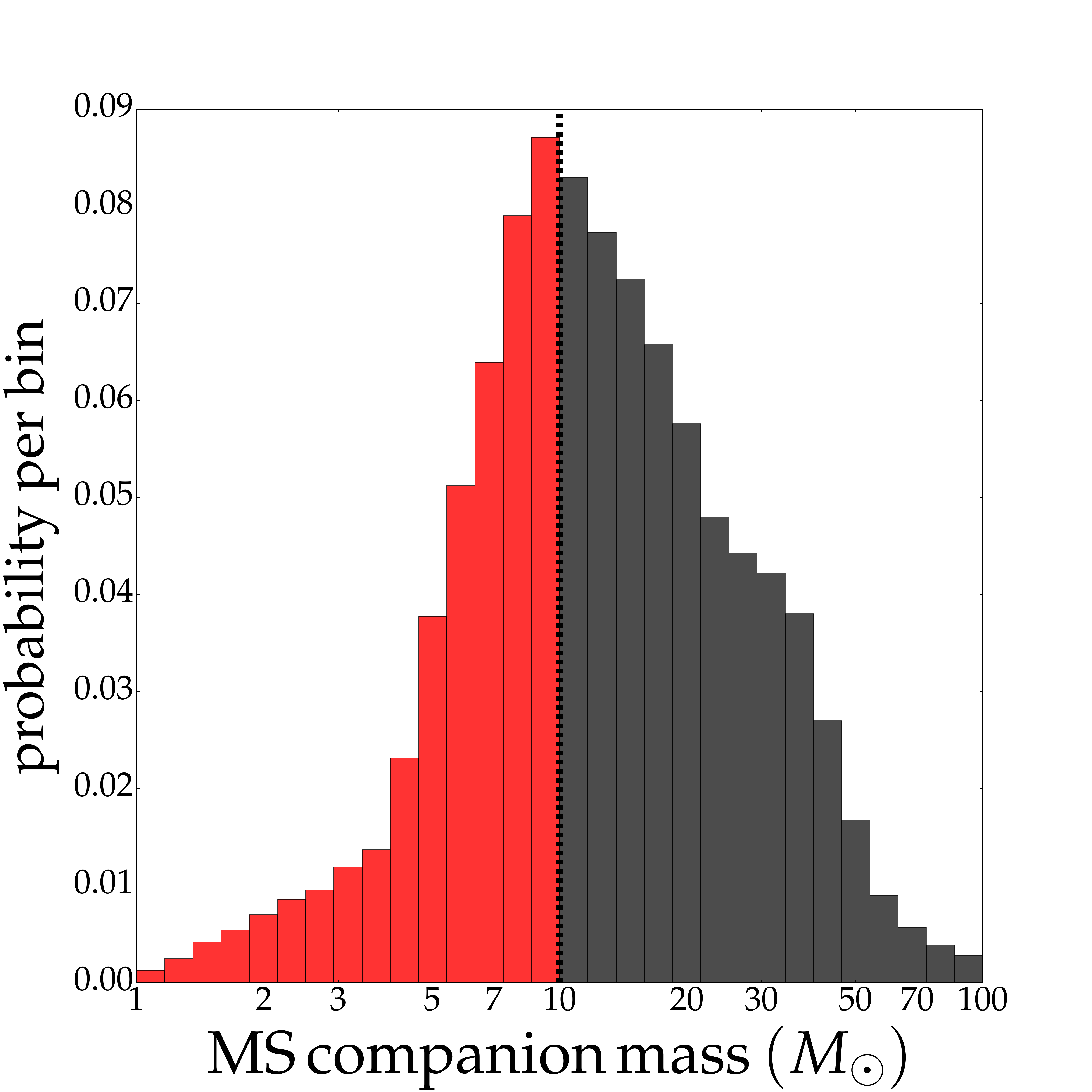}
\includegraphics[width=\onedvariable\textwidth]{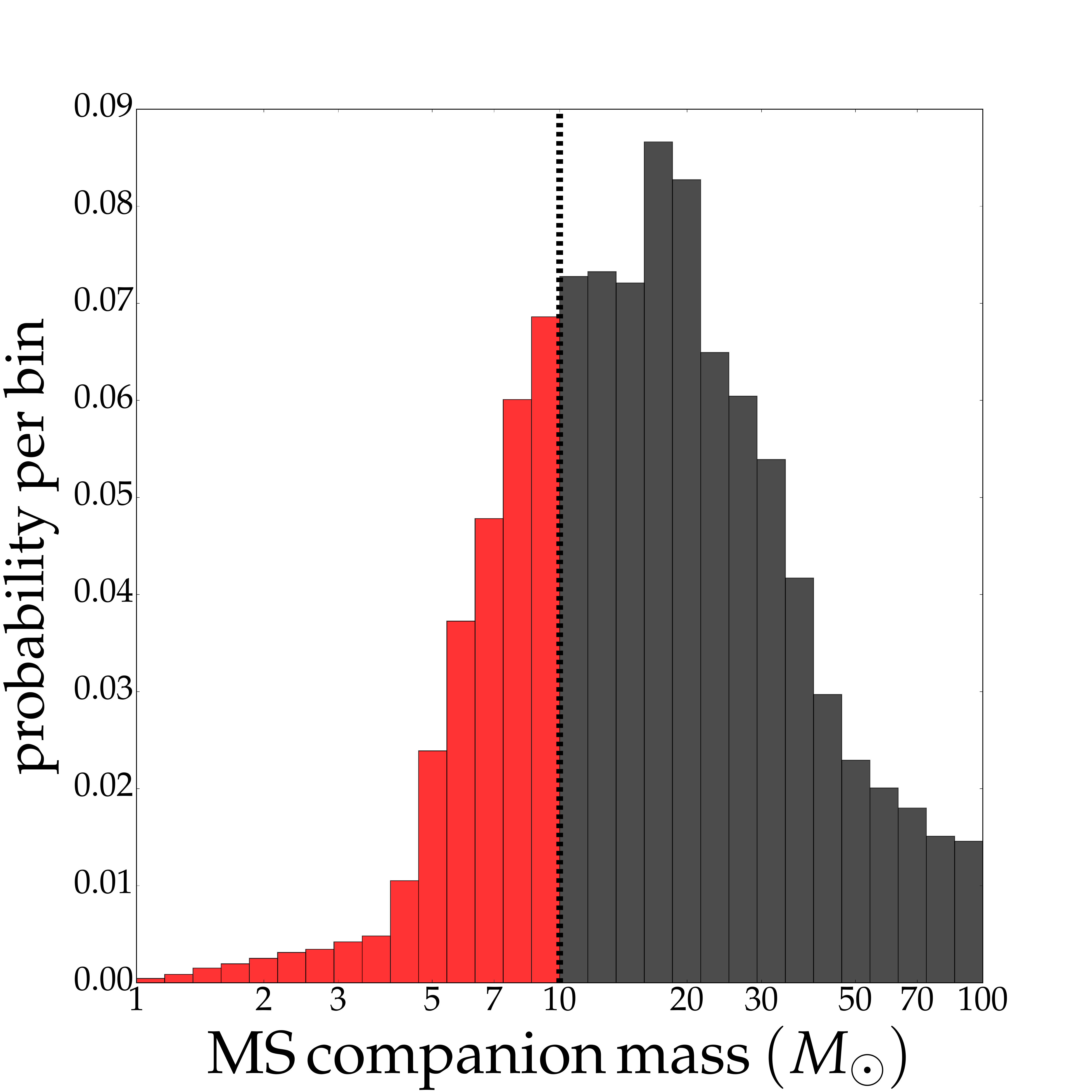}
\includegraphics[width=\onedvariable\textwidth]{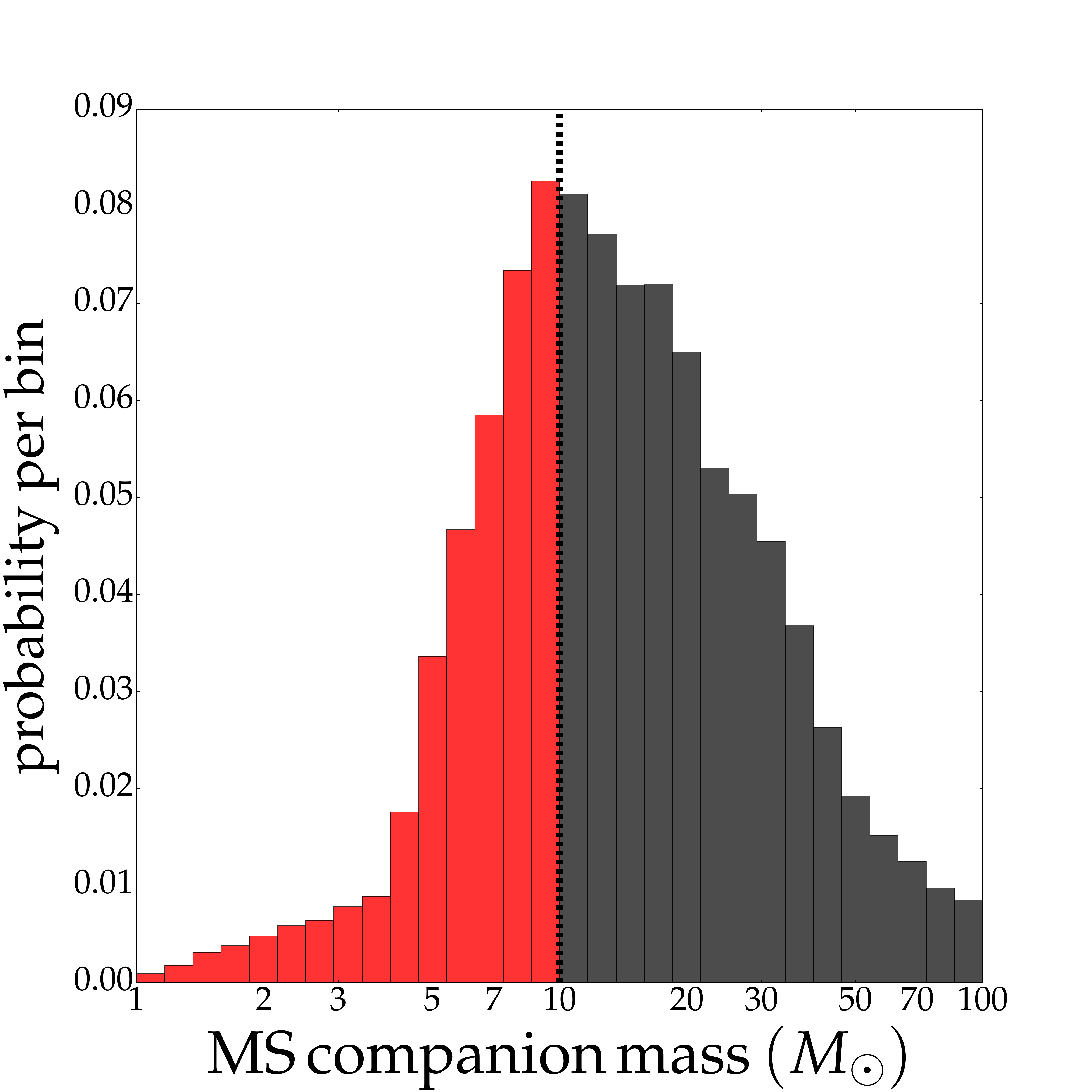}
\includegraphics[width=\onedvariable\textwidth]{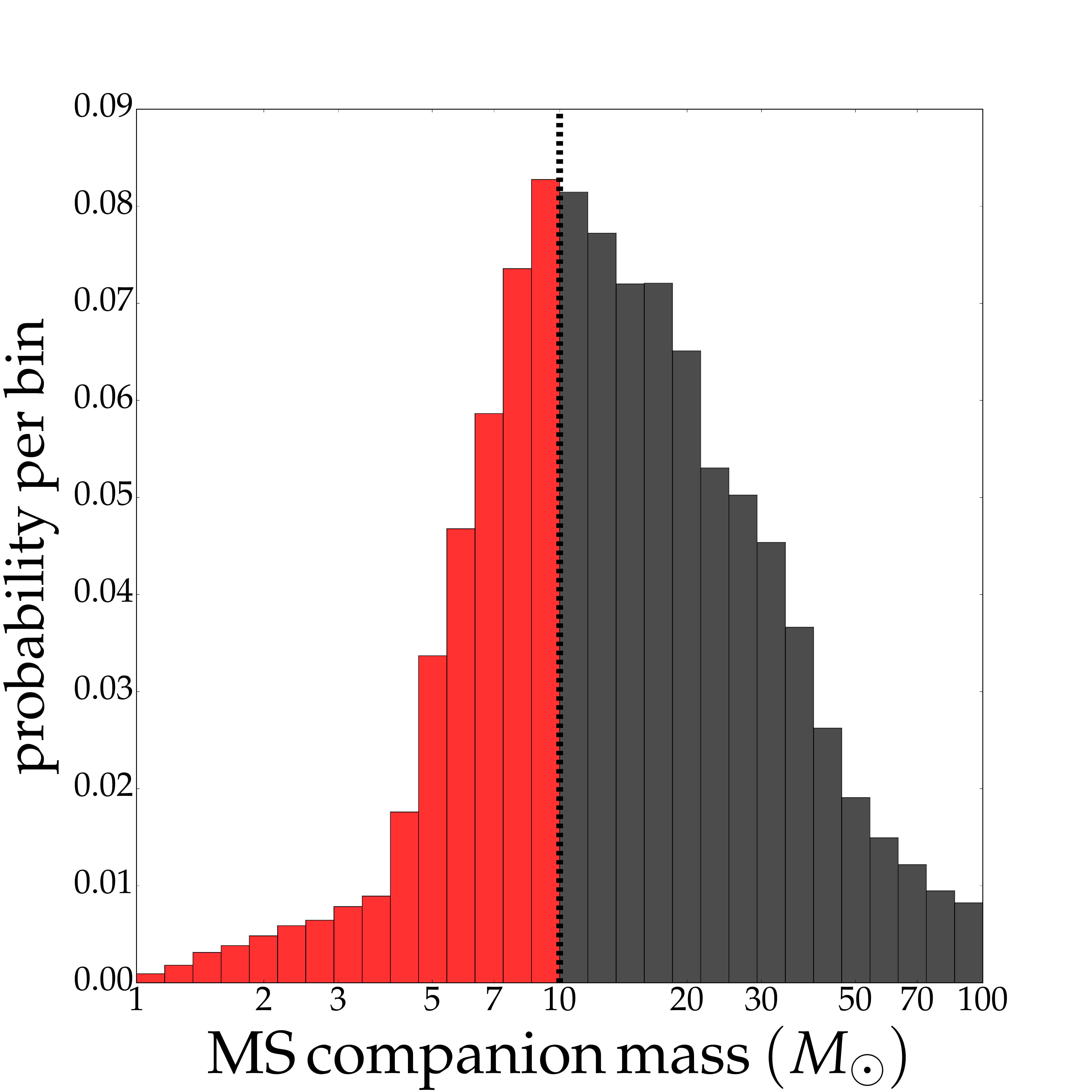}
\includegraphics[width=\onedvariable\textwidth]{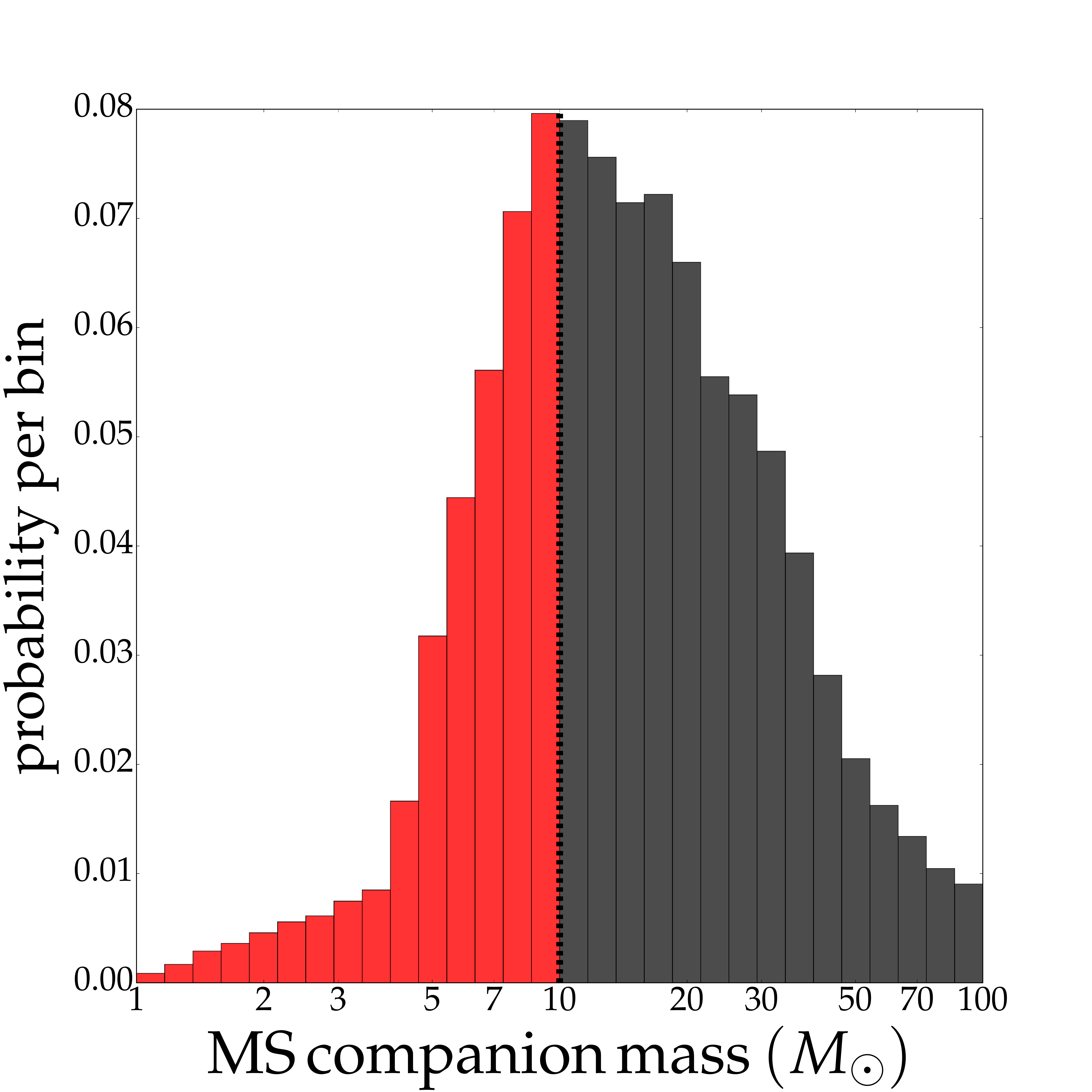}\\
\caption{Same as the reference simulation in Figure \ref{fig:metallicity} ($Z=0.014$; Model 00 of \citetalias{Zapartas+2017}) 
but with different assumed initial conditions.
  }\label{fig:initial_conditions}
\end{center}
\end{figure*}

In this section we run our simulations, changing our model parameters one by one. We determine how our results change for different metallicities, and how robust they are to variations in the assumed physical parameters and initial conditions. Results are shown in Figures~\ref{fig:metallicity} to \ref{fig:initial_conditions} and summarized in Table \ref{table_uncert}.
We discuss the main trends and differences in Section \ref{sec:discussion1}. 

\label{lastpage}

\end{document}